\documentclass[longauth]{aa}
\usepackage[varg]{txfonts}
\usepackage{graphicx}

\usepackage{natbib}
\bibpunct{(}{)}{;}{a}{}{,}
\usepackage[colorlinks,linkcolor=blue,urlcolor=blue,citecolor=blue]{hyperref} 
\usepackage{threeparttablex}
\usepackage{longtable}
\usepackage{lscape}
\usepackage{bm}
\usepackage{verbatim}
\usepackage{makecell}
\usepackage{caption}
\usepackage{subcaption}
\usepackage{multirow}
\usepackage{booktabs}
\usepackage{hyperref}
\usepackage{threeparttable}
\usepackage{colortbl}
\usepackage{xcolor}
\usepackage{booktabs}

\usepackage{graphicx}
\usepackage{txfonts}

\begin{document}

\title{COSMOS Web: Morphological quenching and size-mass evolution of brightest group galaxies from z = 3.7}
\titlerunning{BGG structural evolution and quenching}

\author{
Ghassem Gozaliasl\inst{\ref{Aalto},\ref{Helsinki}}\thanks{email: ghassem.gozaliasl@aalto.fi}\texorpdfstring{\href{https://orcid.org/0000-0002-0236-919X}{\protect\includegraphics{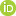}}}{}
\and
Lilan~Yang\inst{\ref{Rochester}} \and
Jeyhan~S.~Kartaltepe\inst{\ref{Rochester}} \and
Greta Toni\inst{\ref{Bologna},\ref{INAF},\ref{Heidelberg}} \and
Fatemeh Abedini\inst{\ref{IASBS}}\and
 Hollis~B.~Akins\inst{\ref{UAT}}\and
Natalie Allen \inst{\ref{DAWN},\ref{DTU}}\and
Rafael~C.~Arango-Toro\inst{\ref{LAM}} \and
Arif Babul\inst{\ref{Victoria},\ref{Infosys}} \and
Caitlin M.~Casey\inst{\ref{ucsb}, \ref{UAT}, \ref{DAWN}} \and
Nima Chartab\inst{\ref{Pasadena}} \and
Nicole E. Drakos \inst{\ref{HawaiiHilo}} \and
Andreas L. Faisst\inst{\ref{Pasadena}} \and
Alexis Finoguenov\inst{\ref{Helsinki}} \and
Carter Flayhart\inst{\ref{Rochester}} \and
Maximilien~Franco\inst{\ref{CEA},\ref{UAT}} \and
Gavin Leroy \inst{\ref{Durham}} \and
Santosh~Harish\inst{\ref{Rochester}} \and
Günther Hasinger\inst{\ref{DZA},\ref{Dresden}} \and
Hossein Hatamnia\inst{\ref{Riverside}} \and
Olivier~Ilbert\inst{\ref{LAM}} \and
Shuowen Jin \inst{\ref{DAWN},\ref{DTU}} \and
Darshan Kakkad\inst{\ref{CAR}}\and
Atousa Kalantari\inst{\ref{IASBS}}\and
Ali Ahmad Khostovan\inst{\ref{Kentucky},\ref{Rochester}} \and
Anton~M.~Koekemoer\inst{\ref{STScI}} \and
Maarit Korpi-Lagg\inst{\ref{Aalto}}\and
Clotilde Laigle\inst{\ref{IAP}} \and
Daizhong Liu\inst{\ref{PurpleMObs}} \and
Georgios Magdis\inst{\ref{DAWN},\ref{DTU}} \and
Matteo Maturi \inst{\ref{Heidelberg},\ref{ITP}} \and
Henry~Joy~McCracken\inst{\ref{IAP}} \and
Jed McKinney\inst{\ref{UAT}} \and
Nicolas McMahon\inst{\ref{Rochester}} \and
Bahram Mobasher\inst{\ref{Riverside}} \and
Lauro Moscardini\inst{\ref{Bologna},\ref{INAF},\ref{INFN}} \and
Jason Rhodes\inst{\ref{Pasadena}} \and
Brant E. Robertson\inst{\ref{Santacruz}}\and
Louise Paquereau\inst{\ref{IAP}} \and
Annagrazia Puglisi \inst{\ref{Southampton}}\and 
Rasha E. Samir \inst{\ref{NRIAG}}\and 
Mark Sargent\inst{\ref{EPFL}} \and
Zahra Sattari\inst{\ref{Pasadena}} \and
Diana Scognamiglio\inst{\ref{JPL}} \and
Nick Scoville\inst{\ref{Caltech}} \and
Marko Shuntov\inst{\ref{DAWN},\ref{NBI},\ref{UG}} \and
David B. Sanders\inst{\ref{Honolulu}} \and
Sina Taamoli\inst{\ref{Riverside}} \and
Sune Toft\inst{\ref{DAWN},\ref{NBI}} \and
Eleni Vardoulaki\inst{\ref{TLS}}
}

\institute{
Department of Computer Science, Aalto University, PO Box 15400, Espoo, FI-00076, Finland \label{Aalto}
\and
Department of Physics, University of Helsinki, P. O. Box 64, FI-00014 Helsinki, Finland \label{Helsinki}
\and
Laboratory for Multiwavelength Astrophysics, School of Physics and Astronomy, Rochester Institute of Technology, 84 Lomb Memorial Drive, Rochester, NY 14623, USA \label{Rochester}
\and
University of Bologna - Department of Physics and Astronomy “Augusto Righi” (DIFA), Via Gobetti 93/2, I-40129 Bologna, Italy \label{Bologna}
\and
INAF - Osservatorio di Astrofisica e Scienza dello Spazio di Bologna, via Gobetti 93/3, 40129 Bologna, Italy \label{INAF}
\and
Zentrum für Astronomie, Universität Heidelberg, Philosophenweg 12, 69120 Heidelberg, Germany \label{Heidelberg}
\and
Institute for Advanced Studies in Basic Sciences (IASBS)
444 Prof. Yousef Sobouti Blvd.,
Zanjan 45137-66731, Iran\label{IASBS}
\and
Aix Marseille Univ, CNRS, LAM, Laboratoire d'Astrophysique de Marseille, Marseille, France \label{LAM}
\and
Department of Physics and Astronomy, University of Victoria, BC V8X 4M6, Canada \label{Victoria}
\and
Infosys Visiting Chair Professor, Indian Institute of Science, Bangalore 560012, India \label{Infosys}
\and
Department of Physics, University of California, Santa Barbara, Santa Barbara, CA 93106 USA \label{ucsb}
\and
The University of Texas at Austin, 2515 Speedway Blvd Stop C1400, Austin, TX 78712, USA \label{UAT}
\and
Cosmic Dawn Center (DAWN), Denmark \label{DAWN}
\and
Department of Physics, University of Hawaii, Hilo, 200 W Kawili St, Hilo, HI 96720, USA \label{HawaiiHilo}
\and
Caltech/IPAC, MS 314-6, 1200 E. California Blvd. Pasadena, CA 91125, USA \label{Pasadena}
\and
Université Paris-Saclay, Université Paris Cité, CEA, CNRS, AIM, 91191 Gif-sur-Yvette, France \label{CEA}
\and
Institute for Computational Cosmology, Department of Physics, Durham University, South Road, Durham DH1 3LE, United Kingdom \label{Durham}
\and
Deutsches Zentrum für Astrophysik, Postplatz 1, 02826, Görlitz, Germany \label{DZA}
\and
TU Dresden, Institute of Nuclear and Particle Physics, 01062, Dresden, Germany; DESY, Notkestrasse 85, 22607, Hamburg, Germany \label{Dresden}
\and
Department of Physics and Astronomy, University of California, Riverside, 900 University Avenue, Riverside, CA 92521, USA \label{Riverside}
\and
School of Physics, Engineering \& Computer ScienceCentre for Astrophysics Research (CAR)Department of Physics, Astronomy and Mathematics \label{CAR}
\and
Department of Physics and Astronomy, University of Kentucky, 505 Rose Street, Lexington, KY 40506, USA \label{Kentucky}
\and
Space Telescope Science Institute, 3700 San Martin Drive, Baltimore, MD 21218, USA \label{STScI}
\and
Institut d'Astrophysique de Paris, UMR 7095, CNRS, and Sorbonne Université, 98 bis boulevard Arago, 75014 Paris, France \label{IAP}
\and
Purple Mountain Observatory, Chinese Academy of Sciences, 10 Yuanhua Road, Nanjing 210023, China \label{PurpleMObs}
\and
DTU-Space, Technical University of Denmark, Elektrovej 327, 2800 Kgs. Lyngby, Denmark \label{DTU}
\and
ITP, Universität Heidelberg, Philosophenweg 16, 69120, Heidelberg, Germany \label{ITP}
\and
INFN – Sezione di Bologna, Viale Berti Pichat 6/2, 40127, Bologna, Italy \label{INFN}
\and
School of Physics and Astronomy, University of Southampton, Highfield SO17 1BJ, UK \label{Southampton}
\and
 Department of Astronomy and Astrophysics, University of California, Santa Cruz, 1156 High Street, Santa Cruz, CA 95064, USA \label{Santacruz}
 \and
National Research Institute of Astronomy and Geophysics (NRIAG), Cairo, Egypt \label{NRIAG}
\and
EPFL Laboratory of Astrophysics (LASTRO), Observatoire de Sauverny, CH – 1290 Versoix, Switzerland \label{EPFL}
\and
Jet Propulsion Laboratory, California Institute of Technology, 4800 Oak Grove Drive, Pasadena, CA 91001, USA \label{JPL}
\and
Astronomy Department, California Institute of Technology, 1200 E. California Blvd, Pasadena, CA 91125, USA \label{Caltech}
\and
Niels Bohr Institute, University of Copenhagen, Jagtvej 128, 2200 Copenhagen, Denmark \label{NBI}
\and
University of Geneva, 24 rue du Général-Dufour, 1211 Genève 4, Switzerland \label{UG}
\and
Institute for Astronomy, University of Hawai’i at Manoa, 2680 Woodlawn Drive, Honolulu, HI 96822, USA \label{Honolulu}
\and
Thüringer Landessternwarte, Sternwarte 5, 07778, Tautenburg, Germany \label{TLS} }

\date{Received April 30, 2025; accepted x xx, xxxx}

 
  \abstract
  {
We present a comprehensive study of the structural evolution of Brightest Group Galaxies (BGGs) from redshift $z\simeq0.08$ to $z=3.7$ using the \textit{James Webb Space Telescope}'s 255h COSMOS-Web program. This survey provides deep NIRCam imaging in four filters (F115W, F150W, F277W, F444W) across $\sim0.54~\mathrm{deg}^2$ and MIRI coverage in $\sim0.2~\mathrm{deg}^2$ of the COSMOS field. High-resolution NIRCam imaging enables robust size and morphological measurements, while multiwavelength photometry yields stellar masses, SFRs, and Sérsic parameters.
We classify BGGs as star forming and quiescent using both rest-frame NUV--$r$--$J$ colors and a redshift-dependent specific star formation rate (sSFR) threshold. Our analysis reveals: (1) quiescent BGGs are systematically more compact than their star-forming counterparts and exhibit steeper size--mass slopes; (2) effective radii evolve as $R_e \propto (1+z)^{-\alpha}$, with $\alpha = 1.11 \pm 0.07$ (star-forming) and $1.40 \pm 0.09$ (quiescent); (3) star formation surface density ($\Sigma_{\mathrm{SFR}}$) increases with redshift and shows stronger evolution for massive BGGs ($\log_{10}(M_\ast/M_\odot) \geq 10.75$); (4) in the $\Sigma_*$--sSFR plane, a structural transition marks the quenching process, with bulge-dominated systems comprising over 80\% of the quiescent population. These results highlight the co-evolution of structure and star formation in BGGs, shaped by both internal and environmental processes, and establish BGGs as critical laboratories for studying the baryonic assembly and morphological transformation of central galaxies in group-scale halos.
}


   \maketitle
%

\section{Introduction}

Galaxy groups occupy a unique position in the hierarchy of cosmic structures, bridging the gap between isolated galaxies and massive clusters. Although clusters (with halo masses $M_{\text{halo}} \gtrsim 10^{14} \, M_{\odot}$) are often the focus of studies in extreme environments, groups (typically $10^{13} \lesssim M_{\text{halo}} \lesssim 10^{14} \, M_{\odot}$) are far more abundant and contribute $\sim$30\%--50\% of the total mass budget in the universe \citep[e.g.,][]{Cui2024,2021MNRAS.500..120P,2019ApJ...872....2B}. Their intermediate mass scale and low velocity dispersion ($\sim100-500 \; km s^{-1}$) make them a crucial environment for probing hierarchical structure formation, as physical processes such as galaxy mergers, gas accretion, and quenching operate in a distinct regime compared to both isolated galaxies and rich clusters.

Brightest Group Galaxies (BGGs), the most massive and luminous galaxies residing at the centers of these groups, serve as key tracers of co-evolution between galaxies and their dark matter halos. Forming early and growing via a combination of gas accretion, satellite mergers, and feedback-driven quenching, BGGs occupy the bottom of their group's potential well, making them sensitive probes of large-scale structure assembly and the environmental processes shaping galaxy evolution \citep{Darragh2019MNRAS.489.5695D, Jung2022, 2023MNRAS.525.5677S,Gozaliasl2016,Gozaliasl2018,Gozaliasl2020,Gozaliasl2024,Einasto2024}.

Although the structural evolution of Brightest Cluster Galaxies (BCGs) has been studied extensively beyond $z>1$ redshifts \citep[e.g.][]{Stott2011,Yang2024}, BGGs have remained comparatively less explored, particularly in the distant universe. This discrepancy may arise because BCGs, typically residing in dense clusters, are more easily identified in observations than BGGs in less massive groups at higher redshifts. However, understanding BGGs is crucial, as they bridge the evolutionary trajectory between central galaxies in isolated halos and those embedded in massive clusters. Studies have shown that the sizes of BCGs and massive elliptical galaxies evolve with redshift, often exhibiting more compact morphologies at earlier epochs \citep{Nelson02,bernardi2009evolution,Ascaso2014}. However, the extent to which BGGs follow similar trends and how their size evolution correlates with star formation activity and stellar mass remains an open question.

Quiescent (or quenched) galaxies---defined by their heavily suppressed specific star formation rates ($\mathrm{SFR}/M_*$) relative to the star-forming ``main sequence'' \citep[e.g.,][]{DADDI07,NOESKE07}---host approximately half of the stellar mass in the local universe \citep{BALDRY04}, and have been observed in large numbers out to $z\sim2$ \citep[e.g.,][]{ILBERT13,MUZZIN13,DAVIDZON17}. These galaxies are characterized not only by their low sSFRs but also by compact morphologies, with effective radii ($R_e$) typically $\sim30$--$50\%$ smaller than star-forming galaxies of comparable mass \citep[e.g.,][]{SHEN03,CIBINEL15}, and exhibit more prominent spheroidal components (S\'{e}rsic indices $n \gtrsim 2$--4). 

BGGs are particularly interesting in this context, as they are predominantly quiescent systems, with studies showing $\sim70$--$90\%$ quiescent fractions at $z \lesssim 1$ \citep{gozaliasl2016brightest}. Their early quenching timescales and structural parameters (e.g., high stellar mass surface densities $\Sigma_* \gtrsim 10^9$ M$_\odot$ kpc$^{-2}$) suggest they may represent the progenitors of today's BCGs, with their size evolution ($R_e \propto (1+z)^{-1.0\pm0.3}$) potentially driven by minor mergers \citep{Lidman12}. This connection makes BGGs crucial for understanding the formation of the most massive galaxies, like BCGs, in the universe. 

At the highest stellar masses ($M_* \gtrsim 10^{11} M_\odot$), this population is dominated by central galaxies in group-scale halos, which exhibit particularly strong quenching signatures \citep{PENG12,WEINMANN06}. The observed properties of these galaxies are shaped by two key processes: (1) the passive evolution of their stellar populations after quenching, characterized by gradual dimming and reddening as stars age \citep[][]{TACCHELLA15,CAROLLO16}, and (2) their distinct formation history through early dissipative collapse \citep{DEKEL09} followed by hierarchical growth via satellite accretion. These evolutionary paths are fundamentally governed by their environmental context, as clearly demonstrated in their observed scaling relations \citep[e.g.,][]{vanDerBurg2014, Kravtsov2018, Gozaliasl2014}. This dual dependence on both internal evolution and environmental factors makes these systems particularly valuable for studying galaxy-halo co-evolution.

Observations indicate that massive galaxies undergo significant structural evolution, with their effective radii increasing by a factor of $\sim2-3$ from $z=2$ to $z=0$ \citep[e.g.,][]{WILLIAMS10,BELLI15}. For star-forming disks, this growth is linked to gradual gas accretion and inside-out assembly \citep[e.g.,][]{MO98,OESCH10,MOSLEH12,SHIBUYA15}, while for quiescent galaxies, gas-poor mergers are the dominant channel of size and mass growth \citep[e.g.,][]{TOFT07,BELLI14,KRIEK09}. Several studies have noted a threshold mass near $\log(M_\star/M_\odot) \sim 11$, above which dissipationless (dry) mergers become increasingly important in size growth \citep{PENG10,POGGIANTI13}, whereas below this mass, the size increase of the quiescent population is primarily due to the continuous addition of larger galaxies that have been quenched at later epochs \citep{CAROLLO13,CASSATA13,SARACCO14}. 
This phenomenon is also supported by the stellar age differences between compact (older) and extended (younger) quiescent galaxies at fixed mass \citep{SARACCO11,BELLI15,FAGIOLI16}.

These structural and demographic changes imply that different quenching mechanisms operate across the mass spectrum. While gradual quenching through gas depletion may preserve galaxy morphology, rapid mechanisms such as gas-rich mergers can induce morphological compaction and starbursts within short ($\sim$100-200 Myr) timescales \citep[e.g.,][]{BARRO13,ZOLOTOV15}. The majority of simulations and models agree that AGN feedback plays a critical role in quenching massive galaxies, both by suppressing star formation and driving the development of elliptical morphologies \citep[e.g.,][]{DUBOIS15,CHOI15,CHOI17}. Identifying the relative contributions of these physical processes (mergers, gas depletion, AGN feedback) remains a major challenge in observational cosmology \citep[e.g.,][]{BIRNBOIM03,CROTON06,DELUCIA12,Tacchella2016,DEKEL14}. Morphological and structural indicators such as provide essential constraints on this evolutionary route.

By combining deep \textit{JWST}/NIRCam imaging (probing rest-frame optical morphologies even at $z > 4$) with multiwavelength photometry and structural modeling, we robustly quantify the size--mass relation for BGGs across cosmic time. Although NIRCam enables detection of galaxies up to $z \sim 7-8$, our analysis is limited to $z < 4$ for BGGs due to: (1) the challenge of reliably identifying group halos and their central galaxies in the epoch of initial group assembly ($z > 4$), and (2) the requirement for sufficient sample sizes to statistically characterize the size--mass relation. This redshift cutoff ensures robust environmental classification while capturing the critical phase of BGG growth from peak star formation to quenching.Yet, this redshift range surpasses what was previously accessible for studying BGGs before the JWST launch. 

Recent work by \citet{Yang2025} used the COSMOS-Web survey to measure rest-frame optical sizes of galaxies from $2 < z < 10$, revealing that star-forming galaxies maintain a nearly constant size--mass slope and surface density relation, while quiescent systems show steeper structural scaling and a clear compactness threshold at $\log \Sigma_* \sim 9.5$–$10\,M_\odot\,\mathrm{kpc}^{-2}$.

In parallel, \citet{Faisst2017} examined the structural evolution of both star-forming and quiescent massive galaxies ($\log(M_\star/M_\odot) > 11.4$) using the COSMOS/UltraVISTA survey. They found a remarkably uniform size evolution across cosmic time, supporting a scenario where rapid quenching below $z \sim 2$ is accompanied by significant structural transformations including central starbursts and compactification. This compact phase is subsequently followed by post-quenching size growth, predominantly through dry minor mergers---a process particularly efficient in massive systems due to their satellite-rich environments and AGN-mediated gas depletion. However, alternative mechanisms such as AGN feedback-induced ``puffing up'' of stellar cores  or tidal stripping in group environments may also contribute to size evolution. The mass-dependence of these processes reveals a critical transition scale above $\log(M_\star/M_\odot) \sim 11.4$, where internal processes (AGN feedback, mergers) and environmental effects dominate over secular evolution.

\cite{vdW2014} used 3D-HST and CANDELS data to study the size–mass relation of galaxies from z = 0 to 3, finding that early-type galaxies are more compact than late-types at all redshifts. They reported a steeper size evolution for early types (R$_{\mathrm{eff}} \propto (1+z)^{-1.48}$) compared to late types ($R_{\mathrm{eff}} \propto (1+z)^{-0.75}$), and a steeper size–mass slope for early types. We compare our BGG sample with their results to explore how group environments influence galaxy structure.

In this study, we build on our COSMOS-Web galaxy group catalog \citep{Toni2025}, which we constructed using deep, high-resolution JWST/NIRCam imaging with covering an area of 0.54 deg$^2$. Excluding masked regions (like bright stars), we utilized the AMICO algorithm \citep{Maturi2019} over 0.45 deg$^2$ to identify 1678 galaxy groups (halo masses < $10^{14} M\odot$) up to $z = 3.7$, achieving a highly pure and complete JWST-based group catalog, the largest and most comprehensive so far. This data set provides a unique foundation for studying BGGs in the $\sim12$ Gyr of cosmic history in a wide range of environments. We investigate how their structural properties evolve over $z\sim 0.08 - 3.7$. We focus on measuring rest-frame optical sizes, separating star-forming and quiescent galaxies based on both color--color diagnostics \citep{Ilbert2013} and specific star-formation rate (sSFR) thresholds \citep{Pacifici2016,Yang2025}. We explore the evolution of the size--mass relation, quantify the size growth as a function of redshift at fixed stellar mass, and study the star formation rate surface density ($\Sigma_{\mathrm{SFR}}$) as a complementary probe of compactness and star formation efficiency.

This paper is organized as follows. In Section~\ref{sec:data}, we describe the data, sample selection, and structural measurements. In Section~\ref{sec:method}, we outline our methodology for classifying the computation of structural properties. In Section~\ref{sec:results}, we present our main results on the size evolution, scaling relations, and $\Sigma_{\mathrm{SFR}}$.  Finally, Section~\ref{sec:conclusions} summarizes our conclusions. Throughout this paper, we adopt a flat $\Lambda$CDM cosmology with parameters $H_0=67.66 $\,km\,s$^{-1}$\,Mpc$^{-1}$, $\Omega_{\rm m,0}= 0.30966 $ and $\Omega_{\Lambda,0}= 0.68884$) consistent with Planck2018 \citep{aghanim2020planck}, and All magnitudes are expressed in the AB system \citep{Oke1974}, for which a flux $f_\nu$ in $\mu$Jy
($10^{-29}$~erg~cm$^{-2}$s$^{-1}$Hz$^{-1}$) corresponds to AB$_\nu=23.9-2.5\,\log_{10}(f_\nu/{\rm \mu Jy})$.
\section{Galaxy and group dataset}\label{sec:data}
\subsection{The COSMOS-Web survey} \label{sec:style}

The COSMOS-Web survey represents the largest observational program undertaken with JWST during Cycle 1, covering a total area of 0.54 deg$^2$ with the Near-Infrared Camera (NIRCam) in four filters: F115W, F150W, F277W, and F444W. In addition, it includes 0.19 deg$^2$ of Mid-Infrared Instrument (MIRI) imaging in the F770W band \citep[PI: Kartaltepe \& Casey;][]{Casey2023}. The NIRCam imaging reaches 5$\sigma$ depths for point sources of 26.6--27.3 mag (F115W), 26.9--27.7 mag (F150W), 27.5--28.2 mag (F277W), and 27.5--28.2 mag (F444W), measured within 0.15\arcsec-radius apertures. MIRI observations achieve depths of 25.33--25.98 mag within 0.3\arcsec-radius apertures.

A detailed summary of the data reduction is provided in \citet{Franco2024}, with full descriptions forthcoming in Franco et al. (in prep) and Harish et al. (in prep).

Observations were carried out across three main epochs: January 2023, April 2023, and December 2023 to January 2024, with additional pointings completed in April/May 2024. The final NIRCam mosaics\footnote{\url{https://cosmos.astro.caltech.edu/page/cosmosweb}} are available at three pixel scales: 20\,mas, 30\,mas and 60\,mas. In this study, we use high-resolution 30\,mas mosaics for structural measurements.

\subsection{COSMOS-Web photometric catalog} \label{sec:catalog}

The construction of the COSMOS-Web multiwavelength photometric catalog is described in detail by Shuntov et al. (in prep). Here we briefly summarize the aspects relevant to our analysis.

Photometric extraction is performed using the \textsc{SourceXtractor++} (\textsc{SE++}) package \citep{Bertin2020, Kummel2020}, an advanced version of the widely adopted \textsc{SExtractor} software \citep{Bertin1996}. \textsc{SE++} applies parametric S\'ersic profile fitting to all detected sources simultaneously across 33 filters spanning ground- and space-based imaging. The fits enforce a consistent structural model across bands, yielding robust flux measurements and average morphological parameters across the full wavelength range.

To complement these measurements and facilitate the analysis of wavelength-dependent morphology, an alternative structural catalog (Yang et al., in prep) provides independent S\'ersic fits in each of the four NIRCam bands. This work uses these measurements to derive rest-frame structural parameters, as described in Section~\ref{sec:method}.

Spectral energy distributions (SED) and physical parameters for all sources are derived using the \textsc{LePHARE} template-fitting code \citep{Arnouts02, Ilbert06}, based on photometry from the \textsc{SE++} extraction. The SED models are built from \citet{BC03} stellar population synthesis templates with a variety of star formation histories, ages, and dust attenuation laws \citep{Calzetti00, Arnouts2013, Salim18}. Emission lines and intergalactic medium (IGM) absorption are modeled using the prescriptions of \citet{Saito20}, \citet{Schaerer09}, and \citet{Madau95}. Photometric redshifts are computed from the redshift probability distribution functions, with the median value adopted for each source. Physical properties such as stellar mass and star formation rate (SFR) are subsequently derived at the fixed redshift.

A comparison with spectroscopic redshifts from the field \citep{Khostovan2025} confirms the high accuracy of the \textsc{LePHARE} redshifts, with a normalized median absolute deviation (NMAD) scatter of $\sigma_{\text{NMAD}} \approx 0.013$ for sources with F444W magnitudes brighter than 25.0:
\[
\sigma_{\text{NMAD}} = 1.48 \times \mathrm{median} \left( \frac{|\Delta z - \mathrm{median}(\Delta z)|}{1 + z_{\mathrm{spec}}} \right), \quad \Delta z = z_{\mathrm{phot}} - z_{\mathrm{spec}}.
\]
Further validation of stellar mass estimates is provided through comparisons with results from \textsc{CIGALE} \citep{Boquien19}, which implements non-parametric star formation histories and alternative dust attenuation models (Shuntov et al. in prep) and \cite{Arango-Toro2024}). These comparisons, presented in \citet{Shuntov2024}, show consistent results across methods.

For this work, we adopt the stellar mass and SFR values derived using \textsc{LePHARE}.

\subsection{The COSMOS-Web galaxy groups catalog} \label{sec:cosmos_groups}

This study uses the recently published COSMOS-Web galaxy group catalog \citep{Toni2025}, the largest and deepest galaxy group sample constructed to date, based on JWST Cycle 1 observations. The catalog spans in the full COSMOS Web area of 0.54 deg$^2$ (an effective area of 0.45 deg$^2$ excluding masked regions) and covers the redshift range , allowing detailed studies of the evolution of galaxies across $\sim$ 12 Gyr of cosmic time.

Group detection was performed using the Adaptive Matched Identifier of Clustered Objects (AMICO) algorithm \citep{Bellagamba2018, Maturi2019MNRAS}, a matched filter technique optimized for extracting clustered galaxy signals from photometric data without reliance on color or spectroscopic priors. The algorithm was applied to a cleaned and high-quality galaxy catalog derived from the COSMOS-Web photometric catalog \citep{Shuntov2024}, which benefits from deep JWST NIRCam imaging (F115W, F150W, F277W, F444W) combined with extensive multi-wavelength ground- and space-based data across over 30 bands.

Photometric redshifts were computed using the \textsc{LePHARE} template-fitting code with high accuracy, achieving a precision of $\sigma_{\mathrm{NMAD}} < 0.03$ up to $z \sim 4$ even at faint magnitudes ($\mathrm{F444W} < 28$). Additional quality cuts were applied to ensure robust structural and SED-based measurements, resulting in a galaxy sample of 389,248 sources used as input to the group detection.

\begin{figure}
    \centering
    \includegraphics[width=0.95\linewidth]{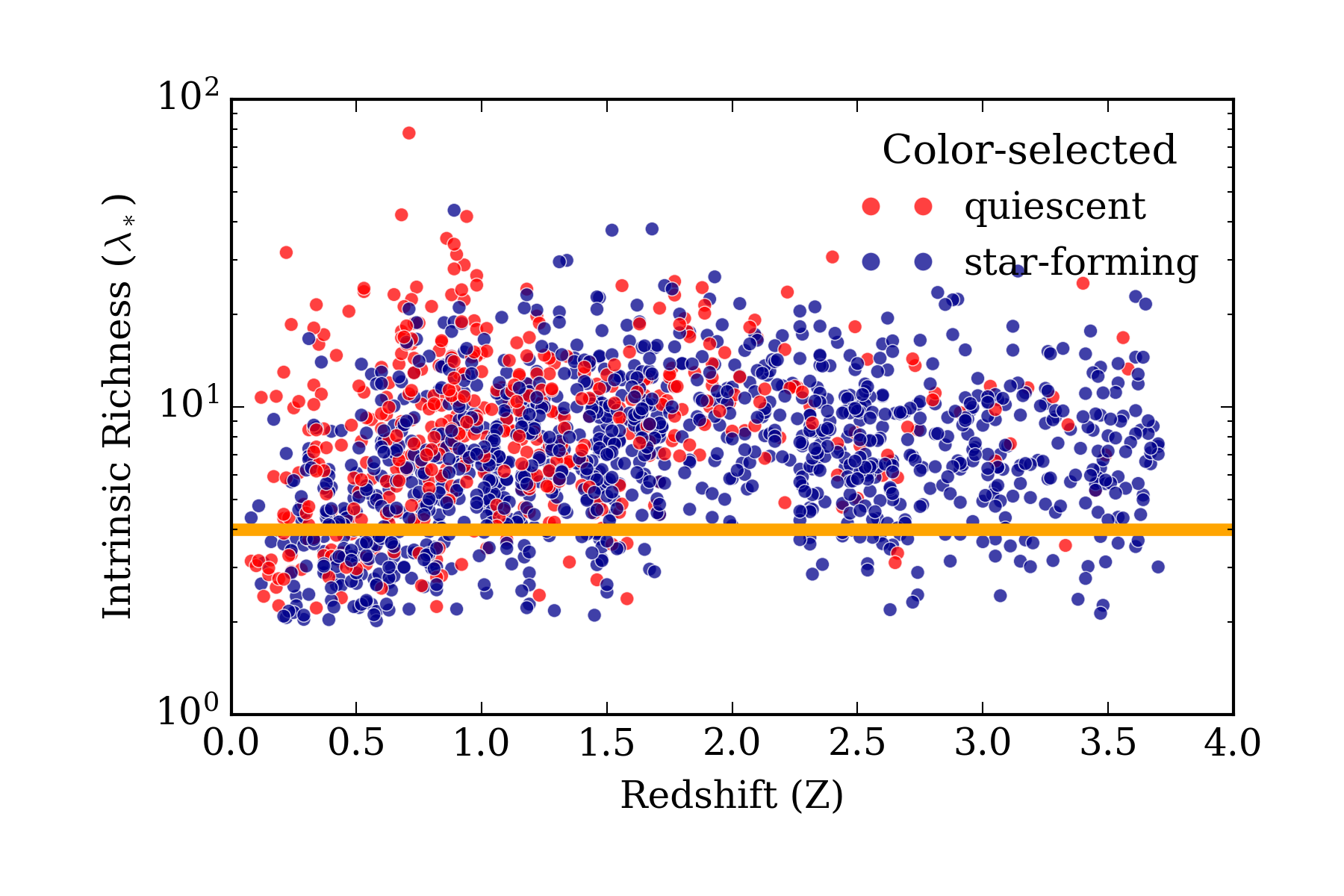}
    \vspace{+0.1cm}
    \caption{The intrinsic richness, $\lambda*$, for the sample of detected groups and its trend with redshift, color-color classified BGGs as quiescent and star-forming are shown in red and blue points, respectively. Dashed horizontal orange line represents the level of $\lambda*$ limit we apply in this study.}
    \label{fig:lambda}
\end{figure}

The final catalog consists of 1678 group detections with a signal-to-noise ratio $\mathrm{S/N_{nocl}} > 6.0$, where $\mathrm{S/N_{nocl}}$ is a refined metric that excludes shot noise from cluster/group members ($nocl$ indicates that no cluster members are included in the S/N measurement). Detections cover a broad range of masses, with intrinsic richness ($\lambda_\star$) values extending to include low-mass systems. Intrinsic richness ($\lambda_\star$) is defined as the sum of the membership probabilities of galaxies, considering only those within the virial radius ($R_{200}$) and restricted to a luminosity range of $m_\star + 1.5$, where $m_\star$ represents the magnitude of the luminosity function of the knee.
The purity and completeness of the catalog were rigorously validated using data-driven mock simulations through the SinFoniA framework \citep{Maturi2023Aa}, demonstrating a purity exceeding 80\% for
$\mathrm{S/N_{nocl}} > 9.5$. Figure \ref{fig:lambda} illustrates the intrinsic richness of the group against the redshift, we apply a $\lambda_\star=4$ limit for our group sample used in this study. Points indicate groups with star-forming and quiescent BGGs, which are discussed later.

Each group entry includes key observables such as photometric redshift, sky position, amplitude, richness proxies, membership probabilities, and flags for spectroscopic validation and data quality. More than 500 groups have been spectroscopically confirmed through cross-matching with a comprehensive compilation of redshifts in the COSMOS field \citep{Khostovan2025}. Importantly, this catalog pushes group detection into the protocluster regime, identifying over 300 new structures in $z > 2$ and assembling over 100 candidate large-scale systems through 3D clustering of high-$z$ protocluster cores.

\subsection{Selection of brightest group galaxies} \label{sec:bgg_selection}
Traditionally, BCGs and BGGs have been identified through either luminosity-based selection (choosing the most luminous galaxy in a given band, typically r-band) or stellar mass-based selection (selecting the most massive galaxy in the system) \citep{DeLucia2007, Koester2007, Gozaliasl2019}. Notably, in defining fossil groups, the r-band magnitude determines the gap between the magnitudes of the two brightest members\citep{Ponman1994}.
We identified BGGs in the COSMOS-Web galaxy group catalog \citep{Toni2025} using a hybrid selection method that combines stellar mass and luminosity criteria. Building on \cite{Gozaliasl2019, Gozaliasl2020}, we address two key limitations of traditional approaches: (1) the spatial offset of BGGs from group centers in low-mass systems, and (2) the contamination by starburst galaxies in luminosity-based selections.

Our methodology processes galaxies within fixed apertures of 250 kpc and 500 kpc (also tested at 750 kpc) from the group centers, implementing a two-stage selection algorithm. First, we filter the galaxy sample by applying a redshift constraint, requiring $\Delta z < 0.05(1+z)$ relative to the group candidate, to ensure the selection remains confined to the vicinity of the group in redshift space. Among this filtered sample, we identify the luminosity-selected BGG as the galaxy with the lowest apparent magnitude $m$ in the F150W band, unless the magnitude difference between the first- and second-brightest galaxies is less than 0.5 mag; in that case, we choose the one with the highest membership probability.  
 
In parallel, we repeat the selection using the median stellar mass derived from \texttt{LePhare} instead of magnitude. Here, we select the most massive galaxy as the mass-selected BGG, unless the logarithmic mass difference between the most massive and the second-most massive galaxy is smaller than 0.25 dex; in that case, we again select the one with the higher membership probability. For all detections, we store both the first and second candidates according to these criteria. This approach ensures that each group has both a mass-selected and a luminosity-selected BGG, providing a robust cross-check of group-centric galaxy identification.

For each group, we evaluated two primary candidates: the galaxy with the highest median stellar mass and the brightest galaxy. We then compare the stellar mass and r-band luminosity of both candidates for a given group. The selection incorporates aperture-dependent corrections, preferentially choosing 250 kpc candidates when: (a) for mass-selected BGGs, the mass difference is $\leq$ 0.1 dex; and (b) for luminosity-selected BGGs, the magnitude difference is $\leq$ 0.15 mag relative to 500 kpc candidates.

The final hybrid selection applies a tiered decision tree:
\begin{enumerate}
    \item Selects the mass-dominant galaxy if its stellar mass exceeds the luminosity-selected candidate by $\geq$ 0.1 dex
    \item Chooses the luminosity-dominant galaxy if it is brighter by an equivalent mass margin
    \item Defaults to the brighter galaxy when mass differences are $<$ 0.1 dex, weighting recent star formation
\end{enumerate}

 Validation includes cross-matching with a master galaxy catalog to eliminate spurious detections, producing three output catalogs: pure mass-selected, pure luminosity-selected, and hybrid-selected BGGs. A total of 1,294 BGGs were selected on the basis of stellar mass, ensuring the identification of the most massive galaxies within each group. Additionally, 384 BGGs were selected using a luminosity-based criterion, which identifies the brightest galaxies. For a more comprehensive selection that accounts for both mass and luminosity, we included 803 BGGs selected within a 250 kpc radius from the group center, and 875 BGGs selected within a 500 kpc radius. This combined approach ensures a more accurate representation of the BGG population, minimizing selection biases and capturing galaxies with both high stellar mass and luminosity.

\begin{figure}[h]
    \centering
    \includegraphics[width=0.95\columnwidth]{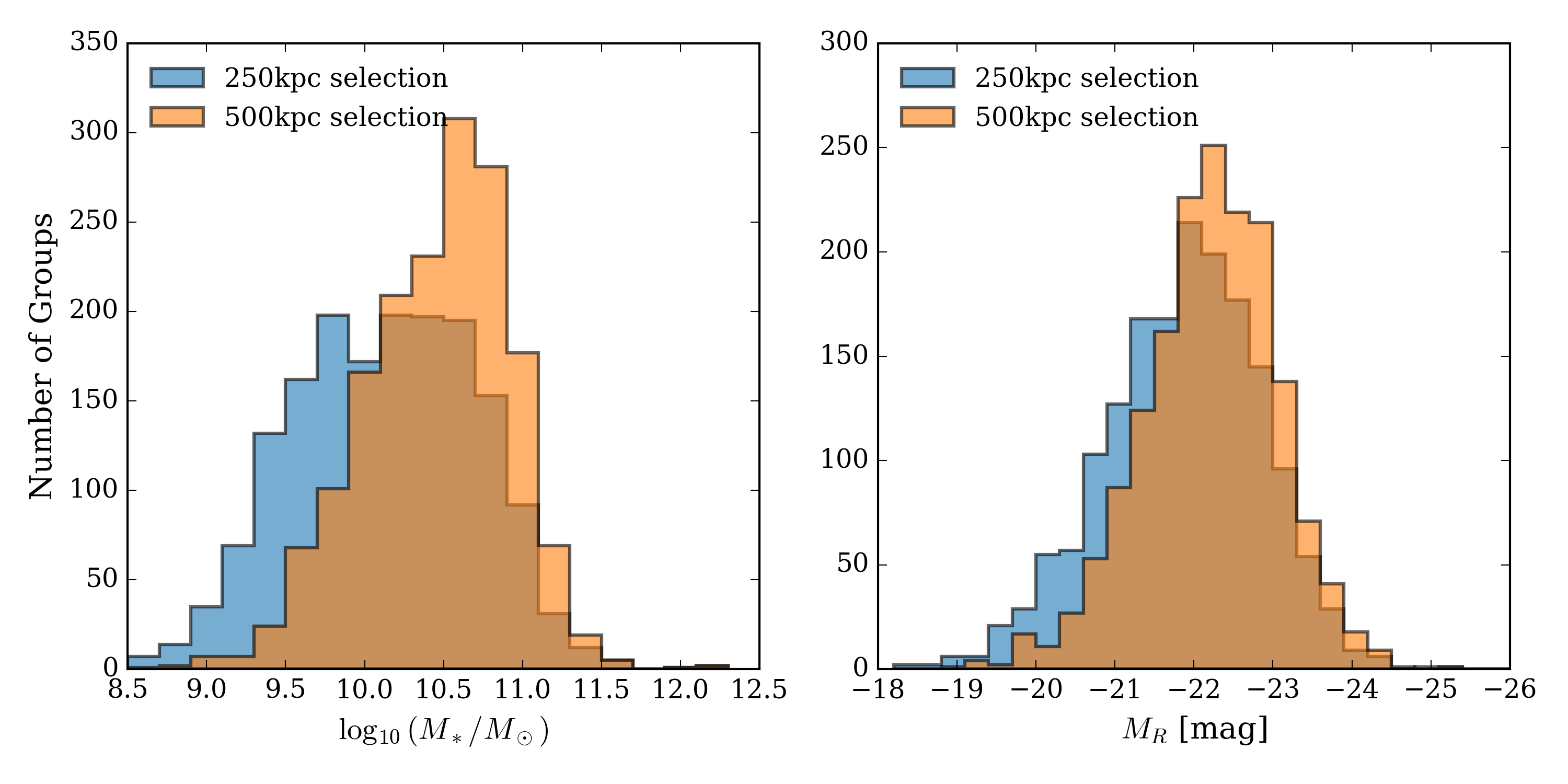}
    \includegraphics[width=0.95\columnwidth]{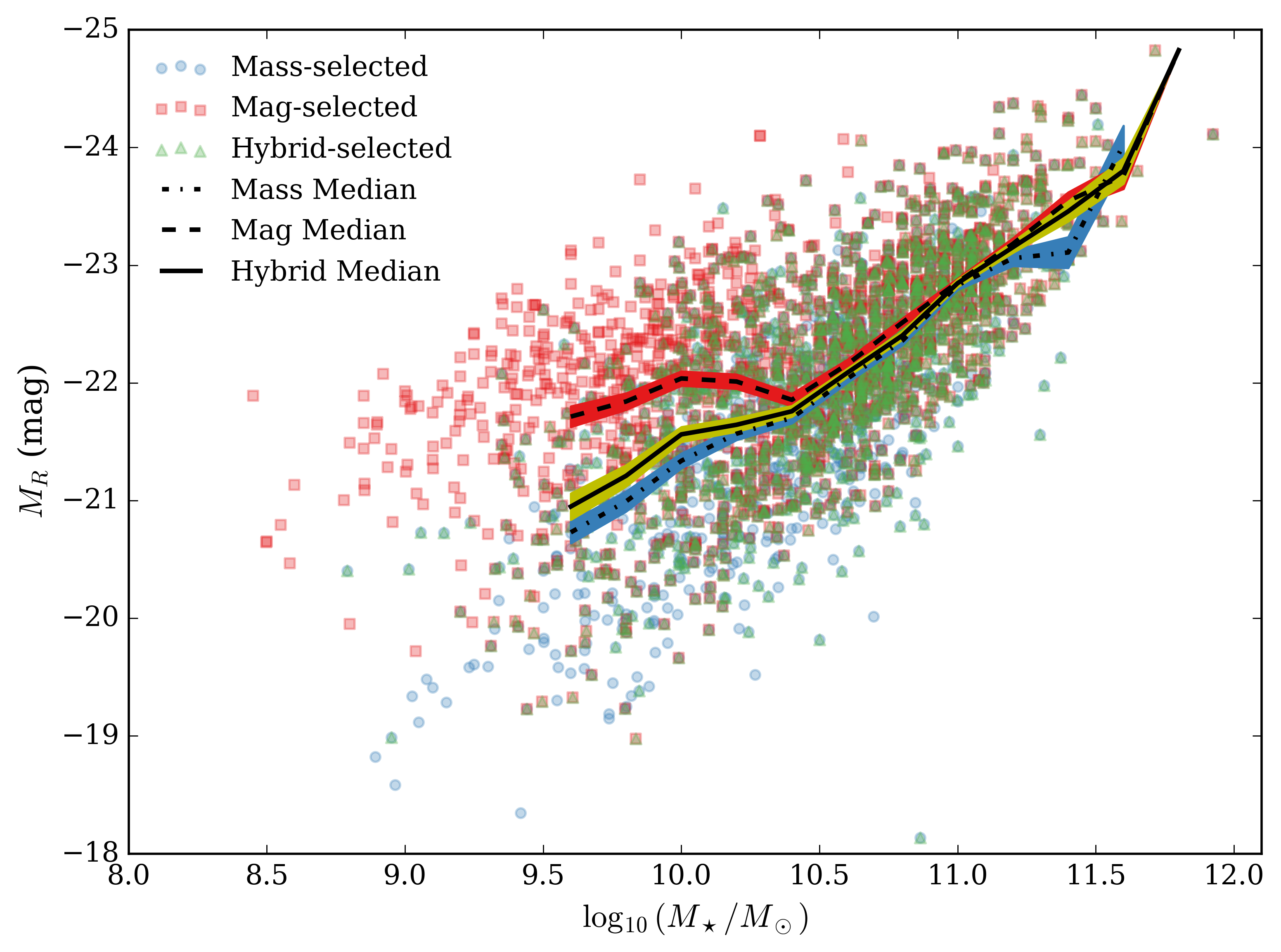}
    \caption{Selection and characterization of BGGs in the COSMOS-Web galaxy group catalog. The top panels show histograms of the number of groups as a function of log stellar mass $(log_{10}(M_*/M_\odot)$ and absolute magnitude ($M_R$), comparing selections based on two different aperture sizes: 250 kpc (blue) and 500 kpc (orange). The bottom panel displays a scatter plot of absolute magnitude ($M_R$) versus log stellar mass $(log_{10}(M_*/M_\odot)$, with points color-coded by selection method: mass-selected (blue circles), magnitude-selected (red squares), and hybrid-selected (green triangles). The solid lines represent the medians for each selection type: mass-selected (blue), magnitude-selected (red), and hybrid-selected (black). The shaded regions around the median lines show the corresponding confidence bands. The figure illustrates how the hybrid selection method balances stellar mass and luminosity to better capture the true BGG population, especially at the high-mass end.}
    \label{fig:mr_vs_z}
\end{figure}

As shown in Fig. \ref{fig:mr_vs_z}, a luminosity-based selection at the low stellar mass end misses some true BGGs, whereas our hybrid method ensures a more complete sample by balancing both criteria. The accompanying histograms demonstrate how smaller apertures (250 kpc) bias selections toward lower luminosity and mass values, frequently misidentifying satellites as BGGs. By combining mass and luminosity metrics across multiple apertures, we significantly reduce selection biases and more accurately represent the true BGG population across diverse group environments, from compact to loosely bound systems.

Each galaxy group in the catalog is detected using the AMICO algorithm, and galaxies are assigned membership probabilities based on their photometric redshifts and spatial distribution. We consider galaxies with a group membership probability, ensuring a high-confidence sample of true group members. For further information on determining the group membership and assigning a membership probability for each group galaxy we refer the reader to \cite{Toni2025}.

To further ensure the robustness of our group and the selection of BGG, we restricted our analysis to groups rich in biodiversity $\lambda_{\mathrm{star}} > 4$ (see also Fig. \ref{fig:lambda} ), corresponding to groups with enough mass members to minimize contamination and projection effects. We also apply a stellar mass threshold of $\log(M_{*}/M_{\odot}) > 9$ to ensure completeness across the redshift range.

\subsection{Classification of BGGs: star-forming and quiescent} \label{sec:bgg_classification}

To investigate the evolutionary trends of BGGs, we classify them into SFGs and QGs using a combination of rest-frame color-color selection and sSFR thresholds. This dual approach allows for a more robust classification by accounting for both photometric and physical star formation indicators.

To investigate the structural and evolutionary differences between star-forming and quiescent BGGs, we classify our sample using three complementary approaches: (i) a rest-frame color--color diagram, (ii) a redshift-dependent sSFR threshold, and (iii) a consensus method combining both criteria. This multitiered classification ensures robustness against contamination by dusty star-forming galaxies and transitional systems.

\subsubsection{Color--color classification: NUV--$r$--$J$ diagram}

We employ the rest-frame $M_{\mathrm{NUV}} - M_r$ versus $M_r - M_J$ (NUV--$r$--$J$) color--color diagram to separate passive from active systems. Following \citet{Ilbert2013}, a galaxy is considered quiescent if it satisfies both of the following conditions:
\begin{align}
M_{\mathrm{NUV}} - M_r &> 3(M_r - M_J) + 1 \nonumber \\
\text{and} \quad M_{\mathrm{NUV}} - M_r &> 3.1.
\end{align}
This method effectively distinguishes red, passively evolving systems from those dominated by ongoing star formation or dust-reddened emission. The classification is calibrated and most reliable within the redshift range $0 < z < 4$, where the photometric bands and derived rest-frame colors are well constrained by the available multiwavelength data.
\subsubsection{sSFR Selection: Redshift-dependent threshold}
\label{sec:ssfr_threshold}

To complement the color-based classification and mitigate contamination from dust-obscured star-forming galaxies, we apply a redshift-dependent threshold on the sSFR, following the formalism of \citet{Pacifici2016}. A BGG is classified as quiescent if:
\begin{equation}
\log_{10}(\mathrm{sSFR}) < \log_{10}\left( \frac{0.2}{t_{\mathrm{obs}}(z)} \right),
\end{equation}
where $t_{\mathrm{obs}}(z)$ is the age of the Universe at the galaxy’s redshift, expressed in Gyr, and sSFR is in units of yr$^{-1}$. This threshold evolves with cosmic time and reflects the declining global star formation rate, providing a more physically motivated and redshift-aware definition of quiescence.

\begin{figure}[ht]
    \centering
    \includegraphics[width=0.9\linewidth]{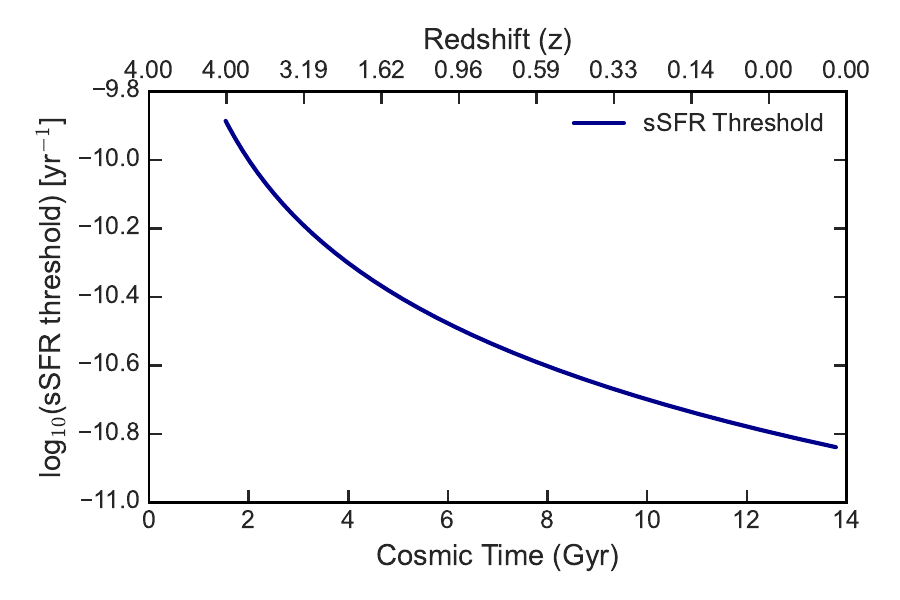}
    \caption{
        Redshift-dependent sSFR threshold used to classify BGGs as quiescent, defined as $\log_{10}(\mathrm{sSFR}) < \log_{10}(0.2 / t_{\mathrm{obs}})$. The bottom x-axis shows cosmic time in Gyr, while the top x-axis shows the corresponding redshift. The threshold decreases smoothly from early to late cosmic epochs, tracing the decline in global star formation efficiency.
    }
    \label{fig:ssfr_threshold}
\end{figure}

Figure~\ref{fig:ssfr_threshold} illustrates the evolution of the sSFR threshold as a function of cosmic time (bottom axis) and redshift (top axis), based on the adopted $\Lambda$CDM cosmology. The threshold becomes more stringent at earlier epochs, reflecting the higher star formation activity of galaxies in the early Universe. At later times (lower redshifts), the sSFR threshold declines steadily, consistent with the cosmic decline in star formation rate density.

\subsubsection{Consensus classification: combined color and sSFR}

Although the NUV--$r$--$J$ diagram and the sSFR threshold each provide effective classification schemes, discrepancies can arise due to photometric uncertainties or atypical dust attenuation. Therefore, we define a final high-confidence classification based on consensus.
\begin{itemize}
    \item \textbf{Quiescent BGGs (QGs):} must satisfy both the color--color and sSFR quiescent criteria.\\
    \item \textbf{Star-Forming BGGs (SFGs):} must fail both criteria.\\
    \item BGGs meeting only one criterion are excluded from the final classification to avoid ambiguity.
\end{itemize}
This conservative approach ensures a clean division between star-forming and passive systems and enhances the reliability of subsequent evolutionary analyses. Table \ref{tab:bgg_counts} compares the number of QGs and SFGs in different redshift bins among different classification methods.

\begin{figure*}[h]
    \centering
    \includegraphics[width=.7\linewidth]{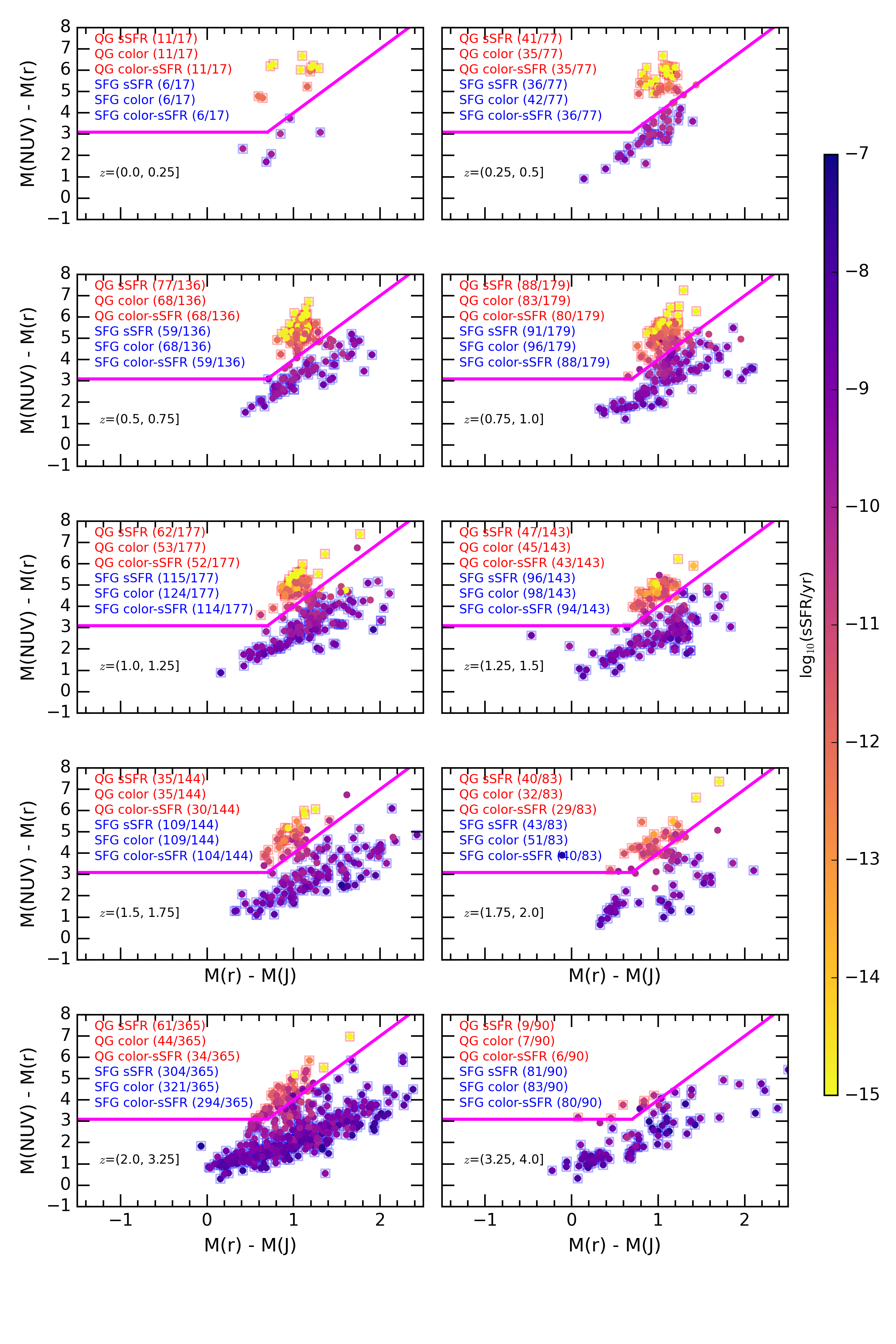}
    \caption{
        Rest-frame NUV--$r$ versus $r$--$J$ color--color diagram for BGGs across eight redshift bins from $z=0$ to $z=4$, each of width $\Delta z=0.5$. BGGs are color-coded by their $\log_{10}(\mathrm{sSFR}/\mathrm{yr}^{-1})$. Magenta lines delineate the quiescent region defined by $M_{\mathrm{NUV}} - M_r > 3(M_r - M_J) + 1$ and $M_{\mathrm{NUV}} - M_r > 3.1$. In each panel, we report the number of quiescent and star-forming BGGs based on sSFR-only, color-only, and the combined (color + sSFR) criteria. Open red and blue squares denote quiescent and star-forming BGGs, respectively, that meet both thresholds. A clear redshift trend emerges: the fraction of star-forming BGGs dominates at high redshift ($z \gtrsim 2$), while quiescent systems increase in prevalence at lower redshifts.
    }
    \label{fig:bgg_nuvrj_ssfr}  
\end{figure*}

\subsubsection{Final classified sample}

Applying the combined classification to our stellar-mass-selected sample of BGGs ($\log M_\ast/M_\odot > 9$ and $\lambda_\mathrm{star} > 4$), we obtain a clean and conservative division into star-forming and quiescent systems across the redshift interval $0.08 < z < 3.7$. This refined classification serves as the basis for our structural analysis in subsequent sections. It allows us to explore the redshift evolution of the size--mass relation separately for SFGs and QGs while minimizing the impact of classification uncertainties.

Figure~\ref{fig:bgg_nuvrj_ssfr} presents the distribution of BGGs in the rest-frame NUV--$r$ versus $r$--$J$ color space across eight redshift bins from $z = 0.0$ to $z= 3.7.0$ with $\Delta z = 0.5$. Each galaxy is color-coded by its $\log_{10}(\mathrm{sSFR})$, providing insight into its recent star formation activity. The magenta lines show the quiescent region, as defined by the NUV--$r$--$J$ criteria.

In each panel, we annotate the number of BGGs classified as quiescent or star-forming according to the three distinct methods.

The open red and blue squares in each panel mark the BGGs that meet both quiescent and star-forming definitions, respectively, and represent the clean sample used for subsequent analysis. We also performed the analysis for two color-color and redshift-dependent sSFR thresholds. 

As seen in the panels:
\begin{itemize}
    \item At higher redshifts ($z \gtrsim 2$), the majority of BGGs reside in the star-forming region of the diagram and exhibit high sSFRs.
    \item Toward lower redshifts ($z \lesssim 1.5$), an increasing number of BGGs populate the quiescent region, indicating a buildup of massive, quenched systems over cosmic time.
    \item The combined classification method (color+sSFR) filters out transitional or dusty systems, ensuring a more secure division of BGGs into star-forming and quiescent types.
\end{itemize}

The total number of BGGs identified as star-forming or quiescent in each redshift bin is as shown in Tab. \ref{tab:bgg_counts}.

\begin{table*}[h]
\centering
\caption{Number of BGGs classified as star-forming or quiescent using color, sSFR, and combined criteria  presented in Sec. \ref{sec:bgg_classification} across redshift bins.}
\begin{tabular}{ccccccc}
\toprule
Redshift bins & SF(sSFR) & SF(color) & SF (color+sSFR) & QG(sSFR) & QG(color) & QG (color+sSFR) \\
\midrule
(0.0, 0.25] & 6 & 6 & 6 & 11 & 11 & 11 \\
(0.25, 0.5] & 36 & 42 & 36 & 41 & 35 & 35 \\
(0.5, 0.75] & 59 & 68 & 59 & 77 & 68 & 68 \\
(0.75, 1.0] & 91 & 96 & 88 & 88 & 83 & 80 \\
(1.0, 1.25] & 115 & 124 & 114 & 62 & 53 & 52 \\
(1.25, 1.5] & 96 & 98 & 94 & 47 & 45 & 43 \\
(1.5, 1.75] & 109 & 109 & 104 & 35 & 35 & 30 \\
(1.75, 2.0] & 43 & 51 & 40 & 40 & 32 & 29 \\
(2.0, 3.25] & 304 & 321 & 294 & 61 & 44 & 34 \\
(3.25, 4.0] & 81 & 83 & 80 & 9 & 7 & 6 \\
\hline
\end{tabular}
\label{tab:bgg_counts}
\end{table*}

This classification framework ensures a high-purity sample of star-forming and quiescent BGGs, free from potential contamination by dusty starbursts or intermediate systems. Robust selection improves the reliability of subsequent structural analyzes, such as the size-mass relation and the size evolution presented in Sections~\ref{sec:size_mass_relation} and~\ref{sec:size_evol_fixed_mass}.

\section{The structural measurement of BGGs}\label{sec:method}
\subsection{The size of BGGs } \label{sec:size_measurement}
To quantify the structural properties of BGGs, we perform two-dimensional S\'ersic profile fitting on JWST/NIRCam imaging. For a comprehensive description of the size measurement pipeline and validation, we refer the reader to \citet{Yang2025}, who conducted a systematic structural analysis of galaxies in the COSMOS-Web survey. Here we summarize the most relevant aspects of their methodology, which we adopt for this work.

The surface brightness distribution of each galaxy is modeled using a single-component S\'ersic function \citep{Sersic1968}:
\begin{equation}
I(r) = I_0 \exp \left[ -b_n \left(\frac{r}{R_e}\right)^{1/n} - 1 \right],
\end{equation}
where $I_0$ is the central surface brightness, $R_e$ is the effective (half-light) radius, $n$ is the S\'ersic index that defines the concentration of the light profile, and $b_n$ is a normalization constant dependent on $n$. The elliptical radius is defined as $r = \sqrt{x^2 + y^2/q^2}$, where $q$ is the axis ratio of the light distribution.

We adopt the measurements from the COSMOS-Web structural catalog generated with the \texttt{Galight} software package \citep{Ding2020}, which is built on the \texttt{Lenstronomy} lens modeling framework \citep{Birrer2021}. This pipeline fits galaxy light profiles while simultaneously modeling neighboring objects and incorporating accurate PSF convolution.

Following \citet{Yang2025}, the PSF for each NIRCam filter is constructed using empirical star stacks from the COSMOS-Web field, achieving spatial precision down to $\sim$0.03 arcsec. Galaxies are fit in all four NIRCam filters (F115W, F150W, F277W, F444W), and rest-frame optical sizes are selected based on redshift to ensure consistency across cosmic time. Postage stamp images are extracted with a size of at least $5 \times R_{e,\mathrm{SE}}$ (measured from Source Extractor) to fully encompass the galaxy light.

Only galaxies with high-quality fits are retained, based on the following criteria from \citet{Yang2025}: (1) no parameters that reach the fitting limits, (2) reduced $\chi^2 < 2$, and (3) clean morphological classification free of neighboring contamination. The structural parameters are validated against high-resolution size measurements from COSMOS HST ACS F814W mosaics \citep{Koekemoer07}, which provide a robust optical benchmark over the $1.64~\mathrm{deg}^2$ COSMOS field. These HST observations offer a median point spread function (PSF) of $\sim0.09''$, allowing precise size determinations to reach faint magnitudes ($i_{\mathrm{AB}} \sim 25$). Extensive image simulations and cross-checks show that the derived sizes from our analysis exhibit a typical uncertainty in $\log(R_e)$ of 0.1--0.2 dex and no significant bias as a function of redshift, size or magnitude. This careful calibration ensures that the measured structural parameters are directly comparable to legacy HST results and suitable for tracing galaxy evolution across cosmic time.

The \citet{Yang2025} catalog provides robust, homogeneous structural measurements across a wide redshift baseline, making it ideal for studying the evolution of galaxy size--mass relations. In our work, we extract BGG sizes directly from this catalog and use them to analyze the morphological evolution of the most massive galaxies in group environments.

The galaxy light profiles are fitted using cutouts from the COSMOS-Web mosaics, adopting a 30\,mas pixel scale. The cutout size is typically five times the SE++-based source radius, with a minimum of 30 pixels and a maximum of 200 pixels to balance computational efficiency and coverage. Each cutout is accompanied by a noise map derived from the ERR extension of the JWST image, which accounts for background, readout, and Poisson noise components.

To ensure robust modeling, contaminating sources near the BGG are either masked or fitted simultaneously with additional S\'ersic components. The parameter space is constrained to physically meaningful ranges: $R_e$ between 0.01 arcsec and the image size, $n$ between 0.3 and 9, and $q$ between 0.1 and 1. These bounds accommodate the diverse morphologies of galaxies while avoiding extreme or unphysical solutions.

The PSF is constructed using \texttt{PSFEx} \citep{Bertin2011} based on empirical stars from the NIRCam images. Accurate PSF modeling is critical for deconvolving the intrinsic light distribution of compact galaxies, particularly at high redshifts.

\begin{figure}[h!]
    \centering

        \includegraphics[width=0.5\textwidth]{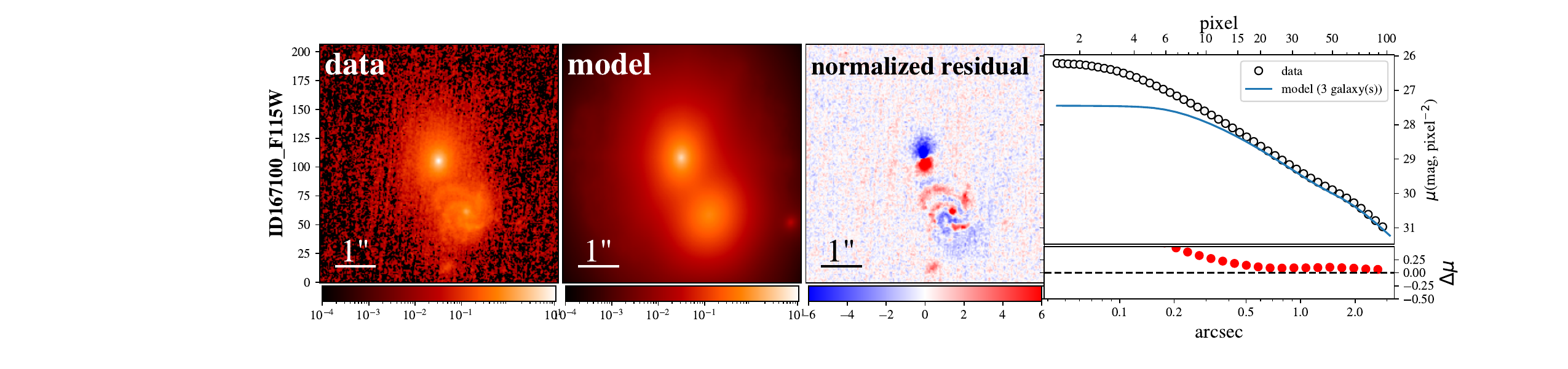} 
        \includegraphics[width=0.5\textwidth]{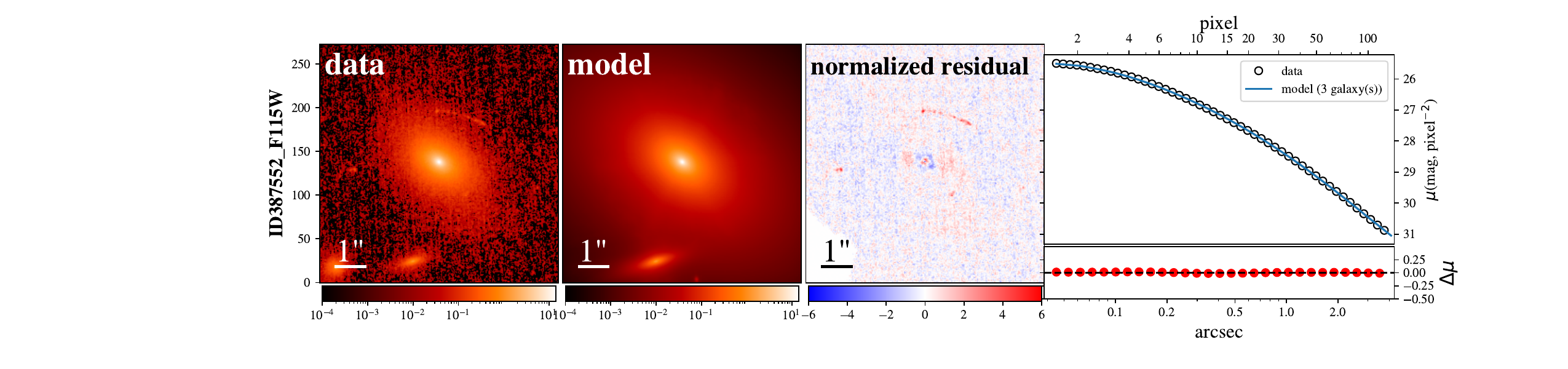}
        \includegraphics[width=0.5\textwidth]{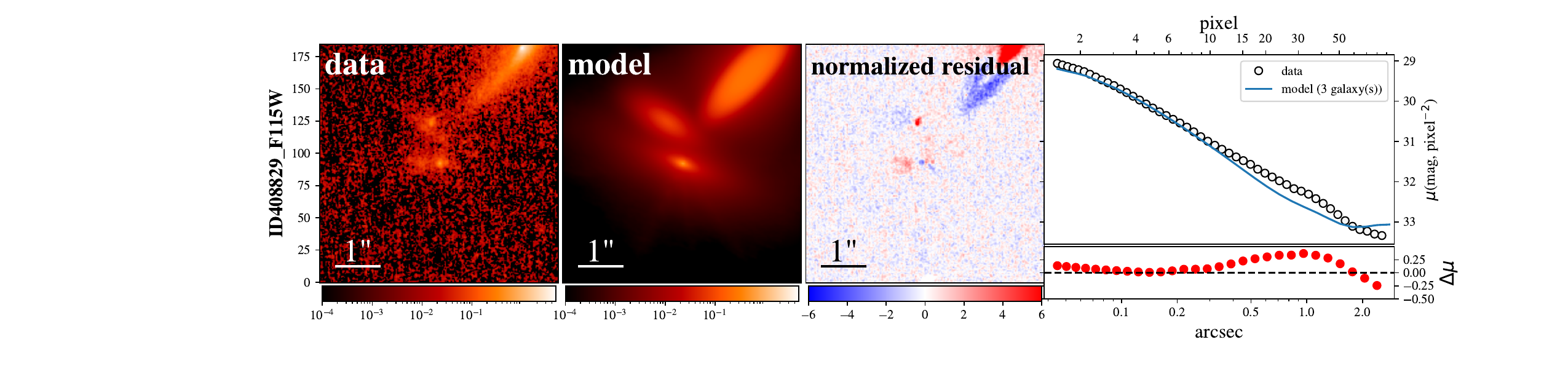} \\
        \includegraphics[width=0.5\textwidth]{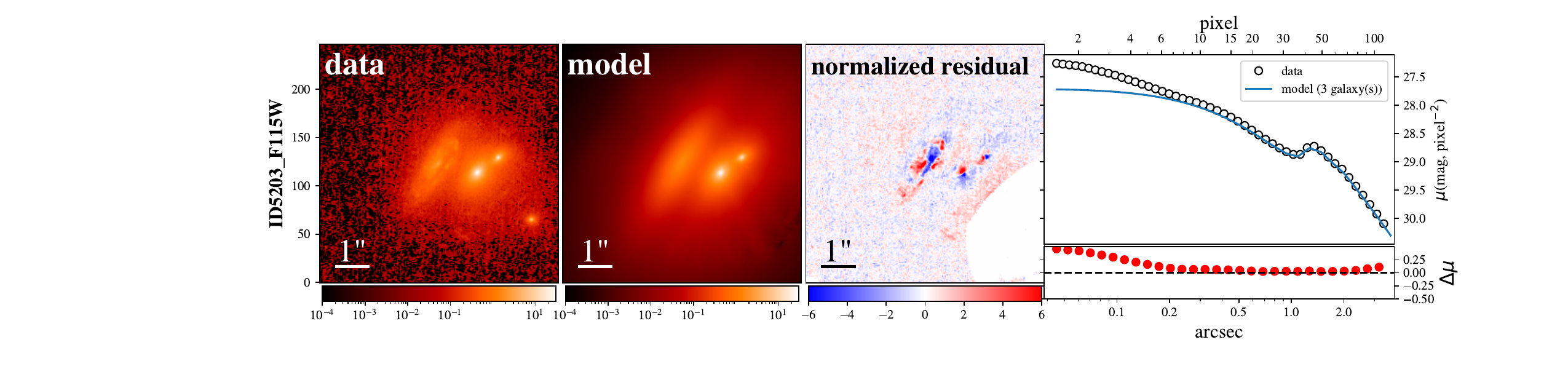}
        \includegraphics[width=0.5\textwidth]{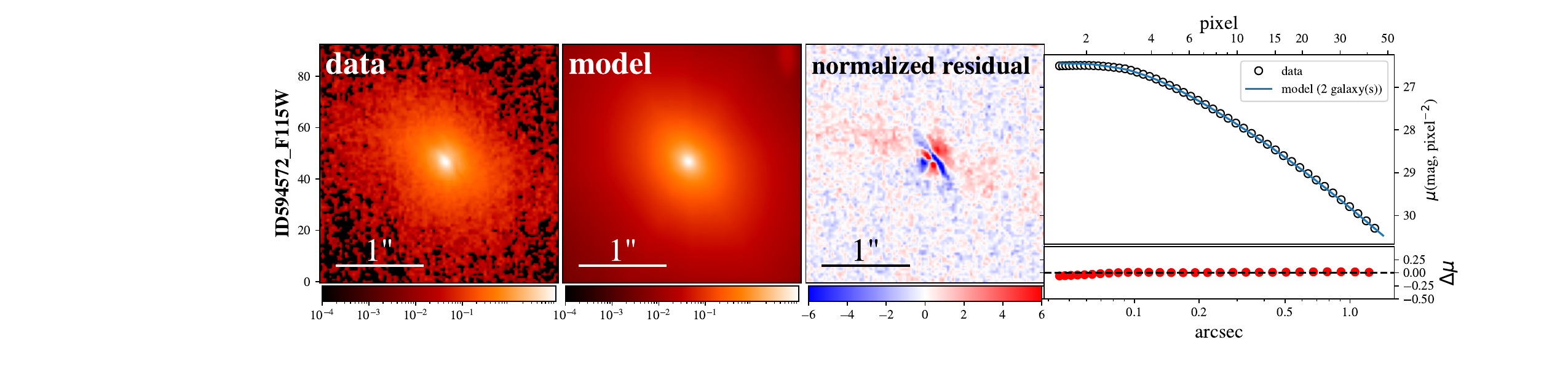} 
        \includegraphics[width=0.5\textwidth]{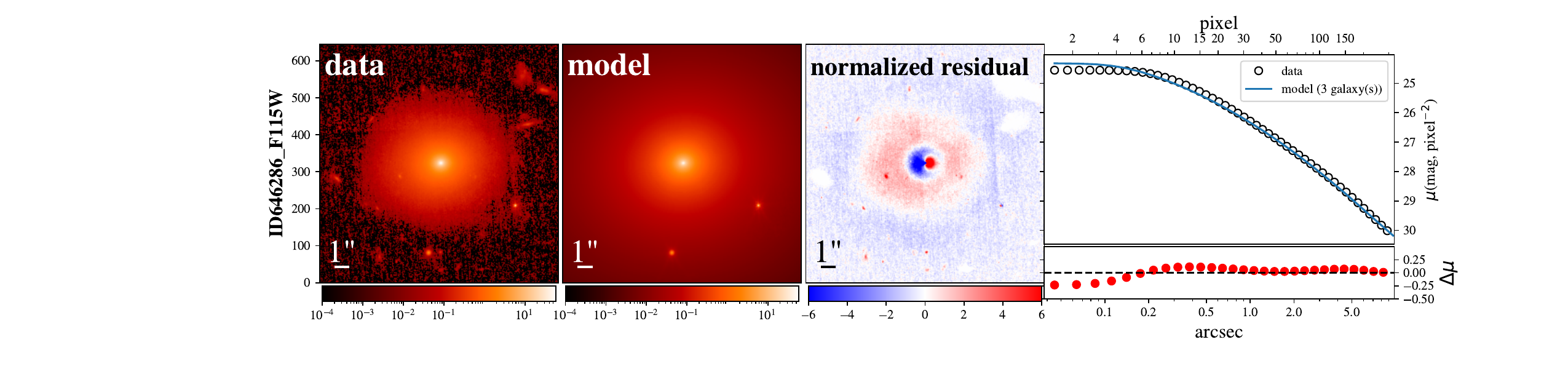}

    \caption{F115W Band: Sersic fit for six galaxies across the band.}
    \label{fig:F115W_band}
\end{figure}

Fits with reduced $\chi^2 > 15$ or those that exceed parameter boundaries are flagged as unreliable and excluded from further analysis.
To visually illustrate the quality and diversity of our structural modeling, Figure~\ref{fig:F115W_band} presents example S'ersic profile fits for six representative BGGs in the F115W band. Each row displays the original data image, the best-fit model, and the normalized residuals, followed by radial surface brightness profiles and fit residuals. These examples highlight the ability of our method to accurately capture the light profiles of both regular and disturbed galaxies across a broad redshift range. We find that the majority of BGGs are well-modeled with a single S'ersic component, while the residuals remain small and symmetric in most cases, confirming reliable fits.

\begin{figure*}
    \centering
    \includegraphics[width=0.9\linewidth]{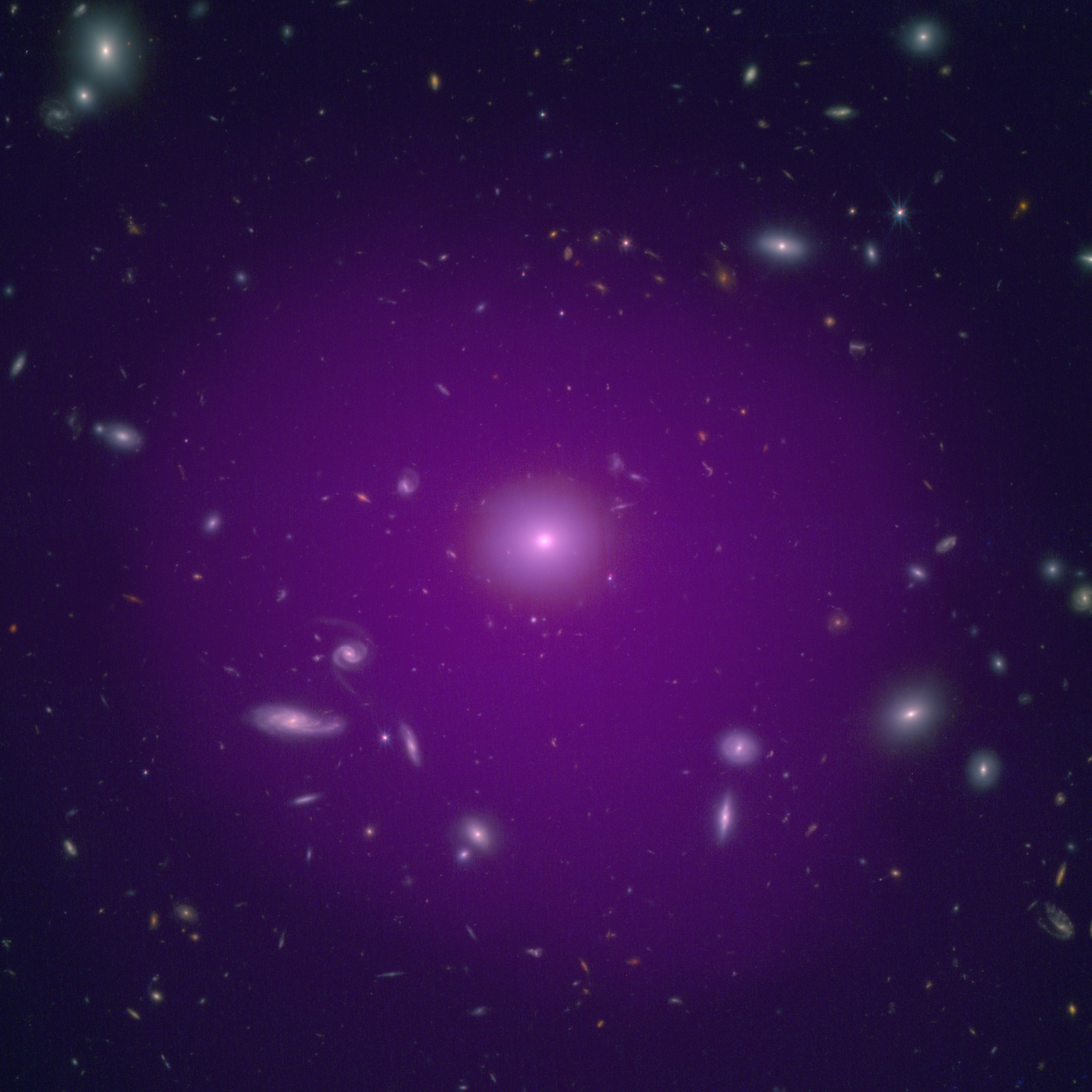}
    \includegraphics[width=0.178\linewidth]{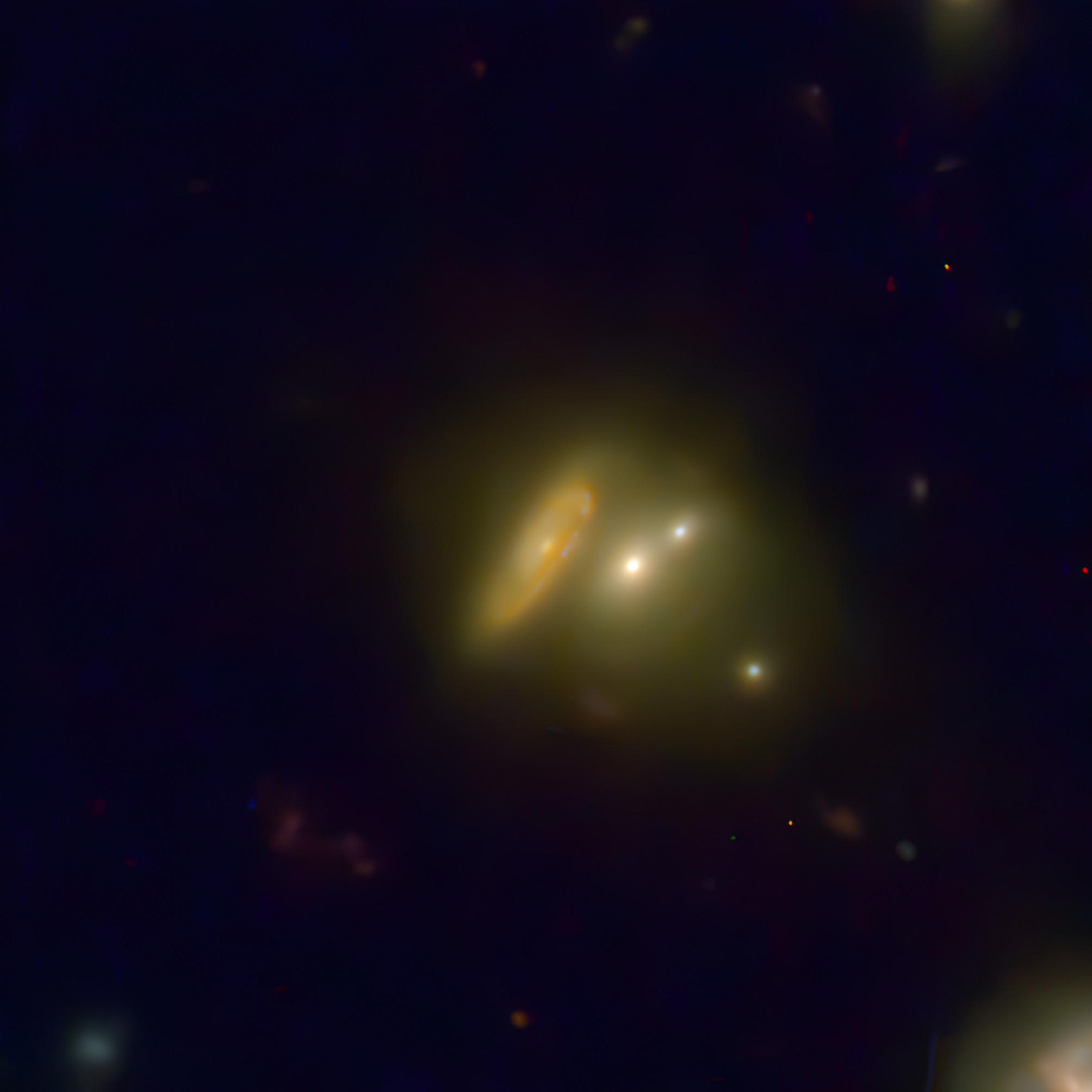}
    \includegraphics[width=0.178\linewidth]{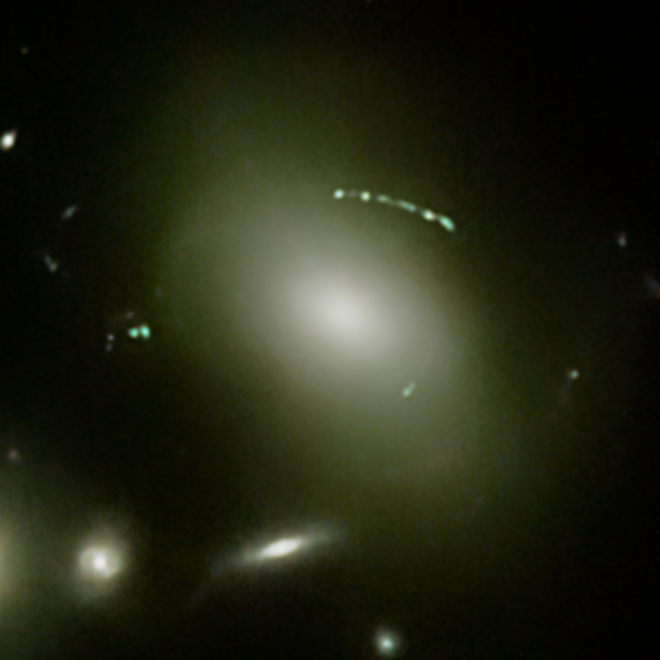}
 \includegraphics[width=0.178\linewidth]{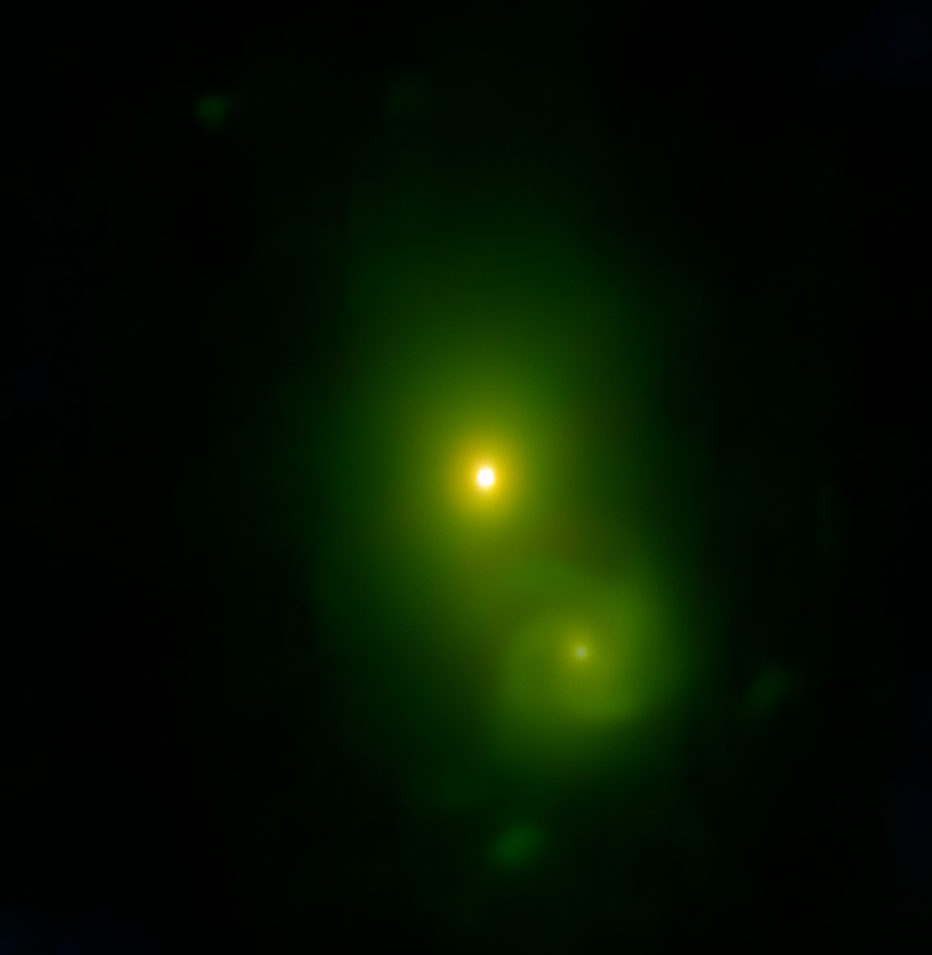}
  \includegraphics[width=0.178\linewidth]{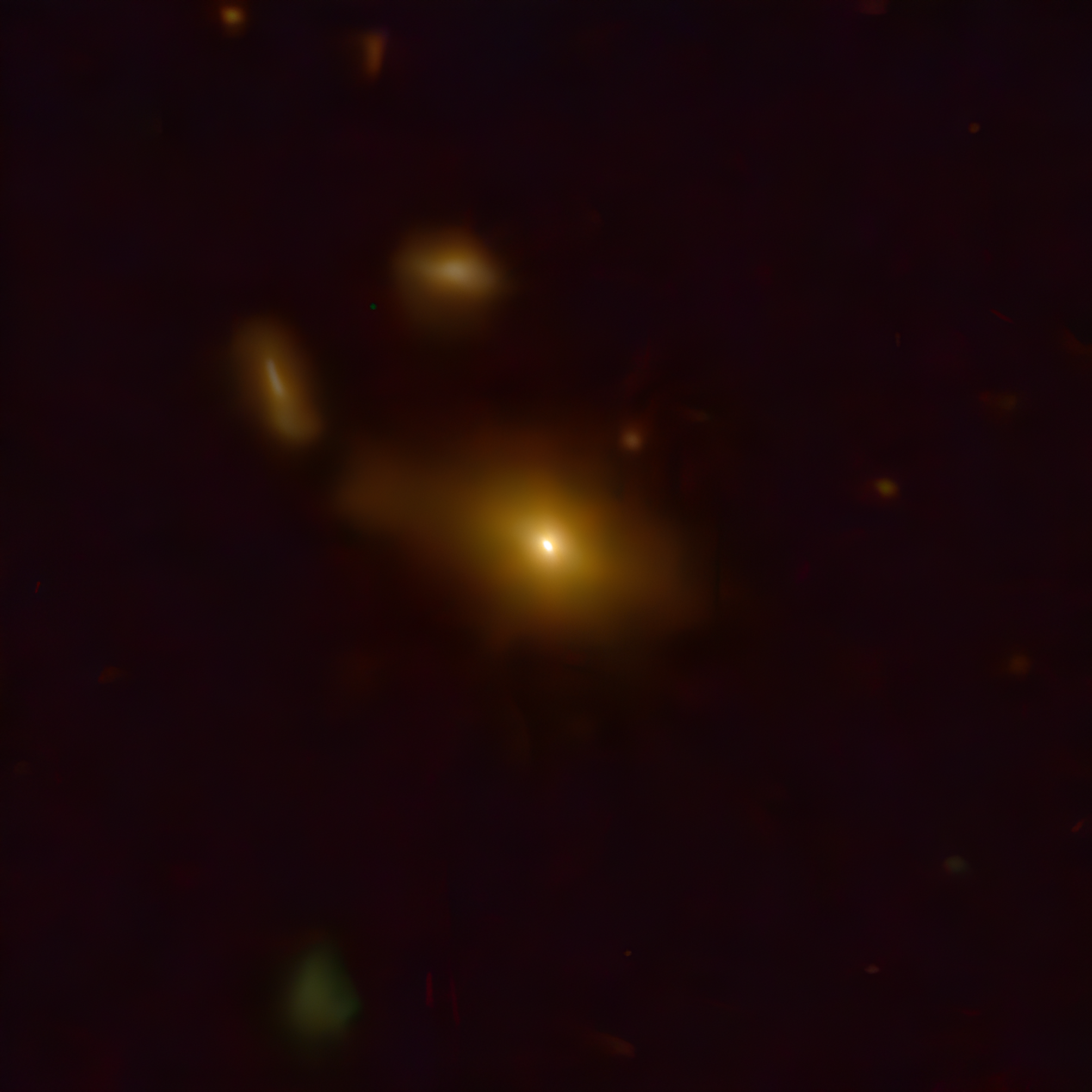}
     \includegraphics[width=0.178\linewidth]{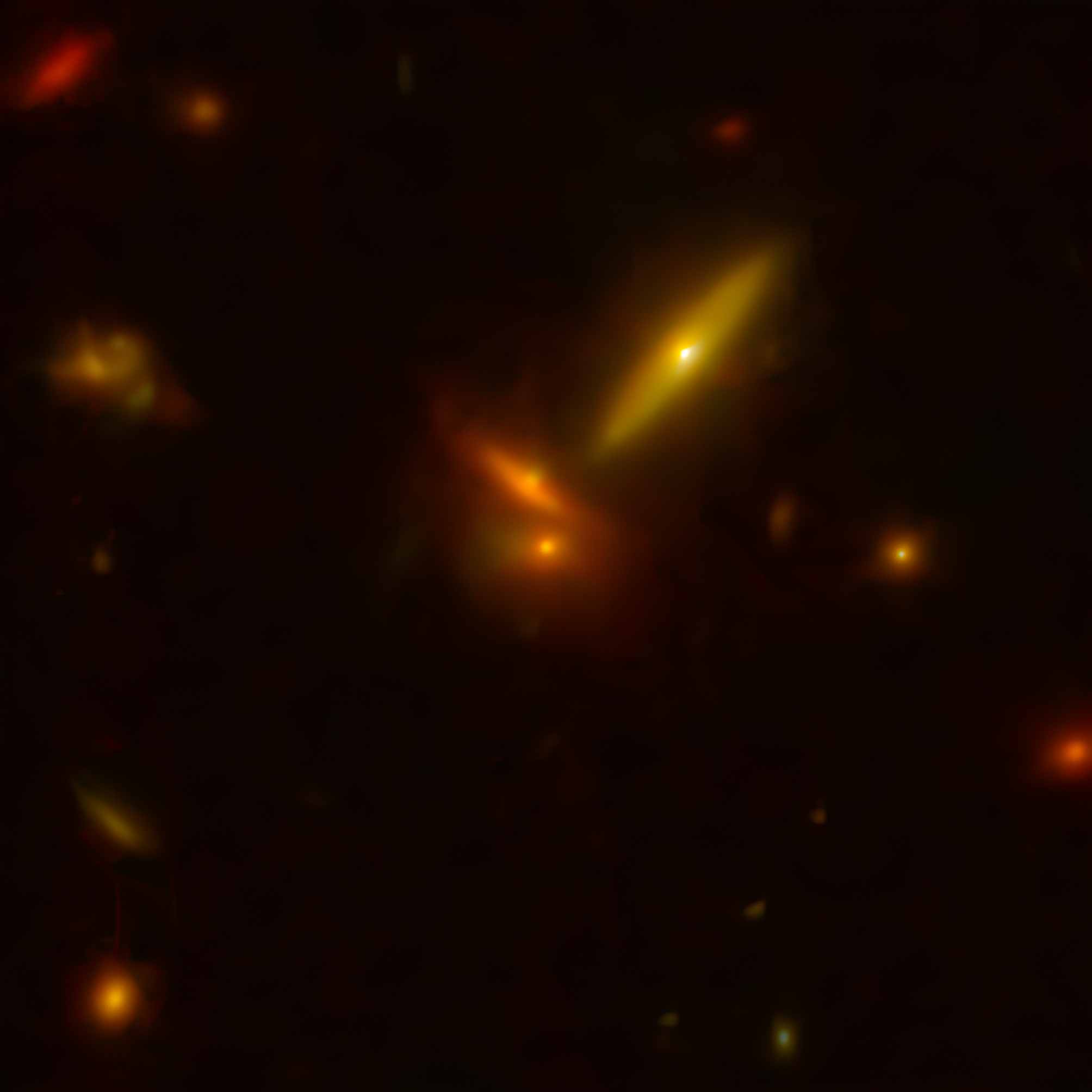}
    \caption{
Examples of JWST/NIRCam color composite of the BGGs in COSMOS-Web groups spanning redshifts from $z=0.22$ to $z=3.09$. The images display diverse morphologies and structural features, including compact spheroids, disturbed or merging systems, and prominent lensing arcs in massive group cores.
}
    \label{fig:enter-label}
\end{figure*}

In Figure~\ref{fig:enter-label}, we show a color composite of the central BGGs in COSMOS-Web groups spanning redshifts from $z=0.22$ to $z=3.09$. (same galaxies presented in Fig.~\ref{fig:F115W_band}) from COSMOS-Web JWST/NIRCam imaging. For the BGG in the top panel, we overlaid diffuse X-ray contours from Chandra and XMM-Newton archival data on the JWST RGB band made using all NIRCam bands (F115W, F150W, F277W, and F444W). The lower panel reveals group cores populated by compact spheroids, edge-on disks, and interacting or lensed systems. In the bottom row, we display five zoomed-in examples that showcase the diversity in morphology, structural concentration, and star-forming features. These panels emphasize the importance of high-resolution imaging in dissecting galaxy structures within group environments.

For completeness, we include in the Appendix the corresponding S'ersic fits for the same six BGGs in three additional JWST/NIRCam bands: F150W (Figure~\ref{fig:F150W_band}), F277W (Figure~\ref{fig:F277W_band}), and F444W (Figure~\ref{fig:F444W_band}). These comparisons provide a consistent view of BGG structures across the near-infrared spectrum, enabling accurate rest-frame optical size estimates for galaxies at different redshifts. The morphological integrity and fit quality observed across these bands further demonstrate the robustness of the Galight pipeline and confirm the fidelity of our structural parameter measurements.

\subsection{Rest-Frame optical size selection by redshift}

\begin{figure}
\includegraphics[width=0.95\columnwidth]{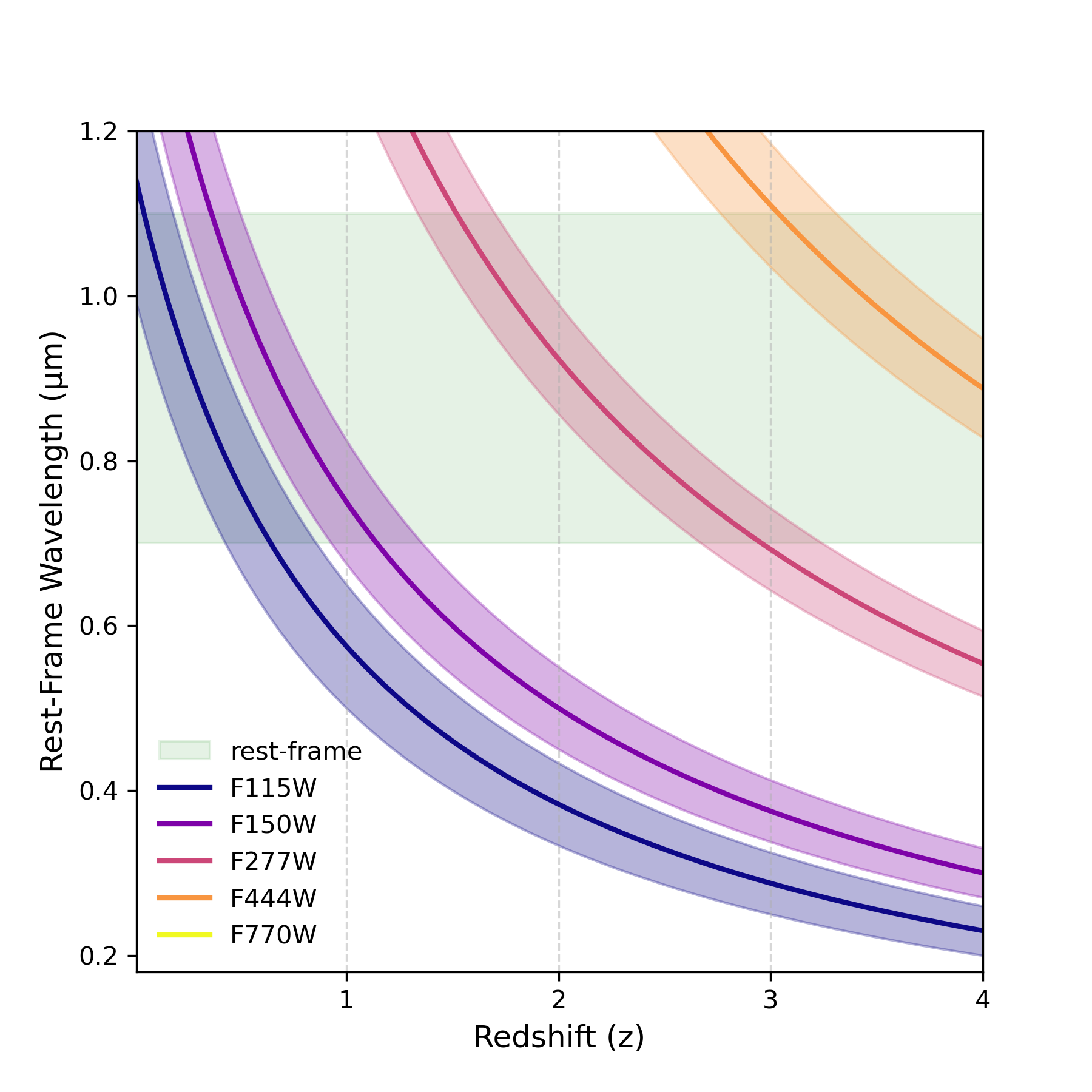}
\caption{Rest-frame wavelength probed by each COSMOS-Web NIRCam filter as a function of redshift. The shaded region indicates the  ($\sim8000$\,\AA, red) ranges.}
\label{img:filters}
\end{figure}

A key objective of this work is to study the evolution of BGG sizes in the rest-frame optical. Since JWST/NIRCam probes different rest-frame wavelengths depending on redshift, we select the appropriate filter per galaxy to ensure consistent rest-frame measurements. We adopt a rest frame wavelength of about 8000,\AA. The filters are specified to redshift as shown in Figure~\ref{img:filters}.

We define rest frame optical-NIR size measurements by selecting the NIRCam filter whose observed wavelength most closely samples the rest-frame $\sim$7000–11000\,\AA\, range at the galaxy’s redshift. Although this choice does not always correspond to the optical spectrum of the canonical rest frame (4000–6000,\AA), it ensures uniform high signal-to-noise morphological measurements across a broad redshift range, utilizing the deepest available imaging in COSMOS-Web. The rest-frame optical is preferred over the UV for structural measurements because it traces the older stellar population and is less affected by bright star-forming clumps or patchy dust, yielding more stable and representative size estimates. Specifically, we adopt:
\begin{itemize}
    \item F115W for $0.05 < z \leq 0.4$
    \item F150W for $0.4 < z \leq 1.0$
    \item F277W for $1.0 < z \leq 3.0$
    \item F444W for $3.0 < z \leq 4.0$
\end{itemize}
These filters correspond to observed frame wavelengths that broadly sample the rest frame optical-NIR transition regime, allowing consistent structural comparisons of BGGs over $\sim$ 12 billion years of cosmic time.

After applying our quality cuts and redshift-dependent filter selection, we obtain a high-quality sample of BGGs with rest-frame optical size measurements across $0.08 < z < 3.7$. The sizes are converted from angular to physical units (kpc) using Planck18 cosmology \citep{aghanim2020planck}. The resulting dataset provides the foundation for our structural and evolutionary analysis of central galaxies in group environments.
 \section{Results}\label{sec:results}
\subsection{Size--Mass relation of star-forming and quiescent BGGs}  
\label{sec:size_mass_relation}

We investigate the size--mass relation of BGGs over cosmic time by dividing them into SF and QG populations using the combined classification scheme (Section~\ref{sec:bgg_classification}), which integrates both rest-frame NUV--$r$--$J$ colors and redshift-dependent sSFR thresholds. This enables a clean comparison of structural scaling relations across stellar mass and redshift, minimizing contamination from transitional systems.

The size–mass relation of BGGs is modeled using a power-law form:

\begin{equation}\label{eq:size-mass}
\log_{10}(R_e/\mathrm{kpc}) = \log A + \alpha \left[\log_{10}(M_\ast / 5 \times 10^{10} M_\odot)\right],
\end{equation}

where $R_e$ is the effective radius (half-light) in kiloparsecs, and $M_\ast$ is the stellar mass. This relation is normalized at a pivot mass of $5 \times 10^{10} M_\odot$, which minimizes the covariance between the slope $\alpha$ and the intercept $\log A$ in the regression.
The slope $\alpha$ quantifies how rapidly the galaxy size scales with stellar mass. A higher value of $\alpha$ indicates a steeper size growth with increasing mass, while the intercept $\log A$ represents the logarithmic size at the pivot mass and captures the characteristic scale of galaxies at that mass.

The best-fit values of $\alpha$ and $\log A$ are obtained by Bayesian MCMC inference, and the shaded bands in the figures represent the 1$\sigma$ posterior uncertainties from the marginalized distributions of these parameters. This approach provides robust estimates of the size–mass relation and its intrinsic scatter in each redshift bin for both SFGs and QGs.

\begin{figure*}[h]
    \centering
    \includegraphics[width=0.75\linewidth]{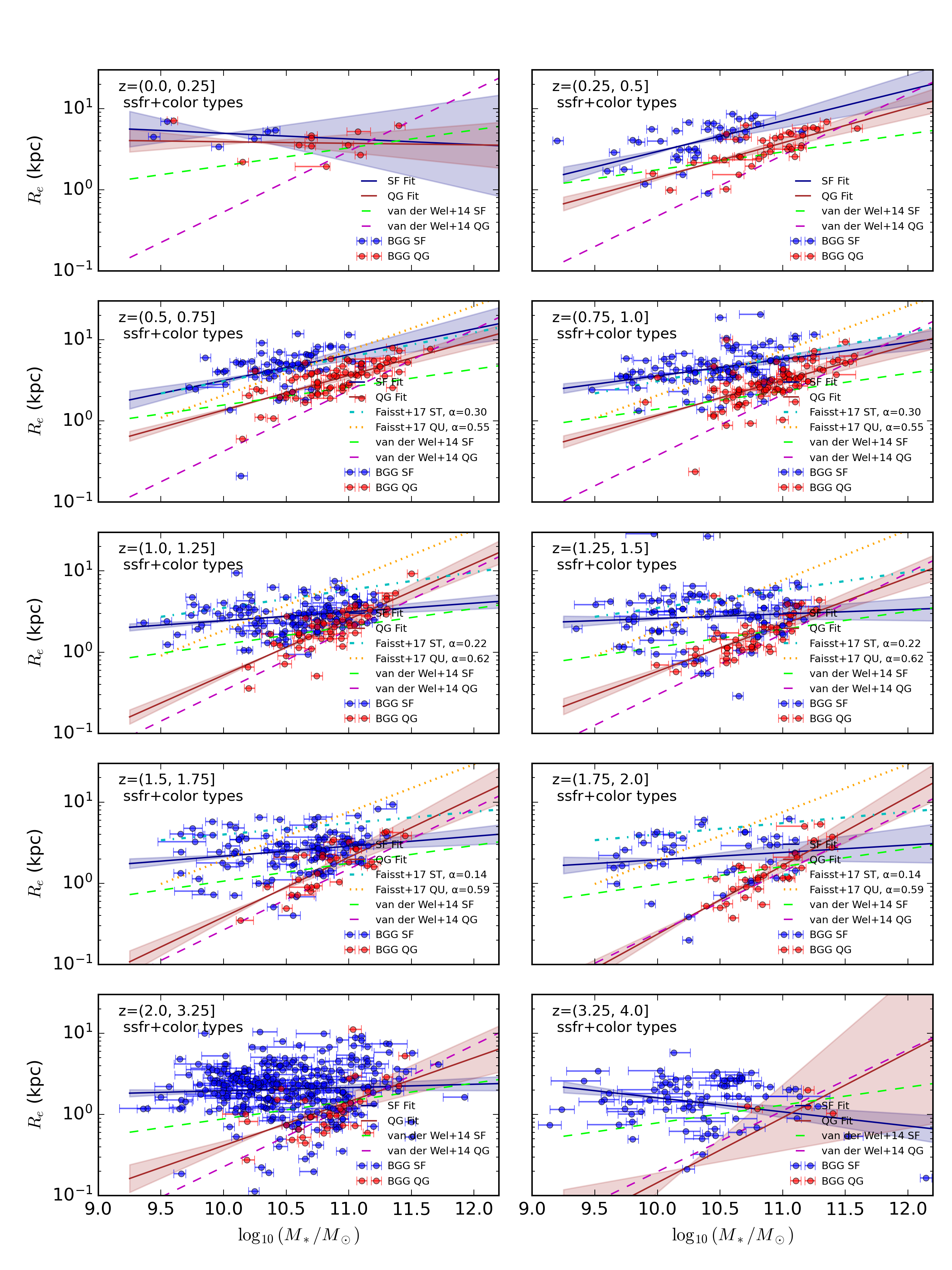}
    \caption{
        Size--mass relation of BGGs classified using the consensus (color+sSFR) method, shown across eight redshift bins from $z = 0$ to $z = 4$ ($\Delta z = 0.5$). Blue circles represent star-forming BGGs, and red hexagons show quiescent BGGs. The dashed and solid lines indicate best-fit power-law relations for SFGs and QGs, respectively, with shaded bands showing $1\sigma$ uncertainties. Cyan dashed-dotted and orange dotted lines show comparison relations from \citet{Faisst2017}, while dashed lime and dashed magenta lines show \citet{vdW2014} results. The BGGs generally follow similar evolutionary trends but are slightly smaller at fixed mass, especially among quiescent centrals, highlighting the impact of group environment on galaxy structure.
    }
    \label{fig:size_mass_consensus}
\end{figure*}

Figure~\ref{fig:size_mass_consensus} presents the size--mass distributions in ten redshift bins from $z = 0$ to $z = 3.7$, with blue circles representing SFGs and red circles indicating QGs. Solid (brown and blue) lines show the best-fit power law relations with shaded bands marking the $1\sigma$ uncertainties from the Bayesian fits.

Across the panels, we observe clear trends:
\begin{itemize}
    \item At low redshift ($z < 1$), SFGs display systematically larger effective radii than QGs at fixed mass, with shallower slopes ($\alpha \sim 0.1$--$0.2$ for SFGs vs.\ $\alpha \sim 0.4$--$0.5$ for QGs), consistent with disk-dominated versus spheroid-dominated structures.
    \item Between $1 < z < 2$, the SFGs continue to show relatively flat slopes, while QGs maintain steeper size--mass relations. The difference in normalization persists, but the scatter increases.
    \item At high redshift ($z > 2$), the trends flatten for both populations, with larger dispersion and overlapping distributions. The small QG sample in the highest bin ($z > 3.25$) limits firm conclusions, but the results hint at compact, early-forming quiescent systems.
\end{itemize}

We compare these trends with measurements from the literature. Cyan dashed-dotted and orange dotted lines show the relations from \citet{Faisst2017}, who focused on ultra-massive galaxies (UMGs; $\log M_\ast/M_\odot > 11.4$) using UltraVISTA/3D-HST, while dashed lime (SFGs) and dashed magenta (QGs) lines show the relations from \citet{vdW2014}, based on CANDELS and 3D-HST data.

Overall, the BGGs in our sample follow broadly similar evolutionary trends, but tend to exhibit slightly smaller sizes at fixed mass compared to field UMGs, especially among the quiescent population. This offset may reflect environmental effects such as earlier assembly, denser merger histories, or suppressed late-time accretion in central group environments. Importantly, while \citet{Faisst2017} focused on the most massive galaxies, our BGG sample extends across a broader mass range ($\log M_\ast/M_\odot \sim 10$--12), allowing a more detailed view of the mass dependence within group central galaxies.

We also compare against the canonical relations from \citet{vdW2014}, widely used as a reference for the global galaxy population. Our SFGs and QGs generally lie close to these reference tracks, although QGs in particular show a tendency toward more compact sizes at $z < 2$, reinforcing the idea that environmental quenching leads to more compact quiescent systems compared to field counterparts.

For completeness, we also compute size--mass relations using the \textit{individual} classification methods—(i) sSFR thresholds and (ii) NUV--$r$--$J$ color cuts—which yield broadly similar slopes and intercepts but with slightly elevated scatter, especially at intermediate redshifts ($1 < z < 2$) where classification uncertainties from dust and photometric noise are higher.  The size mass relation for only the color-color and sSFR threshould based classification of BGGs as SFGS and QGs are presented in the Appendix \ref{app:alternative_classifications}.

Taken together, these results demonstrate that BGGs follow the general size--mass evolutionary patterns seen in the broader galaxy population but show distinct signatures of environment-driven evolution, particularly among quiescent centrals.

\begin{table*}
\centering
\caption{Best-fit parameters of the size--mass relation for BGGs classified using color--color, sSFR, and combined (color+sSFR) criteria. Columns: classification method, redshift bin, galaxy type, intercept ($\log A$), slope ($\alpha$), intrinsic scatter ($\sigma(\log R_e)$), and number of galaxies.}
\label{tab:bgg_fits_compact}
\begin{tabular}{l l l c c c c c c}
\hline
Method & Redshift Bin & Type & $\log A$ & $\Delta \log A$ & $\alpha$ & $\Delta \alpha$ & $\sigma(\log R_e)$ & $N_\text{gal}$ \\
\hline
Color--Color & (0.0, 0.25] & SF & 0.667 & 0.205 & 0.136 & 0.267 & 0.300 & 17 \\
Color--Color & (0.0, 0.25] & QG & 0.519 & 0.043 & 0.227 & 0.094 & 0.210 & 26 \\
Color--Color & (0.25, 0.5] & SF & 0.685 & 0.035 & 0.291 & 0.063 & 0.231 & 80 \\
Color--Color & (0.25, 0.5] & QG & 0.444 & 0.024 & 0.433 & 0.070 & 0.165 & 53 \\
Color--Color & (0.5, 0.75] & SF & 0.675 & 0.026 & 0.202 & 0.054 & 0.220 & 108 \\
Color--Color & (0.5, 0.75] & QG & 0.426 & 0.020 & 0.440 & 0.054 & 0.150 & 75 \\
Color--Color & (0.75, 1.0] & SF & 0.678 & 0.022 & 0.206 & 0.045 & 0.205 & 115 \\
Color--Color & (0.75, 1.0] & QG & 0.364 & 0.023 & 0.456 & 0.061 & 0.199 & 91 \\
Color--Color & (1.0, 1.25] & SF & 0.456 & 0.020 & 0.132 & 0.045 & 0.214 & 137 \\
Color--Color & (1.0, 1.25] & QG & 0.204 & 0.024 & 0.696 & 0.074 & 0.164 & 56 \\
Color--Color & (1.25, 1.5] & SF & 0.445 & 0.034 & 0.071 & 0.064 & 0.306 & 114 \\
Color--Color & (1.25, 1.5] & QG & 0.154 & 0.028 & 0.607 & 0.084 & 0.190 & 47 \\
Color--Color & (1.5, 1.75] & SF & 0.409 & 0.025 & 0.113 & 0.059 & 0.255 & 114 \\
Color--Color & (1.5, 1.75] & QG & 0.129 & 0.038 & 0.594 & 0.109 & 0.219 & 38 \\
Color--Color & (1.75, 2.0] & SF & 0.324 & 0.049 & 0.064 & 0.096 & 0.310 & 51 \\
Color--Color & (1.75, 2.0] & QG & 0.015 & 0.042 & 0.688 & 0.124 & 0.204 & 31 \\
Color--Color & (2.0, 3.25] & SF & 0.304 & 0.019 & 0.069 & 0.040 & 0.331 & 349 \\
Color--Color & (2.0, 3.25] & QG & -0.010 & 0.047 & 0.551 & 0.148 & 0.299 & 46 \\
Color--Color & (3.25, 4.0] & SF & 0.083 & 0.041 & -0.157 & 0.073 & 0.335 & 92 \\
Color--Color & (3.25, 4.0] & QG & -0.214 & 0.165 & 0.637 & 0.352 & 0.403 & 8 \\
\hline
sSFR & (0.0, 0.25] & SF & 0.666 & 0.203 & 0.137 & 0.265 & 0.300 & 17 \\
sSFR & (0.0, 0.25] & QG & 0.518 & 0.043 & 0.227 & 0.094 & 0.210 & 26 \\
sSFR & (0.25, 0.5] & SF & 0.707 & 0.041 & 0.311 & 0.070 & 0.224 & 70 \\
sSFR & (0.25, 0.5] & QG & 0.463 & 0.025 & 0.458 & 0.075 & 0.188 & 63 \\
sSFR & (0.5, 0.75] & SF & 0.684 & 0.032 & 0.214 & 0.063 & 0.226 & 97 \\
sSFR & (0.5, 0.75] & QG & 0.444 & 0.020 & 0.445 & 0.054 & 0.157 & 86 \\
sSFR & (0.75, 1.0] & SF & 0.687 & 0.023 & 0.224 & 0.048 & 0.205 & 109 \\
sSFR & (0.75, 1.0] & QG & 0.369 & 0.023 & 0.473 & 0.061 & 0.200 & 97 \\
sSFR & (1.0, 1.25] & SF & 0.462 & 0.019 & 0.090 & 0.043 & 0.191 & 126 \\
sSFR & (1.0, 1.25] & QG & 0.220 & 0.024 & 0.699 & 0.070 & 0.183 & 67 \\
sSFR & (1.25, 1.5] & SF & 0.439 & 0.037 & 0.084 & 0.069 & 0.324 & 111 \\
sSFR & (1.25, 1.5] & QG & 0.170 & 0.025 & 0.641 & 0.081 & 0.172 & 50 \\
sSFR & (1.5, 1.75] & SF & 0.418 & 0.024 & 0.126 & 0.056 & 0.245 & 114 \\
sSFR & (1.5, 1.75] & QG & 0.083 & 0.037 & 0.748 & 0.114 & 0.205 & 38 \\
sSFR & (1.75, 2.0] & SF & 0.344 & 0.062 & 0.093 & 0.112 & 0.326 & 42 \\
sSFR & (1.75, 2.0] & QG & 0.063 & 0.042 & 0.604 & 0.130 & 0.229 & 40 \\
sSFR & (2.0, 3.25] & SF & 0.305 & 0.020 & 0.052 & 0.041 & 0.326 & 330 \\
sSFR & (2.0, 3.25] & QG & -0.006 & 0.042 & 0.729 & 0.122 & 0.303 & 65 \\
sSFR & (3.25, 4.0] & SF & 0.089 & 0.043 & -0.141 & 0.075 & 0.342 & 89 \\
sSFR & (3.25, 4.0] & QG & -0.165 & 0.126 & 0.627 & 0.357 & 0.357 & 11 \\
 \hline
Color+sSFR & (0.0, 0.25] & SF & 0.667 & 0.203 & 0.136 & 0.265 & 0.299 & 17 \\
Color+sSFR & (0.0, 0.25] & QG & 0.518 & 0.043 & 0.227 & 0.094 & 0.211 & 26 \\
Color+sSFR & (0.25, 0.5] & SF & 0.714 & 0.041 & 0.317 & 0.069 & 0.223 & 69 \\
Color+sSFR & (0.25, 0.5] & QG & 0.443 & 0.025 & 0.436 & 0.072 & 0.167 & 52 \\
Color+sSFR & (0.5, 0.75] & SF & 0.684 & 0.031 & 0.215 & 0.063 & 0.226 & 97 \\
Color+sSFR & (0.5, 0.75] & QG & 0.425 & 0.020 & 0.441 & 0.054 & 0.149 & 75 \\
Color+sSFR & (0.75, 1.0] & SF & 0.699 & 0.023 & 0.223 & 0.047 & 0.197 & 105 \\
Color+sSFR & (0.75, 1.0] & QG & 0.359 & 0.025 & 0.462 & 0.065 & 0.201 & 87 \\
Color+sSFR & (1.0, 1.25] & SF & 0.461 & 0.019 & 0.087 & 0.043 & 0.191 & 125 \\
Color+sSFR & (1.0, 1.25] & QG & 0.200 & 0.024 & 0.692 & 0.073 & 0.162 & 55 \\
Color+sSFR & (1.25, 1.5] & SF & 0.451 & 0.034 & 0.056 & 0.064 & 0.299 & 108 \\
Color+sSFR & (1.25, 1.5] & QG & 0.165 & 0.026 & 0.568 & 0.085 & 0.169 & 44 \\
Color+sSFR & (1.5, 1.75] & SF & 0.423 & 0.026 & 0.120 & 0.058 & 0.247 & 109 \\
Color+sSFR & (1.5, 1.75] & QG & 0.082 & 0.039 & 0.732 & 0.117 & 0.203 & 33 \\
Color+sSFR & (1.75, 2.0] & SF & 0.353 & 0.065 & 0.090 & 0.115 & 0.332 & 40 \\
Color+sSFR & (1.75, 2.0] & QG & -0.036 & 0.042 & 0.846 & 0.126 & 0.182 & 29 \\
Color+sSFR & (2.0, 3.25] & SF & 0.315 & 0.020 & 0.057 & 0.041 & 0.321 & 320 \\
Color+sSFR & (2.0, 3.25] & QG & -0.007 & 0.054 & 0.538 & 0.160 & 0.294 & 36 \\
Color+sSFR & (3.25, 4.0] & SF & 0.093 & 0.043 & -0.151 & 0.073 & 0.338 & 88 \\
Color+sSFR & (3.25, 4.0] & QG & -0.286 & 0.219 & 0.819 & 0.492 & 0.415 & 7 \\
\hline
\end{tabular}
\label{tab:bgg_fits_compact}
\end{table*}

\subsection{Redshift evolution of the size--mass relation slope and intrinsic scatter} \label{sec:slope_scatteR_evolution}

To quantify how the structural scaling of BGGs evolves over cosmic time, we track the slope \(\alpha\) of the size--mass relation and the intrinsic scatter \(\sigma(\log R_e)\) across eight redshift bins from \(z = 0.08\) to \(z = 3.7\). These parameters were extracted from Bayesian fits to the size--mass relation as described in Eq.~\ref{eq:size-mass}. Table~\ref{tab:bgg_fits_compact} presents the best-fit parameter results.

\begin{figure}[h!]
    \centering
    \includegraphics[width=0.95\linewidth]{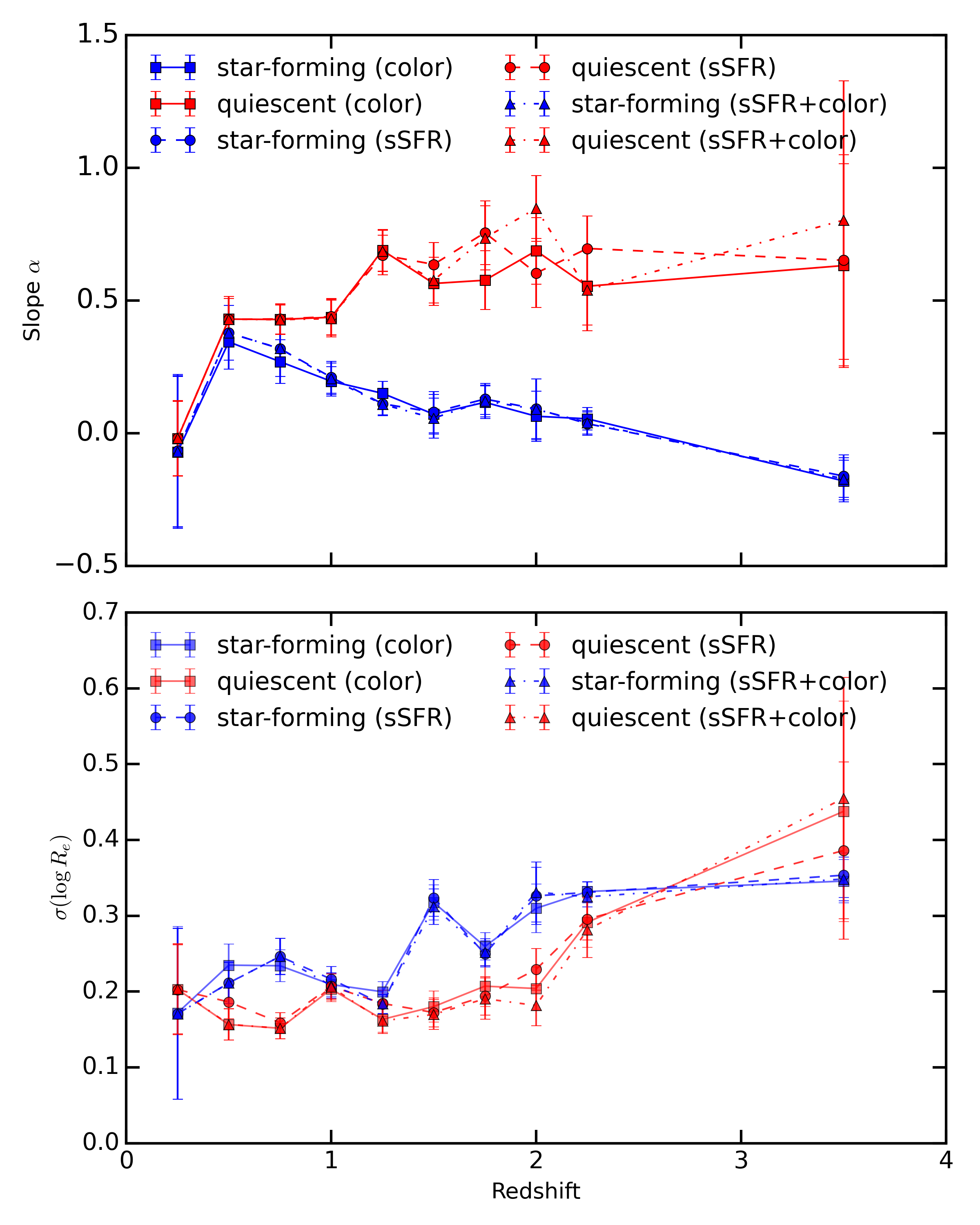}
    \caption{
        Redshift evolution of the size--mass relation slope \(\alpha\) (top panel) and intrinsic scatter \(\sigma(\log R_e)\) (bottom panel) for BGGs classified using color (solid lines), sSFR (dashed lines), and consensus (dotted lines) methods. Star-forming and quiescent BGGs are shown in blue and red, respectively. Quiescent BGGs display consistently steeper slopes, especially at \(z \sim 1\)--3, while star-forming BGGs maintain nearly flat relations across redshift. The scatter increases toward higher redshift for both types, indicating greater structural diversity and formation variability in the early universe.
    }
    \label{fig:slope_sigma_evol}
\end{figure}

Figure~\ref{fig:slope_sigma_evol} shows the redshift evolution of both parameters for star-forming and quiescent BGGs, classified using three methods: color--color, sSFR threshold, and the intersection of both (consensus). The top panel displays the slope \(\alpha(z)\), while the bottom panel presents \(\sigma(\log R_e)(z)\), each as a function of redshift.

\paragraph{Slope evolution:}
Across all classification schemes, quiescent BGGs exhibit significantly steeper size--mass relations than star-forming ones, especially at intermediate to high redshifts (\(z \sim 1\)--3), where \(\alpha_{\mathrm{QG}}\) consistently peaks around 0.5--0.7. This indicates that at these epochs, more massive quiescent BGGs experience relatively stronger size scaling compared to their lower-mass counterparts, likely driven by efficient dry mergers or inside-out growth mechanisms. In contrast, star-forming BGGs show consistently shallow slopes, with \(\alpha \sim 0\)--0.2, across all redshifts, suggesting nearly mass-independent size growth. Interestingly, the slope for SFGs shows a mild decline toward higher redshifts, which may reflect more uniform and disk-dominated structural configurations in the early universe, before significant mass-driven differentiation set in.

\paragraph{Scatter evolution:}
The intrinsic scatter \(\sigma(\log R_e)\) increases with redshift for both galaxy types, reaching \(\sim 0.3\)--0.4 dex at \(z > 2\). Star-forming BGGs consistently show slightly higher scatter than quiescent BGGs, particularly at \(z \gtrsim 2\), likely reflecting greater structural diversity due to clumpy star formation, irregular morphologies, or ongoing gas accretion. For quiescent BGGs, the increase in scatter may arise from mixed evolutionary pathways, including compaction, quenching, and subsequent merger-driven size evolution.

The divergent trends in slope and scatter between star-forming and quiescent BGGs support a picture where quiescent systems undergo more mass-dependent, merger-driven size growth, whereas star-forming systems follow more uniform evolutionary tracks dominated by steady gas inflow and secular processes. The increasing scatter at higher redshifts reinforces the idea that BGGs at early times span a wider range of formation histories and structural states, transitioning toward tighter and more settled scaling relations in the local universe.

\subsection{Size evolution of BGGs at fixed stellar mass} \label{sec:size_evol_fixed_mass}

To investigate the redshift evolution of BGG sizes at fixed stellar mass, we fit the relation:
\begin{equation}\label{r_e}
    \log_{10}(R_e/kpc) = A - \alpha \log(1 + z)
\end{equation}
 in eight redshift bins for galaxies with $\log(M_\ast/M_\odot) = 10.7$. In Eq. \ref{r_e}, $R_e, z$ present the effective radius and redshift of the galaxy and A and $\alpha$ are the intercept and slope, of the relation.
We perform this analysis separately for star-forming and quiescent BGGs, defined using three different classification schemes: (1) rest-frame NUV–r–J color diagram, (2) sSFR, and (3) a combined color+sSFR criterion.

\begin{figure}
    \centering
    \includegraphics[width=1\linewidth]{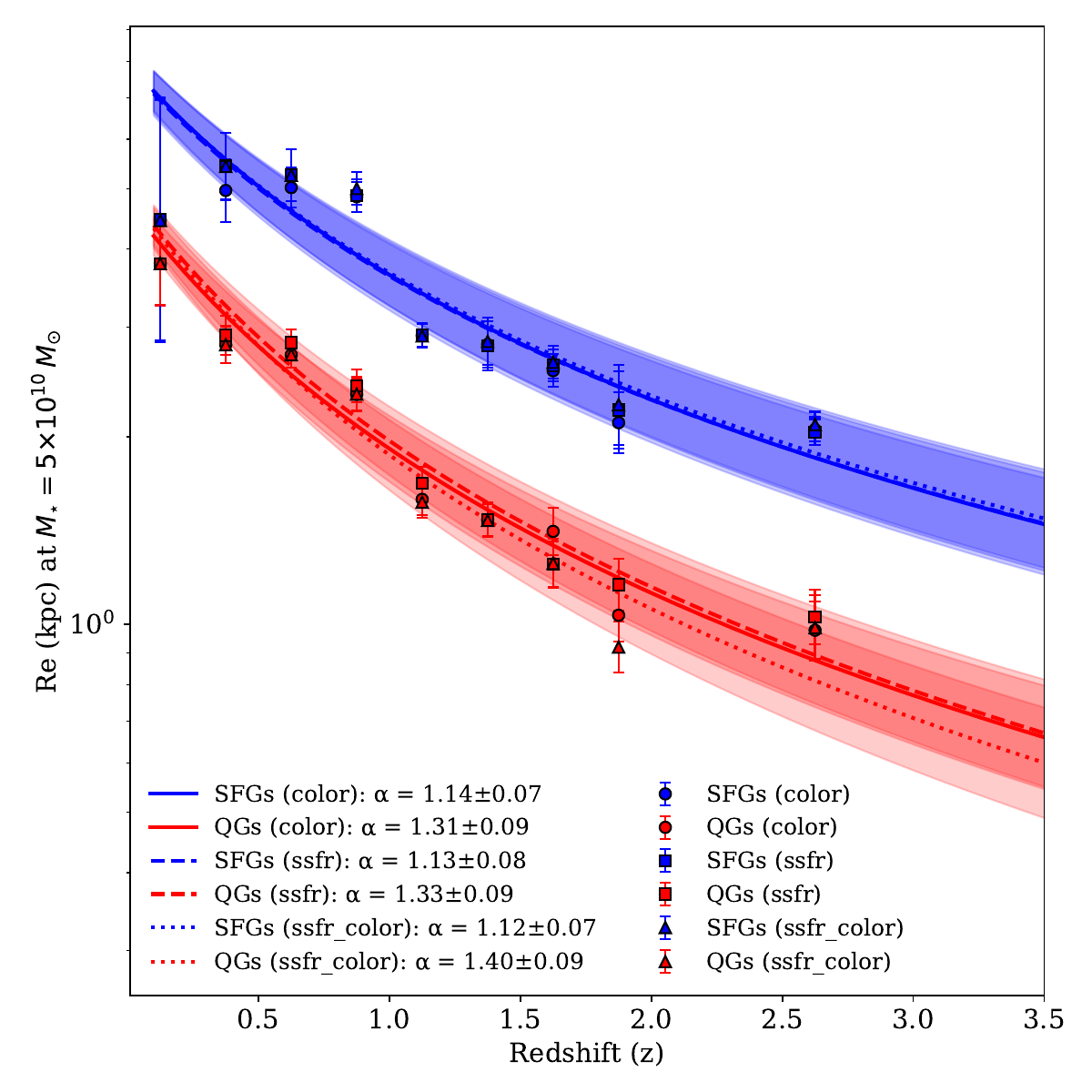}
    \caption{
    Redshift evolution of the effective radius ($R_e$) of star-forming (blue) and quiescent (red) BGGs at fixed stellar mass $M_\ast = 5 \times 10^{10}\,M_\odot$. 
    The BGGs are classified using three criteria: rest-frame NUV–r–J color (solid lines), specific star formation rate (dashed lines), and combined color+sSFR (dotted lines). 
    Curves show the best-fit relation $R_e \propto (1 + z)^{-\alpha}$, with shaded regions indicating $1\sigma$ confidence intervals. 
    Star-forming BGGs show stronger size evolution than quiescent ones in all classification schemes. The slight offset seen at $z \sim 0.6$ may reflect sample scatter or minor overfitting effects in some models.
    }
    \label{fig:size_evol_fixed_mass}
\end{figure}

Figure~\ref{fig:size_evol_fixed_mass} shows the best-fit size–redshift relations for each population, with shaded bands representing the $1\sigma$ uncertainty envelopes from the model fits. 
Our results demonstrate that BGGs experience substantial size growth over cosmic time.
Across all classification schemes, quiescent BGGs show systematically steeper evolutionary slopes (larger $\alpha$) compared to star-forming counterparts, indicating stronger size growth since high redshift.
This suggests that while quiescent BGGs likely underwent an early compaction phase followed by significant size increase - possibly through dissipationless (dry) mergers — star-forming BGGs exhibit a more moderate size evolution, likely reflecting gradual growth through continued star formation and gas accretion.

The best-fit slopes ($\alpha$) for $R_e \propto (1 + z)^{-\alpha}$ are summarized in the tab. \ref{tab:re_growth}. Table \ref{tab:re_growth} also determines the growth factor of the BGG size ($R_e(z_{min})/R_e(z_{max}$)) for both SFs and QGs.
\begin{table*}[h]
    \centering
    \caption{Best-fit slopes (\(\alpha\)) for \( R_e \propto (1 + z)^{-\alpha} \) in both star-forming and quiescent BGGs. Classification is based on color-color, redshift-dependent sSFR, and both criteria, with a fixed stellar mass of \( \log(M_\ast/M_\odot) = 10.7 \). }
\begin{tabular}{llrrrrr}
\hline
 Classification   & Galaxy Type   &    A &   A Error &   Alpha &   Alpha Error &   Growth Factor \\
\hline
 Color            & Star-forming  & 0.9  &      0.03 &    1.13 &          0.07 &            5.68 \\
 Color            & Quiescent     & 0.68 &      0.03 &    1.31 &          0.09 &            7.43 \\
 sSFR             & Star-forming  & 0.9  &      0.03 &    1.13 &          0.07 &            5.65 \\
 sSFR              & Quiescent     & 0.7  &      0.03 &    1.34 &          0.08 &            7.77 \\
 sSFR + color       & Star-forming  & 0.9  &      0.03 &    1.11 &          0.07 &            5.52 \\
sSFR +color       & Quiescent     & 0.69 &      0.03 &    1.4  &          0.09 &            8.54 \\
\hline
\end{tabular}
    
    \label{tab:re_growth}
\end{table*}

We note a slight deviation between the model and the data around $z \sim 0.6$, particularly for star-forming BGGs. This may hint at possible overfitting due to increased scatter or sample variance in that redshift interval. Future work incorporating larger samples and improved error modeling will help clarify this local discrepancy.

\subsection{Size distribution of BGGs across cosmic time} \label{sec:size_distribution}

\begin{figure*}[h]
    \centering
    \includegraphics[width=0.85\linewidth]{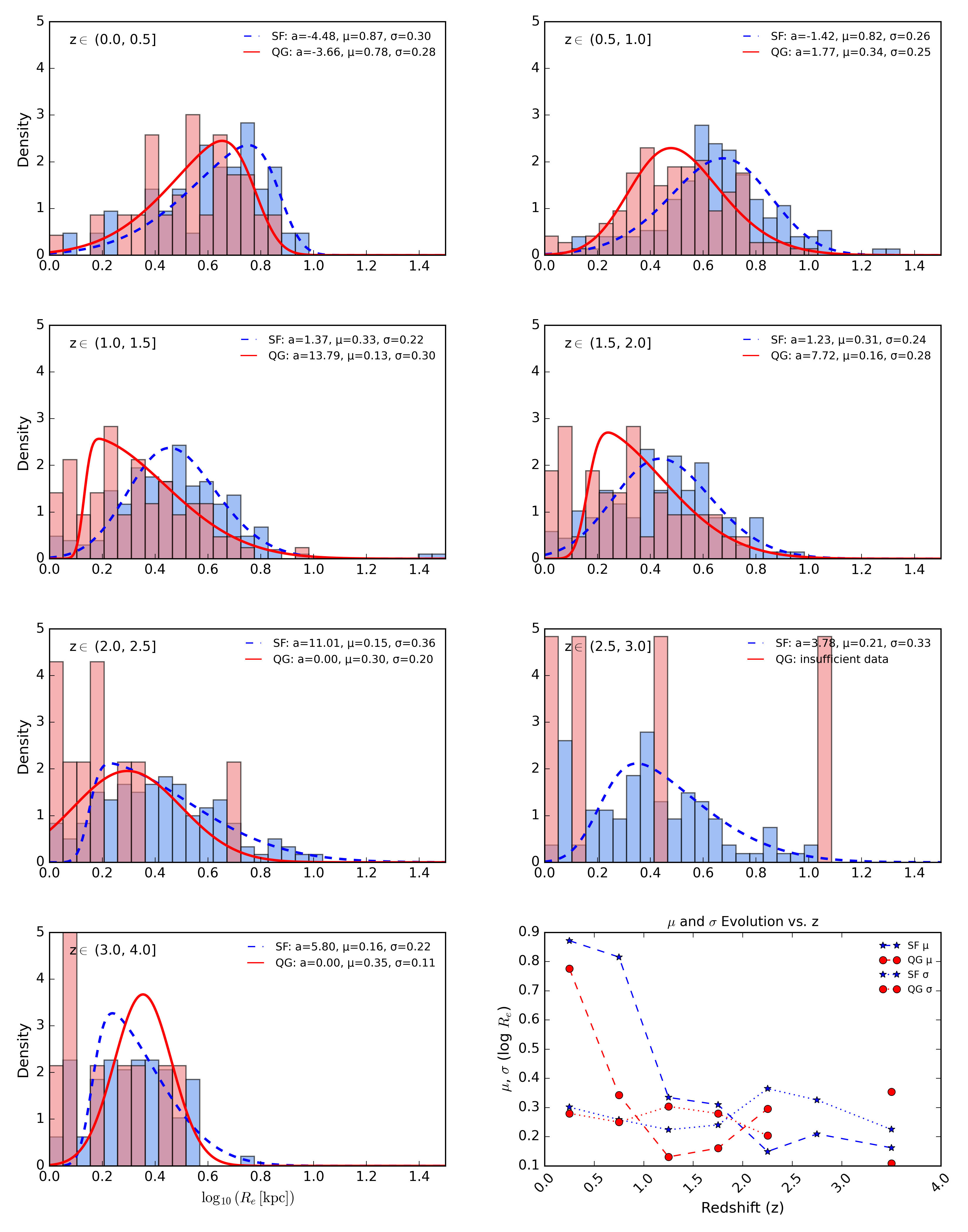}
    \caption{
        Normalized size distributions of BGGs in log scale, shown across seven redshift bins from \(z = 0\) to \(z = 3.7\). Blue and red histograms represent SFGs and QGs, respectively. Dashed lines show best-fit skewed normal functions to each population. Fitted parameters—mean \(\mu\), standard deviation \(\sigma\), and skewness \(a\)—are annotated in each panel. The summary panel tracks \(\mu\) and \(\sigma\) evolution with redshift. At low redshift, QGs are more compact and less scattered, while SFGs show broader, more asymmetric size distributions. At high redshift, the QG population becomes sparse, while SFGs maintain a wide range of structural diversity.
    }
    \label{fig:size_distribution}
\end{figure*}

To better understand the statistical nature and intrinsic scatter of the size distribution of BGGs, we examine the one-dimensional distribution of \(\log_{10}(R_e/\mathrm{kpc})\) for star-forming and quiescent BGGs within seven redshift intervals spanning \(0 < z \leq 3.7\). Figure~\ref{fig:size_distribution} presents normalized histograms for both populations, without binning by stellar mass, ensuring that the full population-wide size distribution is represented at each epoch.

In each panel, SFGs are shown in blue, and QGs in red. A skewed normal function is fitted to each distribution using maximum likelihood estimation, and the best fit parameters - mean (\(\mu\)), standard deviation (\(\sigma\)) and skewness (\(a\)) - are displayed in the legend of each panel. The summary plot in the bottom right panel tracks the evolution of \(\mu\) and \(\sigma\) as a function of redshift.

At low redshift (\(z \lesssim 1.5\)), the quiescent population exhibits a more compact and narrower distribution (\(\mu \sim 0.7\), \(\sigma \sim 0.25\)), while the star-forming BGGs tend to be larger and more broadly distributed (\(\mu \sim 0.8\)--0.9, \(\sigma \sim 0.3\)--0.35). The skewness parameter \(a\) suggests that QGs typically show low or moderate asymmetry, while SFGs display stronger asymmetry or broadening toward larger sizes.

As redshift increases (\(z > 2\)), the number of quiescent BGGs declines rapidly, and their distributions become sparse, making reliable fits more challenging. In contrast, SFGs remain numerous at high redshifts and retain a broad, right-skewed size distribution, reflecting their continued structural diversity and active assembly.

This analysis shows that star-forming and quiescent BGGs occupy distinct structural regimes across cosmic time, with QGs being significantly more compact and tightly distributed, while SFGs maintain broader and more asymmetric size profiles. The width and asymmetry of these distributions provide further evidence of differing evolutionary pathways, likely shaped by differences in gas accretion, star formation activity, and merger histories.

\subsection{Size distributions of BGGs across redshift and stellar mass} \label{sec:size_distributions}

\begin{figure*}[h]
    \centering
    \includegraphics[width=0.95\linewidth]{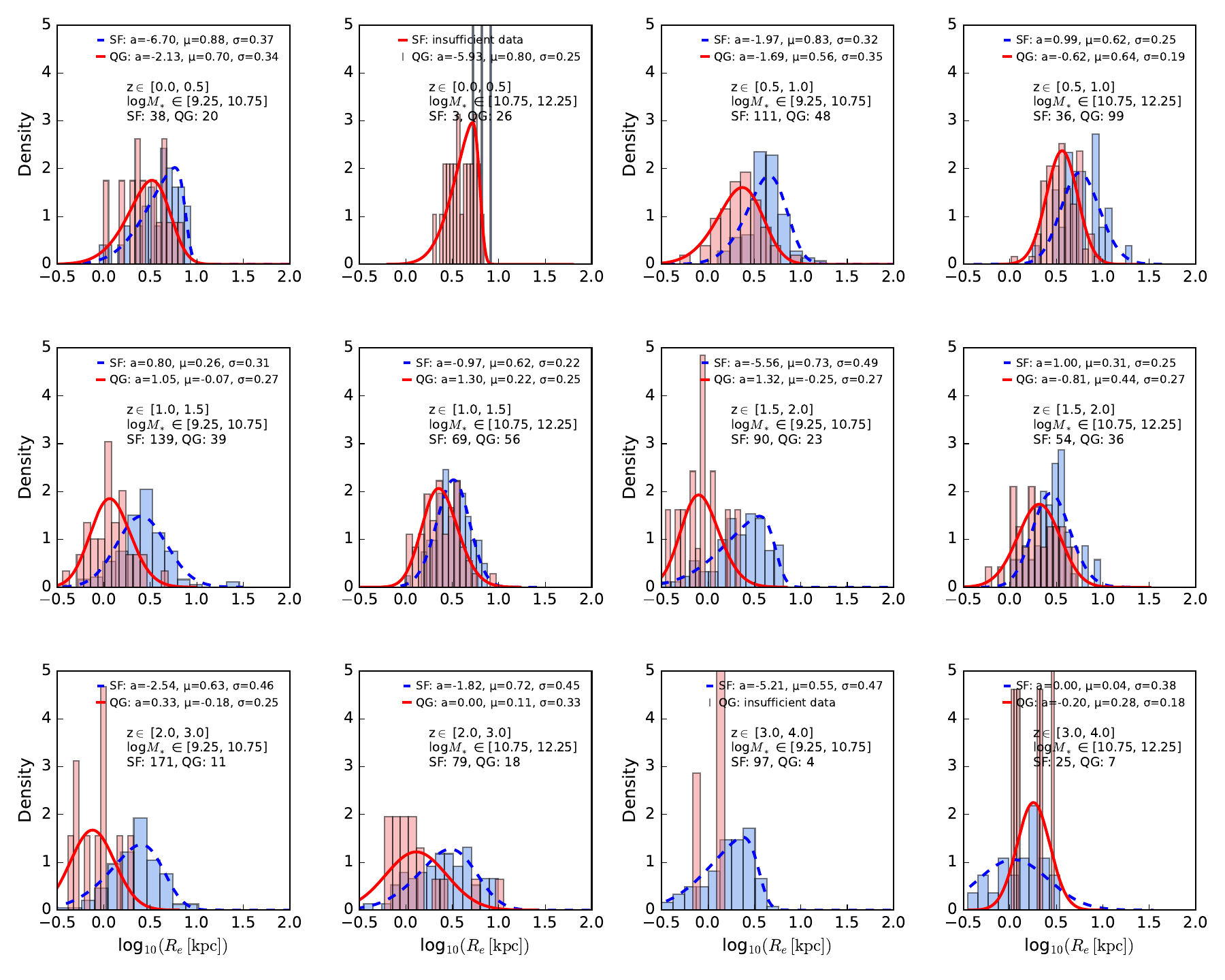}
    \caption{
        Distributions of \(\log_{10}(R_e/\mathrm{kpc})\) for BGGs classified using the combined color–sSFR consensus method. Panels show two stellar mass bins (\(9.25 \leq \log_{10}(M_\ast/M_\odot) < 10.75\) and \(10.75 \leq \log_{10}(M_\ast/M_\odot) < 12.25\)) across eight redshift bins from \(z = 0\) to \(z= 3.7\). Star-forming and quiescent BGGs (SFGs and QGs) are plotted in blue and red, respectively. Skew-normal fits (dashed lines) are overlaid for each distribution, with fitted parameters (\(a\), \(\mu\), \(\sigma\)) annotated. QGs exhibit compact, narrow distributions at low redshift, while SFGs show broader and more skewed profiles, especially at high \(z\). The absence of QGs at \(z > 2\) in low-mass bins reflects the delayed quenching of less massive BGGs.
    }
    \label{fig:size_dist_ssfr_color}
\end{figure*}

To further explore the structural properties of BGGs, we investigated the distribution of galaxy sizes within bins of redshift and stellar mass. Figure~\ref{fig:size_dist_ssfr_color} presents histograms of the effective logarithmic radius (\(\log_{10}(R_e/\mathrm{kpc})\)) for BGG (SFG and QG), classified using the method that combines the color----color-- and redshift--dependent rest frame sSFR criteria. The distributions are shown for two stellar mass intervals (\(9.25 \leq \log_{10}(M_\ast/M_\odot) < 10.75\) and \(10.75 \leq \log_{10}(M_\ast/M_\odot) < 12.25\)), across six redshift bins from \(z = 0\) to \(z = 3.7\).

Each panel displays normalized histograms for SFGs (blue) and QGs (red), overlaid with best-fit skew-normal probability density functions. For each population and bin, we annotate the fitted skewness parameters (\(a\)), location (\(\mu\)) and scale (\(\sigma\)). These provide a detailed statistical summary of the shape, width, and asymmetry of the distribution across cosmic time.

At low redshifts (\(z < 1.5\)), quiescent BGGs show compact and narrow size distributions, with peaks around \(\mu \sim 0.3\)--0.6, consistent across both mass bins. In contrast, star-forming BGGs show broader and more asymmetric distributions, especially in the lower mass bin, often skewed toward smaller sizes. This reflects the structural diversity and clumpy star-forming morphologies in low-mass SFGs.

In the high-mass bin, a noticeable bimodality is observed between SFGs and QGs up to \(z \sim 1.5\), where the separation in the median size becomes most distinct. Beyond \(z > 2\), the quiescent population becomes sparse, particularly in the low-mass bin, reflecting delayed quenching in less massive systems. At these redshifts, SFGs dominate the sample and maintain broad, right-skewed size distributions, reflecting active disk growth and gas accretion.

\begin{figure}[h]
    \centering
    \includegraphics[width=1\linewidth]{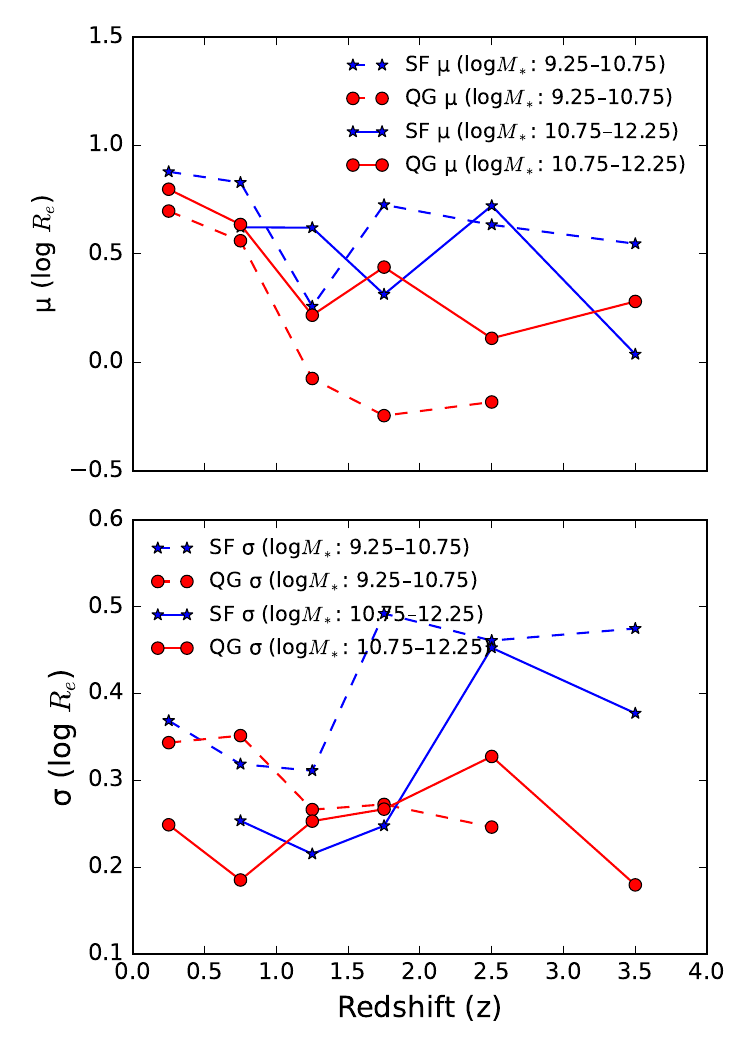}
    \caption{
        Summary of the redshift evolution of the fitted mean (\(\mu\), top) and scatter (\(\sigma\), bottom) of \(\log_{10} R_e\) distributions for BGGs in two stellar mass bins (\(9.25 \leq \log_{10}(M_\ast/M_\odot) < 10.75\) and \(10.75 \leq \log_{10}(M_\ast/M_\odot) < 12.25\)), classified using the consensus method. Star-forming (SFG) and quiescent (QG) populations are shown with blue and red lines, respectively. SFGs display systematically higher mean sizes and larger scatter compared to QGs at all epochs. Both populations show decreasing \(\mu\) toward higher redshift, but scatter increases more strongly in SFGs, reflecting enhanced structural diversity in the early universe.
    }
    \label{fig:summary_logRe_distributions_mass_zbins}
\end{figure}

Figure~\ref{fig:summary_logRe_distributions_mass_zbins} summarizes the redshift evolution of the fitted \(\mu\) (mean \(\log_{10} R_e\)) and \(\sigma\) (scatter) parameters for each mass bin and galaxy type. For both mass bins, the SFGs show systematically higher mean sizes than the QGs at all redshifts. Both populations show a decline in \(\mu\) with increasing redshift, but the decline is steeper for SFGs, particularly in the high-mass bin. The scatter \(\sigma\) increases modestly with redshift for both SFGs and QGs, but the increase is more noticeable for SFGs, especially at \(z > 2\), indicating enhanced structural diversity in earlier epochs. In contrast, QGs maintain consistently narrower scatter across cosmic time, reflecting their more homogeneous and settled morphologies.
In summary, these findings emphasize the evolution of the size of BGGs based on mass and type, showing a diverse, extended growth in SFGs, and initial compaction with later passive size change in QGs.

For a complementary overview using two broad redshift bins to boost statistical power, see Appendix~\ref{app:size_twozbins}.

\subsection{Evolution of star formation surface density} \label{sec:sigma_sfR_evolution}

The star formation surface density, $\Sigma_{\mathrm{SFR}}$, serves as a key diagnostic of the compactness and efficiency of star-forming regions in galaxies. Defined as
\begin{equation}
\Sigma_{\mathrm{SFR}} = \frac{0.5 \times \mathrm{SFR}}{\pi R_e^2},
\end{equation}
this parameter traces the local gas surface density and provides insight into the available fuel for star formation \citep{Kennicutt1998}. High $\Sigma_{\mathrm{SFR}}$ values typically mark compact, intense starburst regions, whereas lower values correspond to extended, disk-like star-forming areas. Moreover, $\Sigma_{\mathrm{SFR}}$ regulates feedback processes such as stellar winds, supernova-driven outflows, and radiation pressure, shaping the interplay between gas depletion, star formation efficiency, and quenching.

\begin{figure}[h]
    \centering
    \includegraphics[width=1\linewidth]{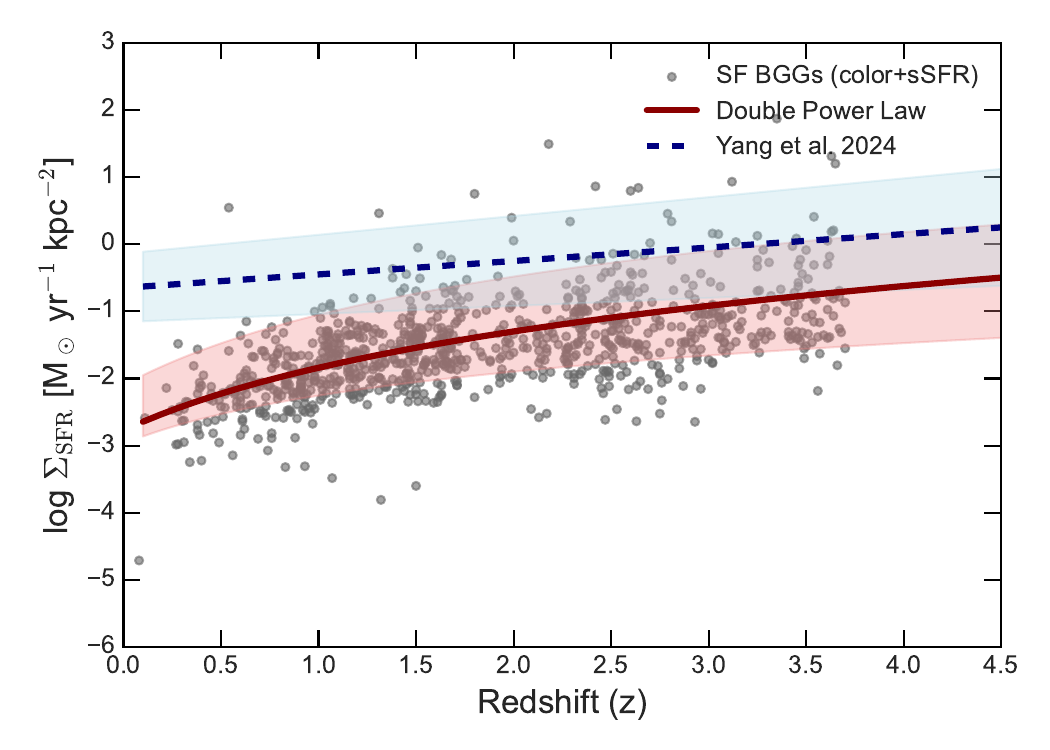}
    \caption{
    Redshift evolution of the star formation surface density, $\log \Sigma_{\mathrm{SFR}}$ [M$_\odot$ yr$^{-1}$ kpc$^{-2}$], for BGGs classified as star-forming using the combined color and sSFR criteria. Gray points show individual BGGs, while black points and error bars mark the median and scatter in redshift bins. The solid dark red line and shaded band show the double power-law best fit with its 1$\sigma$ uncertainty. The dashed dark blue line and shaded band represent the linear fit from \citet{Yang2025}. The results indicate compact, intense star formation at high redshift, transitioning to more extended and diffuse star-forming regions at later cosmic times.
    }
    \label{fig:sigma_sfR_evolution}
\end{figure}

Figure~\ref{fig:sigma_sfR_evolution} shows the redshift evolution of $\log \Sigma_{\mathrm{SFR}}$ for star-forming BGGs, classified using the joint color--sSFR selection. Individual BGGs are plotted as gray points.

We fit the global trend using a double power-law function \citep{madau2014} of the form:
\begin{equation}
\log \Sigma_{\mathrm{SFR}} = \log \left( \frac{(1 + z)^A}{1 + \left[(1 + z)/z_0\right]^B} \right) + C,
\end{equation}
which captures the rising and flattening behavior of $\Sigma_{\mathrm{SFR}}$ across cosmic time. The best-fit parameters are:
\begin{align*}
A &= 2.82 \pm 0.97, \\
B &= -0.54 \pm 1.97, \\
z_0 &= 1.99 \pm 1.14, \\
C &= -2.40 \pm 0.37.
\end{align*}
This model describes an initial rise in $\Sigma_{\mathrm{SFR}}$ toward higher redshift, reflecting more compact and intense star formation at early epochs, followed by a plateau or gentle decline around $z \sim 2$--$3$.

For comparison, we overlay the linear relation from \citet{Yang2025}:
\begin{equation}
\log \Sigma_{\mathrm{SFR}} = (0.20 \pm 0.08) z + (-0.65 \pm 0.51),
\end{equation}
derived from a broader sample of SFGs with stellar masses spanning $\log M_\ast/M_\odot \sim 8$--12. While the two trends qualitatively align, BGGs systematically fall below the general population, highlighting the environmental suppression of star formation surface densities in massive group environments.

Overall, the observed evolution of $\Sigma_{\mathrm{SFR}}$ in BGGs supports a picture of declining star formation efficiency over cosmic time, likely tied to decreasing gas fractions, growing stellar masses, and environmentally driven quenching within dense group halos.

To quantify how BGGs differ from the general population of star-forming galaxies, we compare our best-fit double power-law model for $\Sigma_{\mathrm{SFR}}$ against the linear relation from \citet{Yang2025}, evaluated at five key redshifts. Table~\ref{tab:delta_sigma_sfr} summarizes the predicted $\log \Sigma_{\mathrm{SFR}}$ from both models, as well as the offset:
\begin{equation}
\Delta \log \Sigma_{\mathrm{SFR}} = \log \Sigma_{\mathrm{SFR}}^{\mathrm{BGG}} - \log \Sigma_{\mathrm{SFR}}^{\mathrm{Yang+2024}}.
\end{equation}
\begin{table}[h]
    \centering
    \caption{Comparison between our BGG best-fit double power-law model (DPL) and the \citet{Yang2025} linear fit, evaluated at representative redshifts. The offset $\Delta \log \Sigma_{\mathrm{SFR}}$ indicates how much lower the BGG star formation surface densities are relative to general SFGs.}
    \label{tab:delta_sigma_sfr}
    \begin{tabular}{rrrr}
        \toprule
        Redshift $z$ & DPL Fit (dex) & Yang+2025 (dex) & $\Delta \log \Sigma_{\mathrm{SFR}}$ (dex) \\
        \midrule
        0.50 & -2.24 & -0.55 & -1.69 \\
        1.00 & -1.85 & -0.45 & -1.40 \\
        2.00 & -1.31 & -0.25 & -1.06 \\
        3.00 & -0.93 & -0.05 & -0.88 \\
        4.00 & -0.64 &  0.15 & -0.79 \\
        \bottomrule
    \end{tabular}
\end{table}

These results show that at low redshift ($z = 0.5$), the BGGs are strongly suppressed, lying about $1.7$ dex (a factor of $\sim 50$) below the general SFG relation. This offset decreases with increasing redshift, reducing to $1.4$ dex at $z = 1$, $1.1$ dex at $z = 2$, and $0.8$--$0.9$ dex at $z = 3$--$4$. 

This trend suggests that, while BGGs consistently exhibit lower $\Sigma_{\mathrm{SFR}}$ compared to typical star-forming galaxies across cosmic time, the suppression is strongest at late epochs and becomes less significant at earlier times. This supports an evolutionary scenario where BGGs at high redshift were still building up similarly to field galaxies, but progressively diverged due to mass quenching, gas depletion, and environmental suppression within their growing group-scale halos.
\subsection{Redshift evolution of the $\Sigma_{\mathrm{SFR}}$--$M_\ast$ relation}
\label{sec:sigmasfr_mstar}

To investigate the dependence of star formation surface density on stellar mass and cosmic time, we examine the relation between $\Sigma_{\mathrm{SFR}}$ and stellar mass ($M_\ast$) for BGGs across eight redshift bins from $z = 0$ to $z = 3.7$. We divide the sample into two mass bins, $9.25 \leq \log_{10}(M_\ast/M_\odot) < 10.75$ and $10.75 \leq \log_{10}(M_\ast/M_\odot) < 12.25$, to explore possible mass-dependent evolutionary trends.

\begin{figure}[ht]
    \centering
    \includegraphics[width=1.\linewidth]{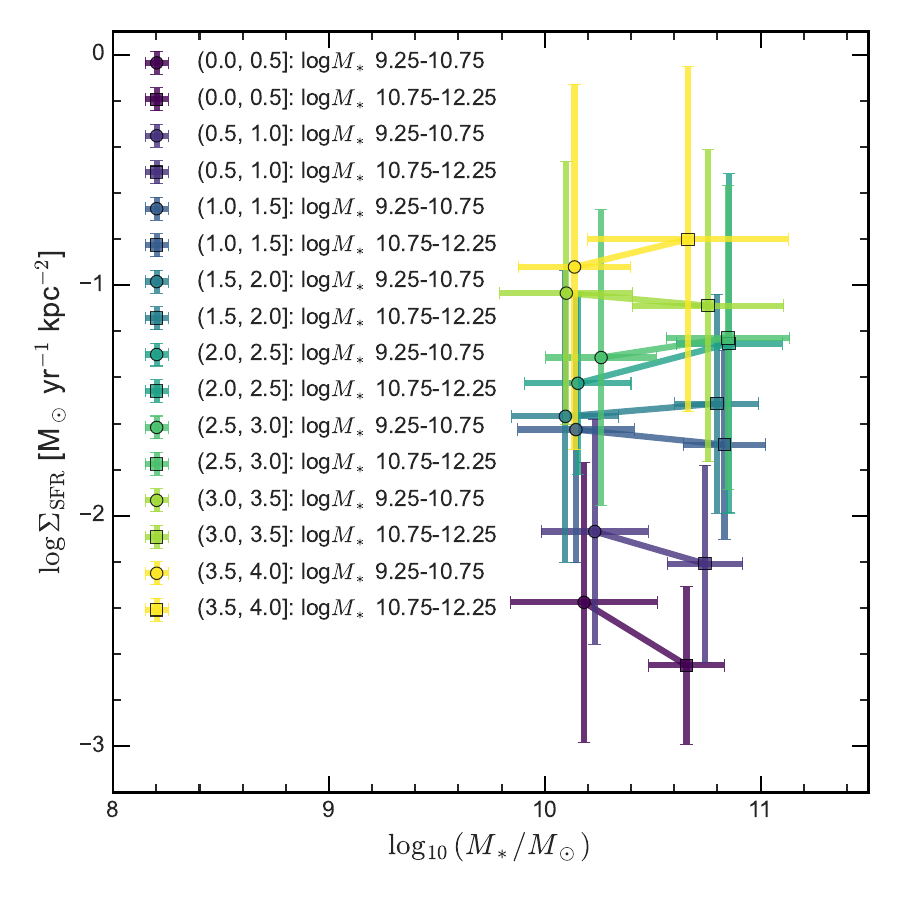}
    \caption{
        Relation between stellar mass surface density of star formation ($\Sigma_{\mathrm{SFR}}$) and stellar mass ($M_\ast$) for BGGs in eight redshift bins from $z=0$ to $z=4$. Circles and squares represent median values for low- and high-mass BGGs, $9.25 \leq \log_{10}(M_\ast/M_\odot) < 10.75$ and $10.75 \leq \log_{10}(M_\ast/M_\odot) < 12.25$, respectively. Error bars denote the standard deviation in both $\log_{10}(M_\ast/M_\odot)$ and $\log_{10}(\Sigma_{\mathrm{SFR}})$. Data points are color-coded by redshift. An increasing trend of $\Sigma_{\mathrm{SFR}}$ with redshift is observed at fixed stellar mass, especially for low-mass BGGs.}
    \label{fig:sigmasfr_mstar_redshift}
\end{figure}

Figure~\ref{fig:sigmasfr_mstar_redshift} shows the median $\log_{10}(\Sigma_{\mathrm{SFR}})$ as a function of $\log_{10}(M_\ast/M_\odot)$ for BGGs, with data points color-coded by redshift bin. Circles represent low-mass BGGs, while squares correspond to high-mass BGGs. Error bars reflect the 1$\sigma$ scatter in both $\log_{10}(\Sigma_{\mathrm{SFR}})$ and $\log_{10}(M_\ast)$. 

We find a clear redshift dependence: at fixed stellar mass, $\Sigma_{\mathrm{SFR}}$ increases with redshift, consistent with the global rise in cosmic star formation activity at earlier times. Notably, the low-mass BGGs exhibit a steeper redshift evolution compared to their high-mass counterparts, indicating more rapid growth in star formation surface density over cosmic time. In contrast, the high-mass BGGs show relatively flat or even declining $\Sigma_{\mathrm{SFR}}$ trends at the highest masses, suggesting that their star formation efficiency has been more strongly suppressed, possibly due to earlier quenching or feedback processes.

The separation between the two mass bins becomes particularly pronounced at $z > 2$, where low-mass BGGs continue to rise sharply in $\Sigma_{\mathrm{SFR}}$, while high-mass BGGs appear to plateau. This behavior may reflect differences in gas accretion, halo assembly bias, or environmental regulation of star formation, with lower-mass BGGs remaining more sensitive to gas supply and feedback at late epochs.

Overall, these results highlight the importance of jointly considering stellar mass, halo mass, and redshift when analyzing the structural and star-forming evolution of BGGs, pointing to distinct evolutionary pathways for systems across the mass spectrum.

\subsection{Redshift evolution of $\Sigma_{\mathrm{SFR}}$ in two stellar mass regimes}
\label{sec:sigmasfr_redshift_massbins}

To investigate the cosmic evolution of the star formation surface density ($\Sigma_{\mathrm{SFR}}$) in different mass regimes, we split the BGG population into two stellar mass bins: $\log_{10}(M_\ast/M_\odot) < 10.75$ (low-mass) and $\log_{10}(M_\ast/M_\odot) \geq 10.75$ (high-mass). We calculate the median $\log_{10}(\Sigma_{\mathrm{SFR}})$ in each redshift bin, with uncertainties estimated from the standard error of the mean.

\begin{figure}[ht]
\centering
\includegraphics[width=0.95\linewidth]{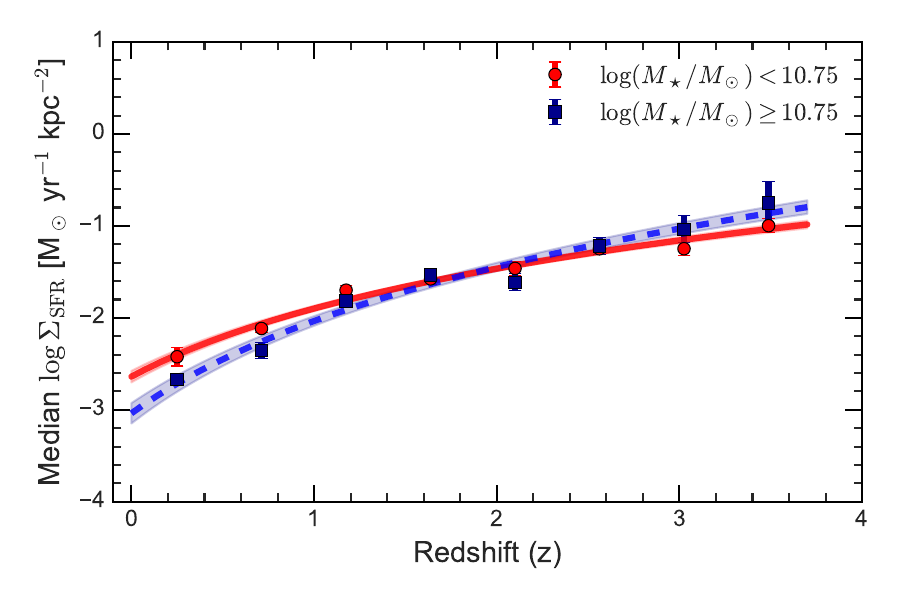}
\caption{
Redshift evolution of the median star formation surface density ($\log\Sigma_{\mathrm{SFR}}$) for BGGs split by stellar mass: $\log_{10}(M_\ast/M_\odot) < 10.75$ (red) and $\log_{10}(M_\ast/M_\odot) \geq 10.75$ (blue). Solid and dashed curves show best-fit power-law relations, with shaded areas representing $1\sigma$ confidence bands. The best-fit slopes and intercepts are reported in the text. While overall trends are similar, KS test results (Table~\ref{tab:ks_test_results}) reveal significant differences at $z \approx 0.7$–1.2, indicating possible mass-dependent effects during this period.
}
\label{fig:sigmasfr_redshift_massbins}
\end{figure}

Figure~\ref{fig:sigmasfr_redshift_massbins} shows the redshift evolution of $\log\Sigma_{\mathrm{SFR}}$ for the two mass bins, with red and blue curves representing the low- and high-mass BGGs, respectively. We fit a power-law relation of the form $\log\Sigma_{\mathrm{SFR}} = a \log(1 + z) + b$ to both populations, shown as solid (low-mass) and dashed (high-mass) lines, with shaded bands representing the $1\sigma$ confidence intervals derived from the fit covariances.

The best-fit parameters are:
\begin{itemize}
    \item \textbf{Low-mass bin} ($\log_{10}(M_\ast/M_\odot) < 10.75$): $a = 2.46 \pm 0.14$, $b = -2.64 \pm 0.06$
    \item \textbf{High-mass bin} ($\log_{10}(M_\ast/M_\odot) \geq 10.75$): $a = 3.34 \pm 0.24$, $b = -3.04 \pm 0.11$
\end{itemize}

These fits indicate that while both mass bins show an overall increase in $\Sigma_{\mathrm{SFR}}$ with redshift, the high-mass BGGs exhibit a somewhat steeper evolution compared to their low-mass counterparts.

To statistically assess the difference between the two populations, we performed Kolmogorov–Smirnov (KS) tests comparing the $\log\Sigma_{\mathrm{SFR}}$ distributions between the low- and high-mass BGGs in each redshift bin. 
These results suggest that while the overall trends are similar across mass bins, significant differences emerge at intermediate redshifts ($z \approx 0.7$–1.2), potentially indicating transient mass-dependent effects during that epoch.

\begin{table*}[ht]
\centering
\caption{
    Summary of the median star formation surface density ($\log\Sigma_{\mathrm{SFR}}$) for BGGs separated by stellar mass at $\log_{10}(M_\ast/M_\odot) = 10.75$ in each redshift bin. Reported are the median $\log\Sigma_{\mathrm{SFR}}$ (in M$_\odot$ yr$^{-1}$ kpc$^{-2}$) for low-mass ($\log_{10}(M_\ast/M_\odot) < 10.75$) and high-mass ($\log_{10}(M_\ast/M_\odot) \geq 10.75$) BGGs, along with their standard errors. The Kolmogorov–Smirnov (KS) test compares the $\log\Sigma_{\mathrm{SFR}}$ distributions between mass bins, reporting the KS statistic and p-value. P-values below 0.05 are considered statistically significant.
}
\begin{tabular}{c c c c c c c}
\toprule
$z_{\mathrm{bin}}$ & Low-mass median & High-mass median & Low-mass err & High-mass err & KS stat & KS p-val \\
\midrule
0.25 & -2.42 & -2.67 & 0.10 & 0.02 & 0.71 & 0.21 \\
0.71 & -2.12 & -2.36 & 0.05 & 0.08 & 0.31 & 0.03 \\
1.18 & -1.70 & -1.82 & 0.05 & 0.04 & 0.22 & 0.03 \\
1.64 & -1.58 & -1.54 & 0.06 & 0.05 & 0.16 & 0.21 \\
2.10 & -1.46 & -1.61 & 0.08 & 0.09 & 0.24 & 0.21 \\
2.56 & -1.25 & -1.22 & 0.06 & 0.09 & 0.16 & 0.33 \\
3.03 & -1.25 & -1.04 & 0.07 & 0.15 & 0.25 & 0.20 \\
3.49 & -1.00 & -0.75 & 0.07 & 0.24 & 0.27 & 0.26 \\
\bottomrule
\end{tabular}
\label{tab:ks_test_results}
\end{table*}

\subsection{Structural transition and morphology-quenching connection in BGGs}  
\label{sec:compaction_sersic}  

To explore the connection between galaxy structure and star formation activity in BGGs, we examine their distribution in the $\log_{10}(\Sigma_*)$ vs. $\log_{10}(\mathrm{sSFR})$ plane, with points color-coded by Sérsic index $n_{\text{s\'ersic}}$. Here, $\Sigma_* = M_\ast / (2\pi R_e^2)$ is the stellar mass surface density within the effective radius $R_e$, and sSFR is derived from SED-based measurements. Galaxies with $n_{\text{s\'ersic}} > 2.5$ are classified as bulge-dominated. We divide the galaxy sample into four redshift bins: $0.08 < z < 1.0$, $1.0 < z < 2.0$, $2.0 < z < 3.0$, and $3.0 < z < 4.0$.  

\vspace{0.2cm}  
\noindent  
\textbf{Morphological classification criteria.}  
We further classify BGGs into five morphological categories based on their structural parameters, using a combination of Sérsic index ($n_{\text{s\'ersic}}$), axis ratio ($q_{\mathrm{ratio}}$), and effective radius ($R_e$ in arcseconds):  
\begin{itemize}
    \item \textbf{Irregular Clumpy}: Galaxies with $n_{\text{s\'ersic}} < 1.5$ that show asymmetric or irregular features, identified by either $q_{\mathrm{ratio}} < 0.4$ or $R_e > 0.5''$.  
    \item \textbf{Disk-like}: Galaxies with $n_{\text{s\'ersic}} < 1.5$ but regular shapes ($q_{\mathrm{ratio}} \geq 0.4$ and $R_e \leq 0.5''$), as well as those with $1.5 \leq n_{\text{s\'ersic}} < 2.5$.  
    \item \textbf{Compact Spheroid}: Spheroid-dominated systems with $2.5 \leq n_{\text{s\'ersic}} < 3.0$ and compact sizes ($R_e < 0.3''$).  
    \item \textbf{Intermediate Spheroid}: Systems with $2.5 \leq n_{\text{s\'ersic}} < 3.0$ and $R_e \geq 0.3''$, or $3.0 \leq n_{\text{s\'ersic}} < 4.0$.  
    \item \textbf{Elliptical}: Bulge-dominated galaxies with $n_{\text{s\'ersic}} \geq 4.0$, consistent with the profiles of classical ellipticals.  
\end{itemize}

These definitions distinguish between irregular star-forming disks, stable disk systems, compact quenching candidates, intermediate spheroid–disk hybrids, and fully quenched ellipticals. The axis ratio and size thresholds refine the separation, particularly among low- and intermediate-Sérsic galaxies.

\begin{figure*}[t]
    \centering
    \includegraphics[width=\linewidth]{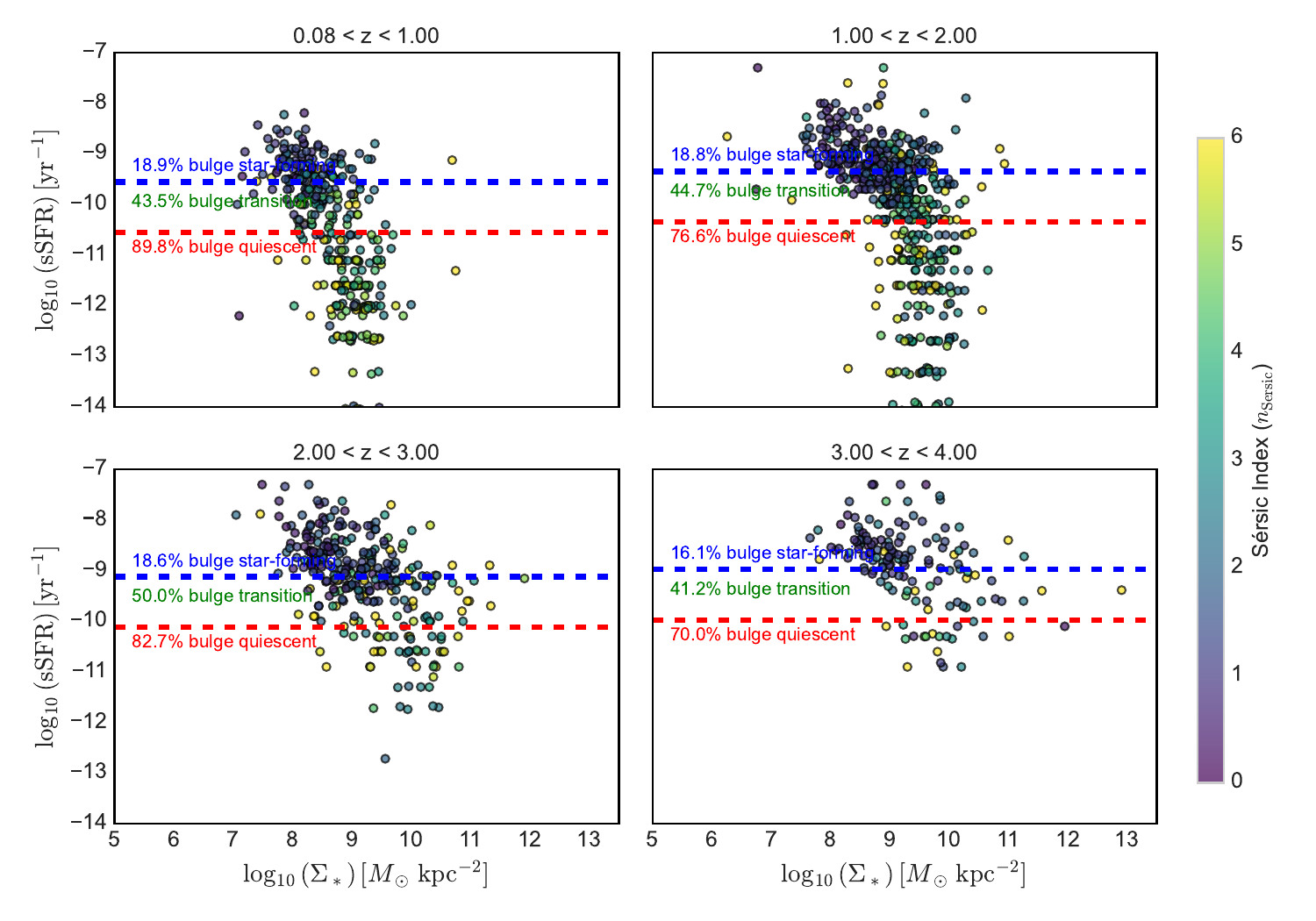}
    \caption{
        Stellar mass surface density ($\log_{10} \Sigma_*$) versus specific star formation rate ($\log_{10} \mathrm{sSFR}$) for BGGs in four redshift bins ($0.08 < z < 1.0$, $1.0 < z < 2.0$, $2.0 < z < 3.0$, $3.0 < z < 4.0$). Points are color-coded by Sérsic index $n_{\text{s\'ersic}}$, with the colorbar at right. The red dashed line marks the quiescent threshold, and the blue dashed line marks the transition range. Percentages indicate the fraction of bulge-dominated systems ($n_{\text{s\'ersic}} > 2.5$) in each phase and redshift bin, revealing a progressive increase in bulge fraction from star-forming to quiescent galaxies.
    }
    \label{fig:sigmae_sersic_plane}
\end{figure*}

\vspace{0.2cm}  
\noindent  
In Figure~\ref{fig:sigmae_sersic_plane}, we show the $\log_{10}(\Sigma_*)$--$\log_{10}(\mathrm{sSFR})$ relation for BGGs in the four redshift bins. The red dashed line marks the quiescence threshold $\log_{10}(\mathrm{sSFR}) = \log_{10}(0.2 / t_{\mathrm{obs}})$, and the blue dashed line identifies the transition region offset by +1 dex. Bulge-dominated fractions are annotated within each phase (star-forming, transition, quiescent), providing a quantitative view of structural transformation over time.

We observe the expected L-shaped sequence, consistent with the evolutionary compaction scenario proposed by \citet{Barro2017}. Star-forming BGGs predominantly occupy the lower $\Sigma_*$, low-$n_{\text{s\'ersic}}$ region, while quiescent systems are more compact and bulge-dominated. Transitioning galaxies bridge these regimes, indicating ongoing morphological transformation accompanying quenching.

\begin{figure}[t]
    \centering
    \includegraphics[width=0.95\linewidth]{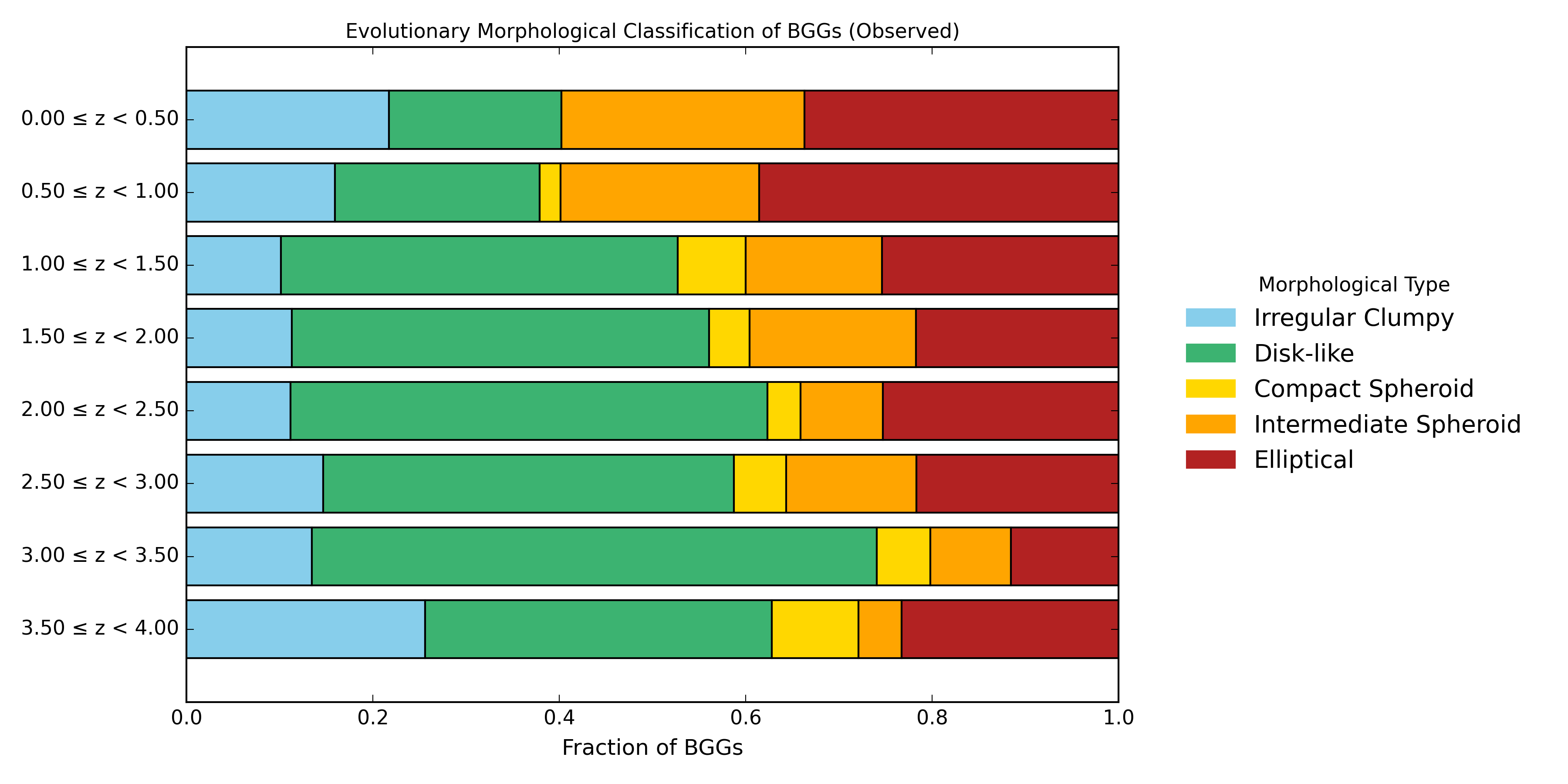}
    \caption{
        Evolution of the morphological fraction of BGGs across redshift, separated into five classes: Irregular Clumpy, Disk-like, Compact Spheroid, Intermediate Spheroid, and Elliptical. Stacked bar heights represent the fractional contribution within each redshift bin ($\Delta z = 0.5$). The increasing fraction of spheroid-dominated and elliptical systems toward lower redshifts reflects the cumulative impact of morphological transformation and quenching processes in the BGG population.
    }
    \label{fig:morph_fraction_redshift}
\end{figure}

To complement this, Figure~\ref{fig:morph_fraction_redshift} presents the redshift evolution of morphological fractions across the five classes. We calculate the fractional contribution of each morphological type in redshift bins of size $\Delta z = 0.5$. We find that:
\begin{itemize}
    \item Disk-like systems dominate across all redshifts but show modest decline at lower redshift.
    \item Compact and intermediate spheroids increase in representation at intermediate redshifts ($1 < z < 3$), reflecting morphological transformation pathways.
    \item Ellipticals rise in fraction toward lower redshifts, consistent with cumulative quenching and structural settling.
\end{itemize}

These results support a picture of gradual transformation from irregular and disk-dominated systems toward spheroid- and elliptical-dominated morphologies, driven by compaction, internal instabilities, and quenching processes. Our findings align with simulations by \citet{Shen2024} and observations by \citet{Huertas-Company2024} and \citet{Carnall2023}, which highlight the role of central density buildup and morphological change in regulating the shutdown of star formation in massive galaxies.

\subsection{BGG Catalog}
\label{sec:bgg_catalog}

We construct the BGG catalog by identifying the most massive and luminous galaxy within each group in the COSMOS-Web group sample as discussed in Sec. \ref{sec:bgg_selection}. Each group is characterized by its central coordinates (RA\_GR, DEC\_GR), redshift (Z\_GR) from \cite{Toni2025}, and an associated group radius defined as the hybrid radius derived from the AMICO detection algorithm. The galaxy membership within each group is determined using proximity in both spatial and redshift dimensions.

The full physical properties of BGGs can be obtained from the COSMOS2025 multiwavelength catalog (Shuntov et al. subm.) through cross-matching with the SE++ catalog. We use the unique galaxy identifier \texttt{ID\_COSMOS2025\_SE++} for matching.  

The BGG catalog includes the following columns:
\begin{itemize}
    \item \texttt{Group\_ID}: Unique identifier of the galaxy group.
    \item \texttt{RA\_GR, DEC\_GR}: Right ascension and declination of the center of the group (in degrees).
    \item \texttt{Z\_GR}: Redshift of the group.
    \item \texttt{Hybrid\_radius}: Estimated physical radius of the group in arcminutes.
    \item \texttt{ID\_COSMOS2025\_SE++}: BGG source identifier from the COSMOS2025 SE++ catalog.
    \item \texttt{RA\_DETEC, DEC\_DETEC}: Right ascension and declination of the detected BGG (in degrees).
    \item \texttt{logM}: Stellar mass of the BGG, in units of $\log_{10}(M_\odot)$.
    \item \texttt{MR}: Absolute magnitude of the rest frame in the R-band.
    \item \texttt{flag\_type}: A quality or classification flag indicating the BGG selection type or ambiguity in group membership.
\end{itemize}

The BGG catalog is accessible at:\\
\url{https://github.com/gozaliasl/COSMOS-WEB-Brightest-Groups-Galaxies-.git}. 

This catalog serves as the basis for our subsequent structural and star-forming analyses of BGGs across cosmic time.


\section{Conclusion}
\label{sec:conclusions}

We have analyzed the structural and star-forming evolution of $\sim$1700 BGGs over $0.08 < z < 3.7$ using COSMOS-Web NIRCam imaging. Our main conclusions are as follows.

\begin{enumerate}
    \item \textbf{Size–mass scaling:} BGGs follow distinct size–mass relations. Quiescent BGGs are more compact and exhibit steeper slopes compared to star-forming counterparts, suggesting different growth mechanisms and quenching routes.

    \item \textbf{Size growth with redshift:} At fixed stellar mass ($(M_\ast = 5\times 10^{10} M_\odot$), the effective radius evolves as $R_e \propto (1+z)^{-\alpha}$, with $\alpha = 1.31 \pm 0.071$ for star-forming galaxies and $\alpha = 1.40 \pm 0.09$ for quiescent galaxies. This differential growth suggests gas-driven expansion in star-forming BGGs and merger-driven evolution in quenched systems.

    \item \textbf{Redshift-dependent $\Sigma_{\mathrm{SFR}}$:} Star formation surface density increases with redshift for all BGGs, but low-mass systems maintain elevated $\Sigma_{\mathrm{SFR}}$ longer, implying more extended star formation histories compared to their high-mass counterparts.

    \item \textbf{Mass-dependent quenching and structural transition:} Quiescent BGGs exhibit a high concentration in the $\Sigma_*$–sSFR plane, with bulge-dominated morphologies making up 80\% of the population. This supports a compaction–quenching sequence linked to morphological transformation.

    \item \textbf{Robustness across classification methods:} The observed trends in morphology, $\Sigma_{\mathrm{SFR}}$, and size evolution persist across NUV–$r$–$J$, sSFR-based, and joint classification schemes, confirming the physical consistency of our results.
\end{enumerate}

These findings emphasize the role of BGGs as testbeds for understanding the interplay between environment, structure, and star formation in galaxy evolution. The unprecedented depth and resolution of JWST/COSMOS-Web open a new window into the high-redshift group regime, enabling future studies of feedback, satellite accretion, and baryonic assembly in group-scale halos.
\begin{acknowledgements}  We acknowledge the contribution of the COSMOS collaboration, consisting of more than 200 scientists. More information about the
COSMOS survey can be found at https://cosmos.astro.caltech.edu/.
This work was made possible by using the CANDIDE cluster at the Institut d’Astrophysique de Paris. The cluster was funded through grants from
the PNCG, CNES, DIM-ACAV, the Euclid Consortium, and the Danish National Research Foundation Cosmic Dawn Center (DNRF140). It is maintained
by Stephane Rouberol.
      Part of this work was supported by the German
      \emph{Deut\-sche For\-schungs\-ge\-mein\-schaft, DFG\/} project
      number Ts~17/2--1. Part of this research was carried out at the Jet Propulsion Laboratory, California Institute of Technology, under a contract with the National Aeronautics and Space Administration (80NM0018D0004).  French COSMOS team members are partly supported by the Centre National d’Etudes Spatiales (CNES). We acknowledge the funding of the French Agence Nationale de la Recherche for the project iMAGE (grant ANR-22-CE31-0007). LM acknowledges the financial contribution from the PRIN-MUR 2022 20227RNLY3 grant “The concordance cosmological model: stress-tests with galaxy clusters” supported by Next Generation EU and
from the grant ASI n. 2024-10-HH.0 “Attività scientifiche per lamissione Euclid – fase E”.
\end{acknowledgements}

%
 \bibliographystyle{aa} 
   \bibliography{main} 

\appendix

\section{Multi-band Sérsic Fits for BGGs}

To complement the structural modeling presented in the main text, we show in this appendix the Sérsic profile fits for the same six BGGs across three additional JWST/NIRCam filters: F150W, F277W, and F444W. These figures provide visual confirmation of the fitting quality and structural consistency across different wavelengths.

\begin{figure}[h!]
    \centering
        \includegraphics[width=0.5\textwidth]{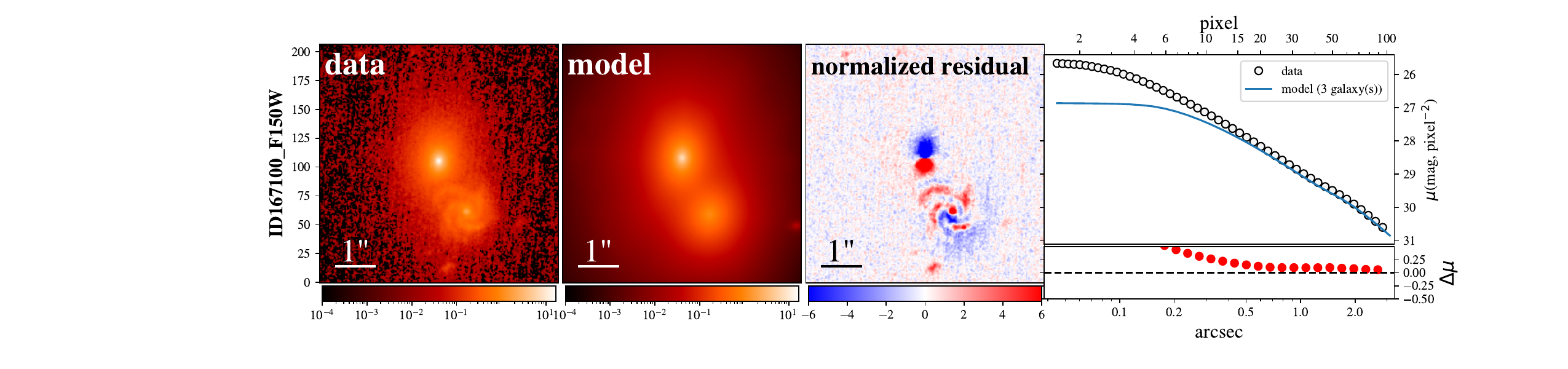} 
        \includegraphics[width=0.5\textwidth]{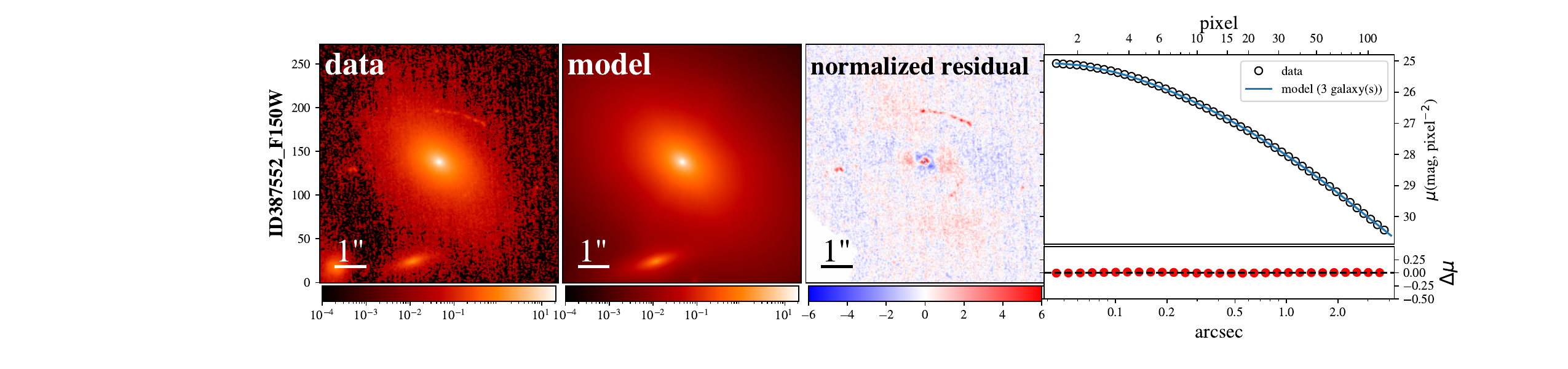}
        \includegraphics[width=0.5\textwidth]{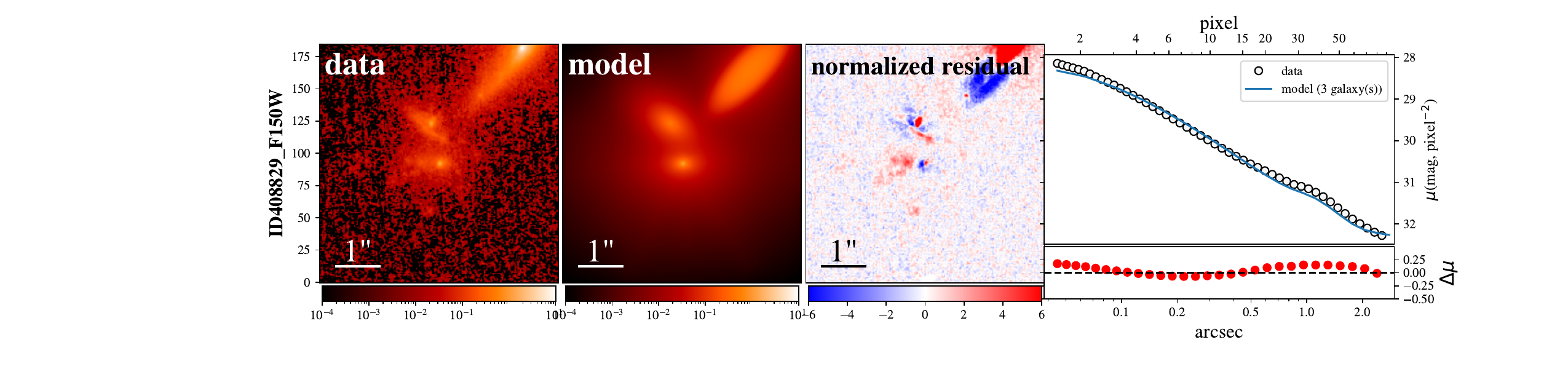} \\
        \includegraphics[width=0.5\textwidth]{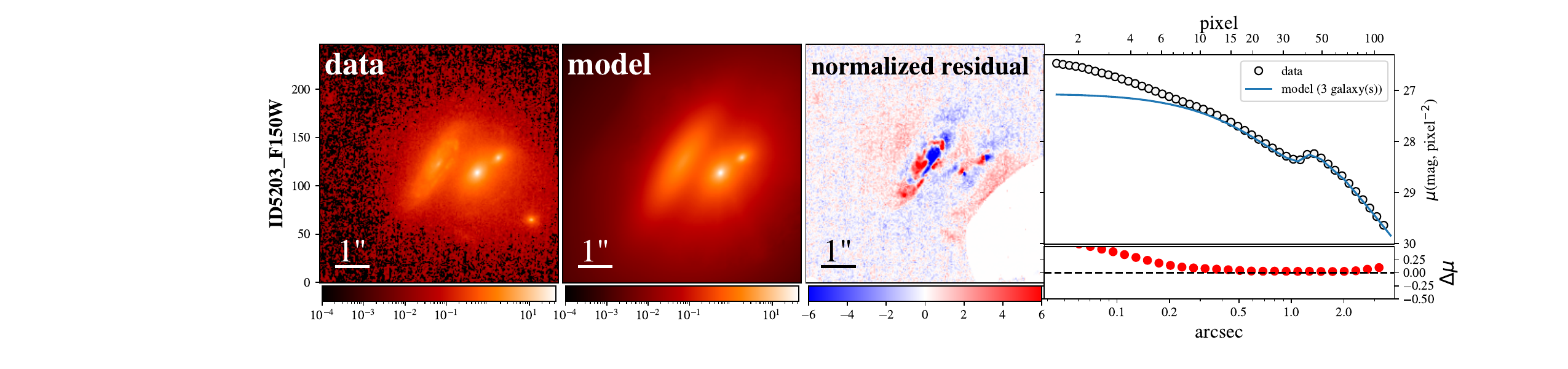}
        \includegraphics[width=0.5\textwidth]{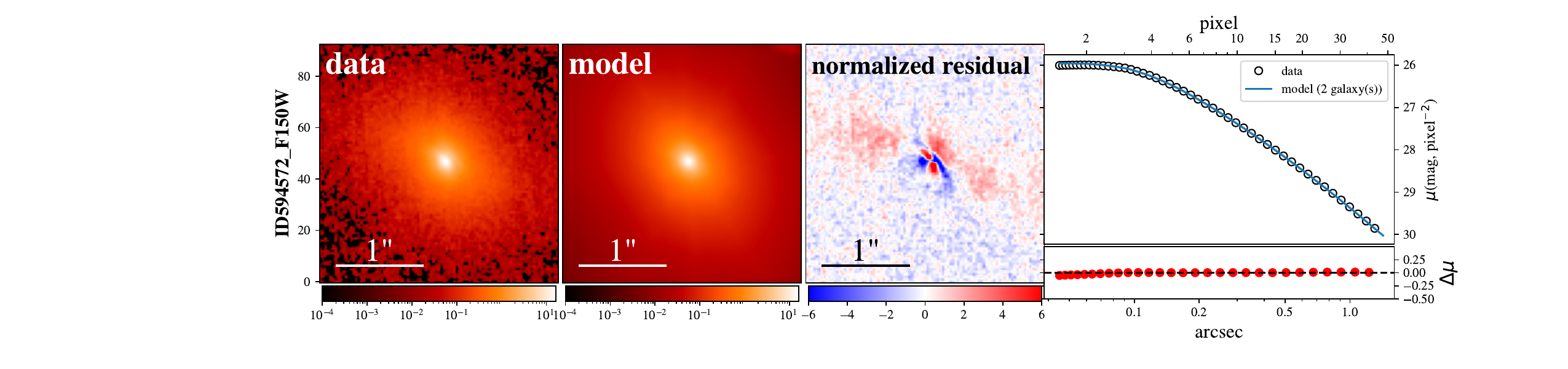} 
        \includegraphics[width=0.5\textwidth]{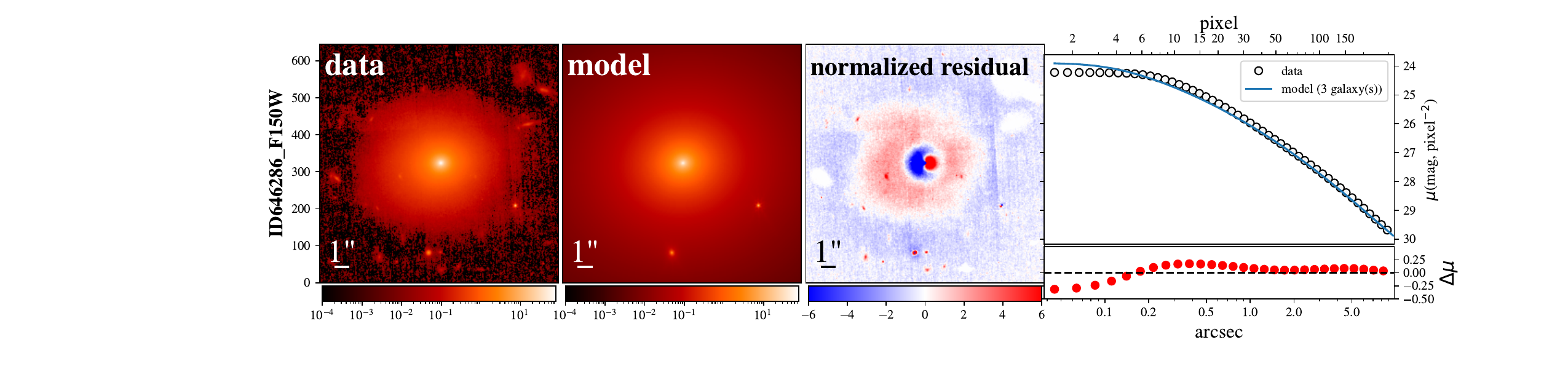}
    \caption{F150W Band: Sersic fit for six galaxies across the band.}
    \label{fig:F150W_band}
\end{figure}

\begin{figure}[h!]
    \centering

        \includegraphics[width=0.5\textwidth]{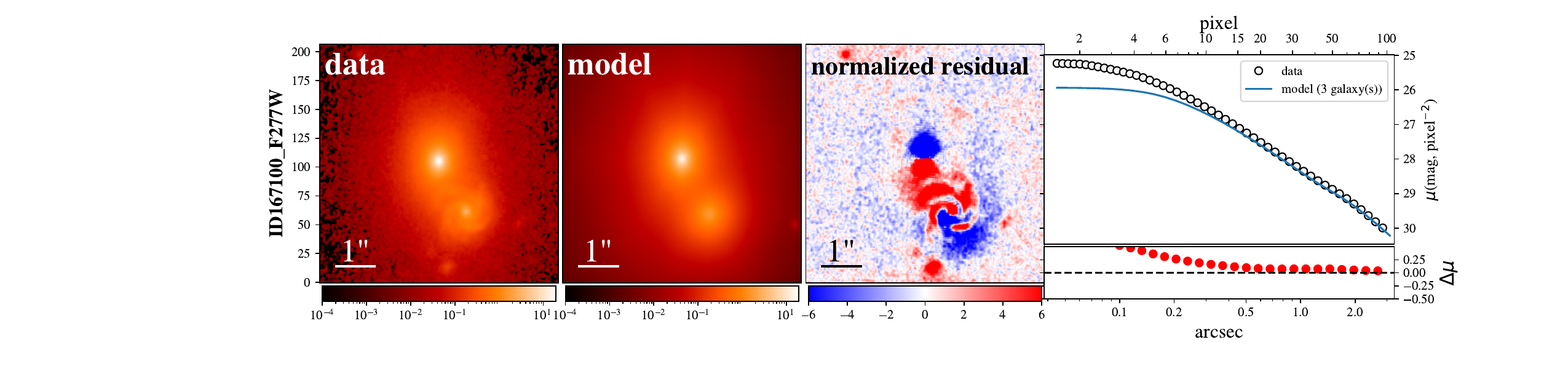} 
        \includegraphics[width=0.5\textwidth]{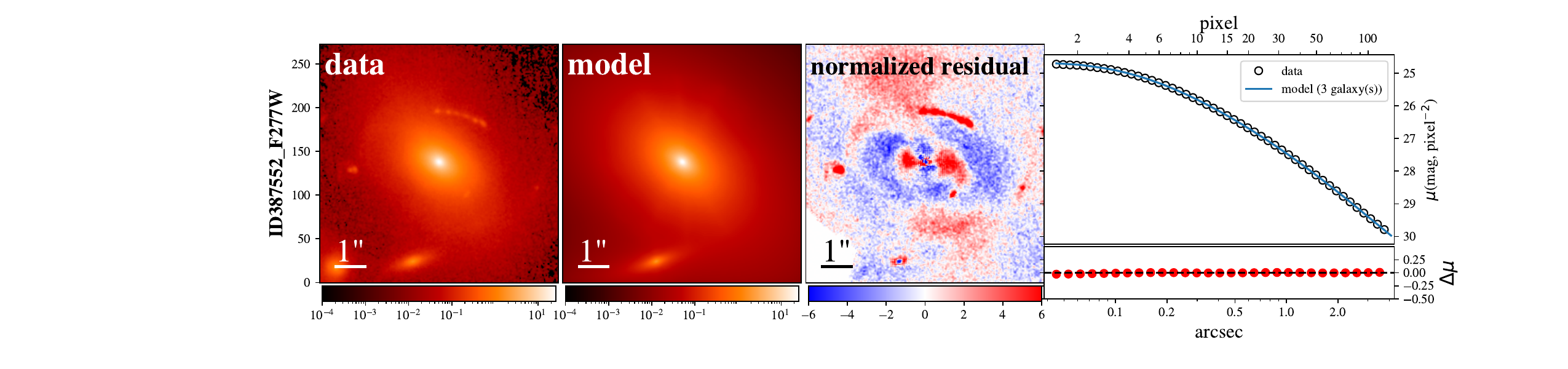}
        \includegraphics[width=0.5\textwidth]{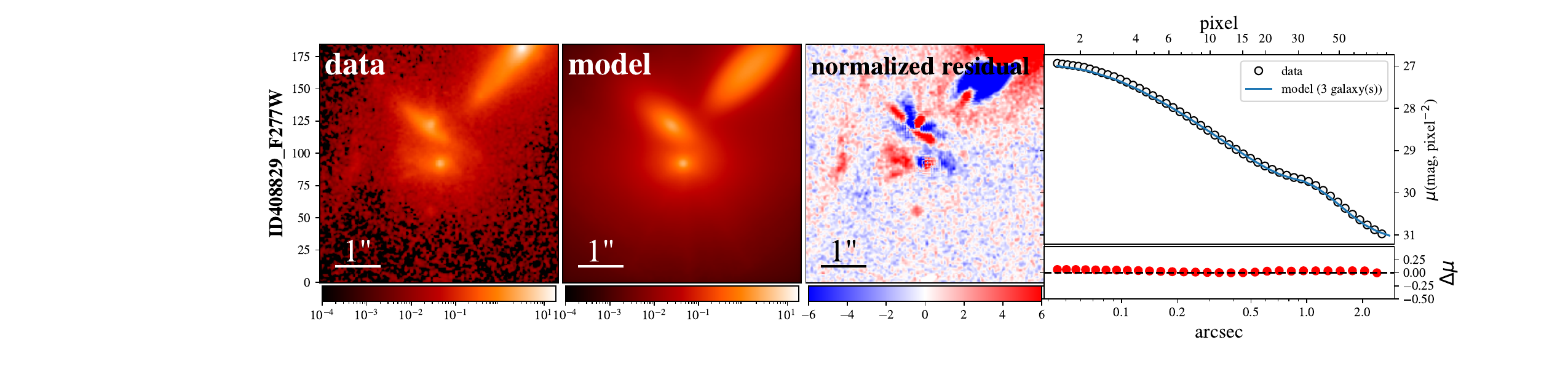} \\
        \includegraphics[width=0.5\textwidth]{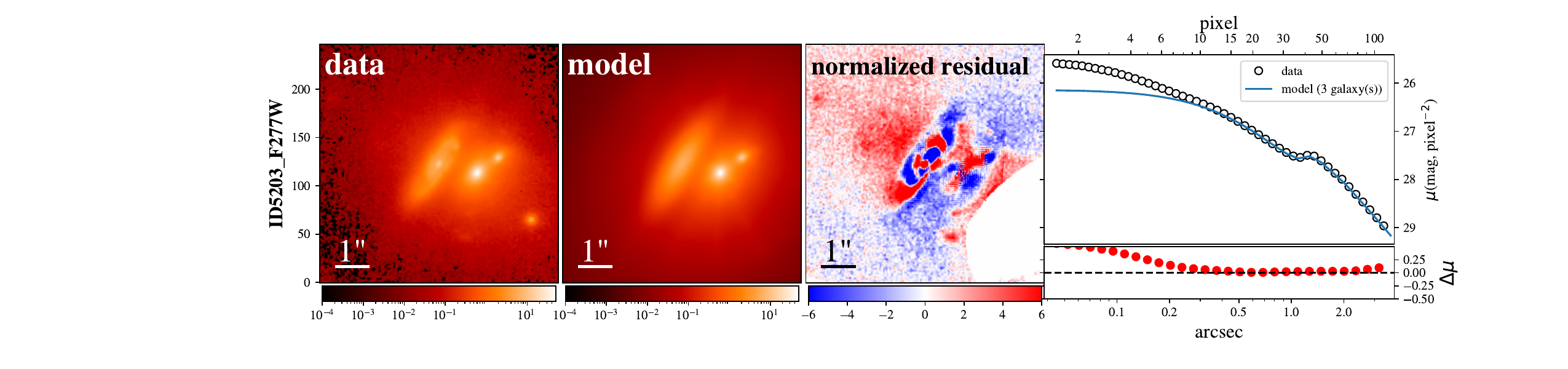}
        \includegraphics[width=0.5\textwidth]{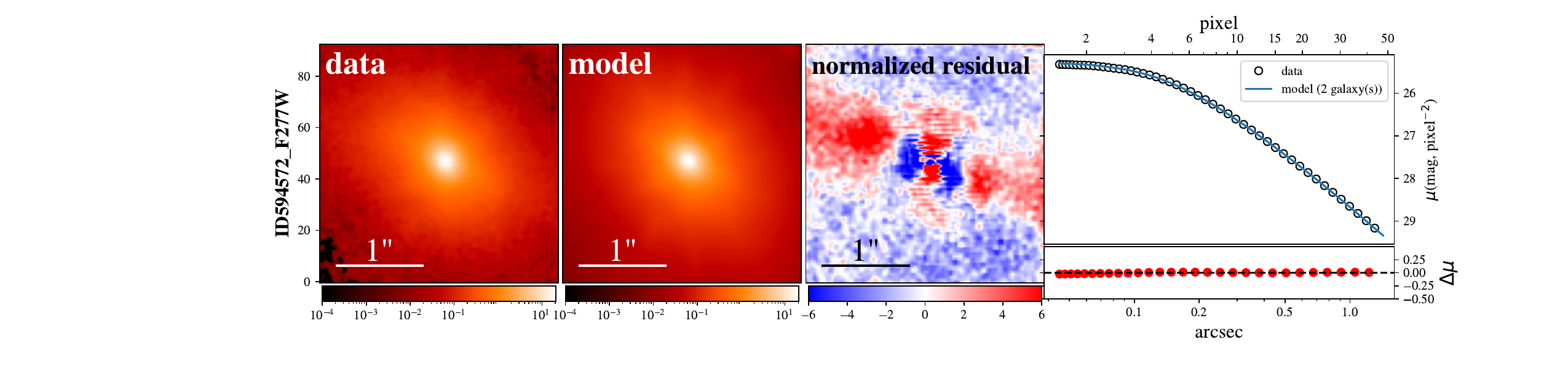} 
        \includegraphics[width=0.5\textwidth]{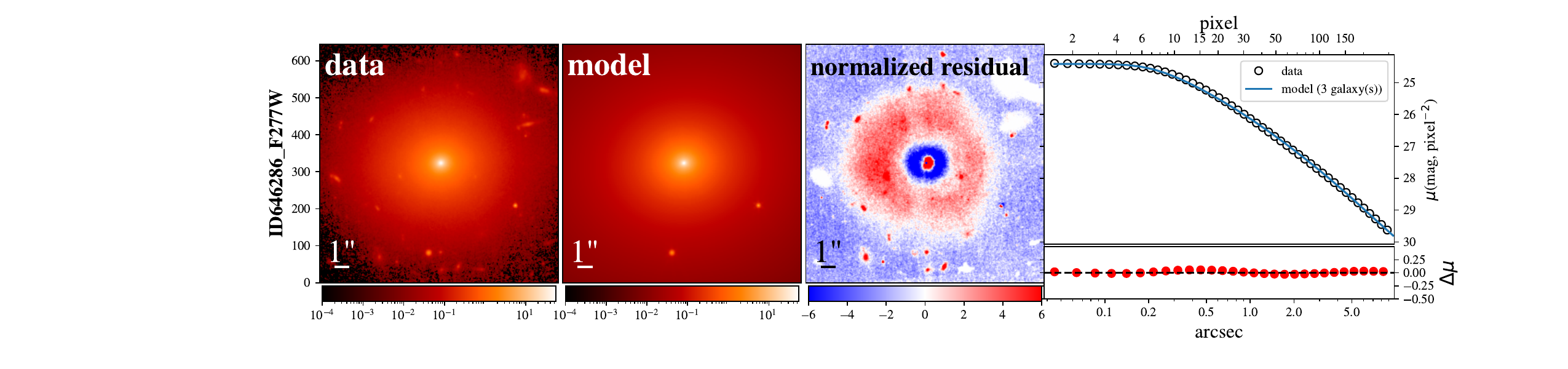}

    \caption{F277W Band: Sersic fit for six galaxies across the band.}
    \label{fig:F277W_band}
\end{figure}

\begin{figure}[h!]
    \centering

        \includegraphics[width=0.5\textwidth]{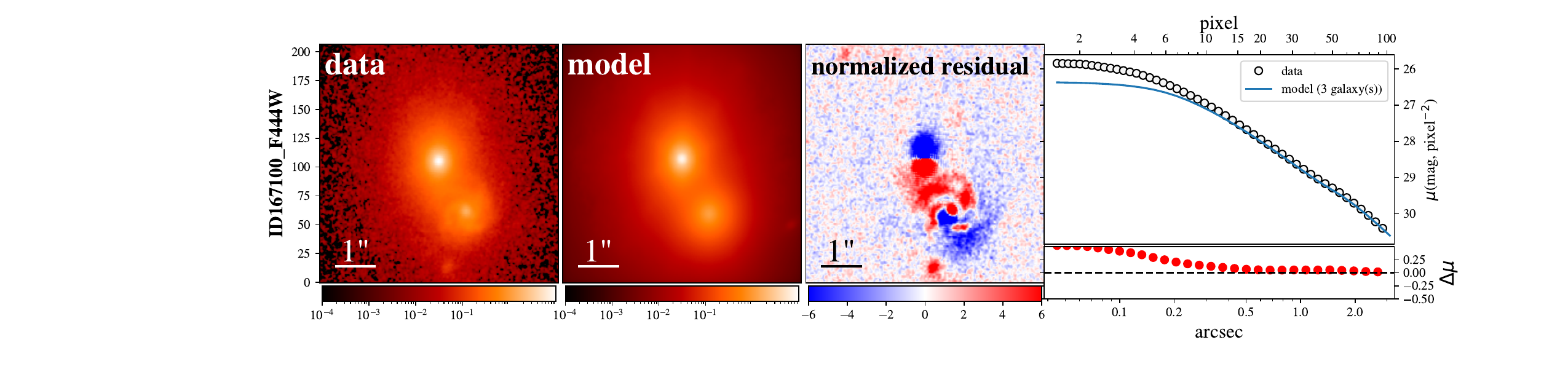} 
        \includegraphics[width=0.5\textwidth]{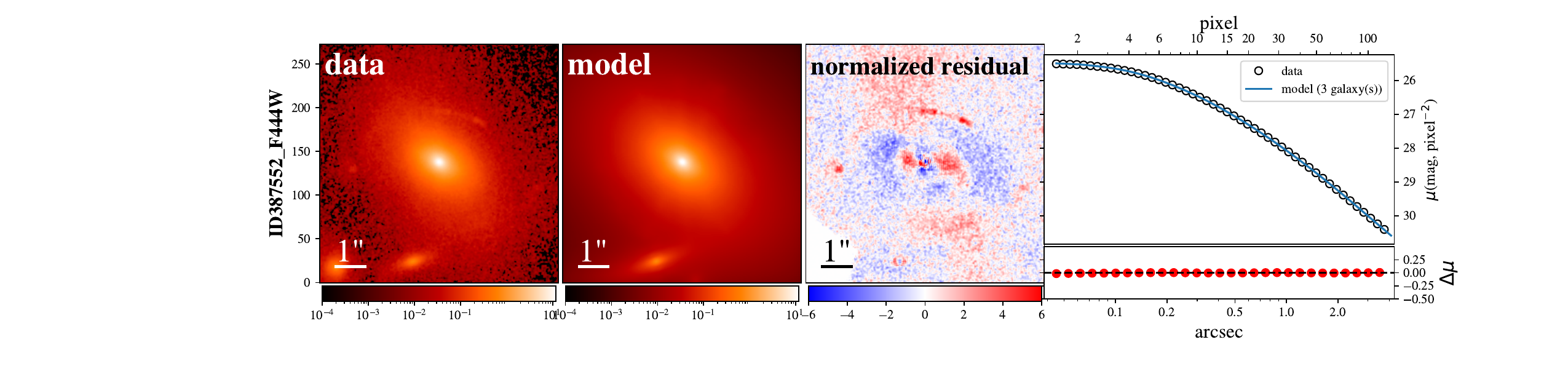}
        \includegraphics[width=0.5\textwidth]{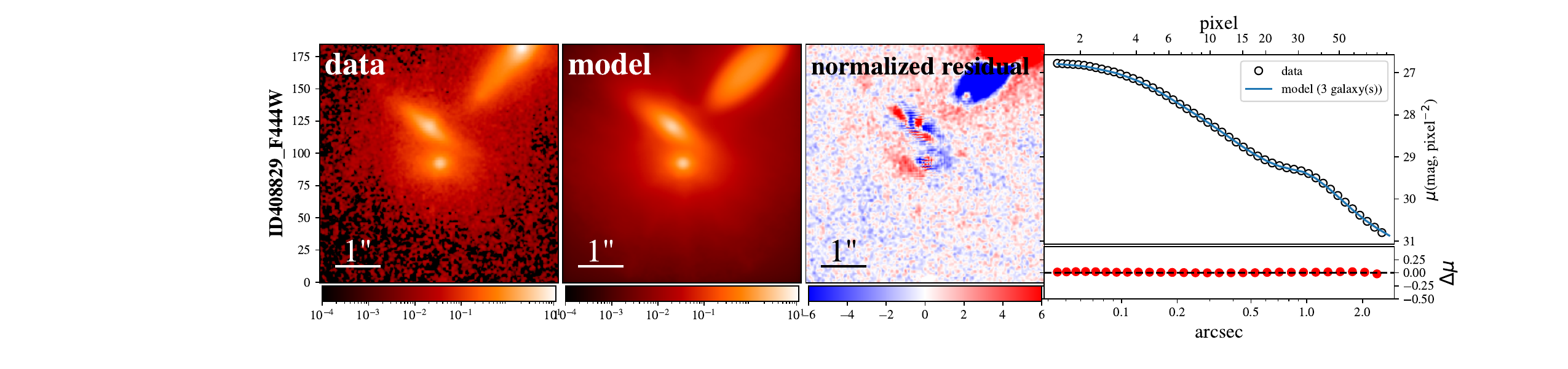} \\
        \includegraphics[width=0.5\textwidth]{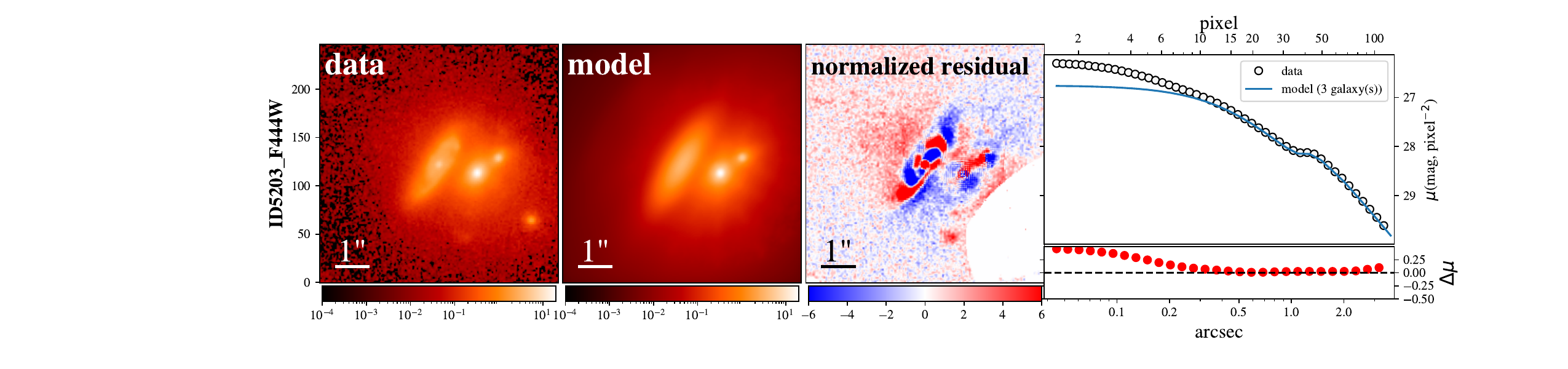}
        \includegraphics[width=0.5\textwidth]{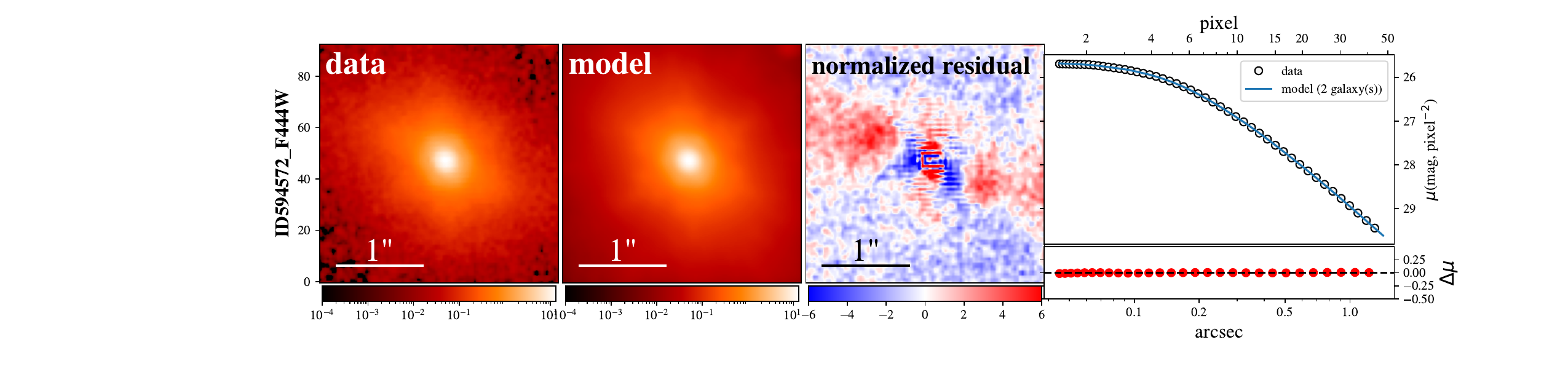} 
        \includegraphics[width=0.5\textwidth]{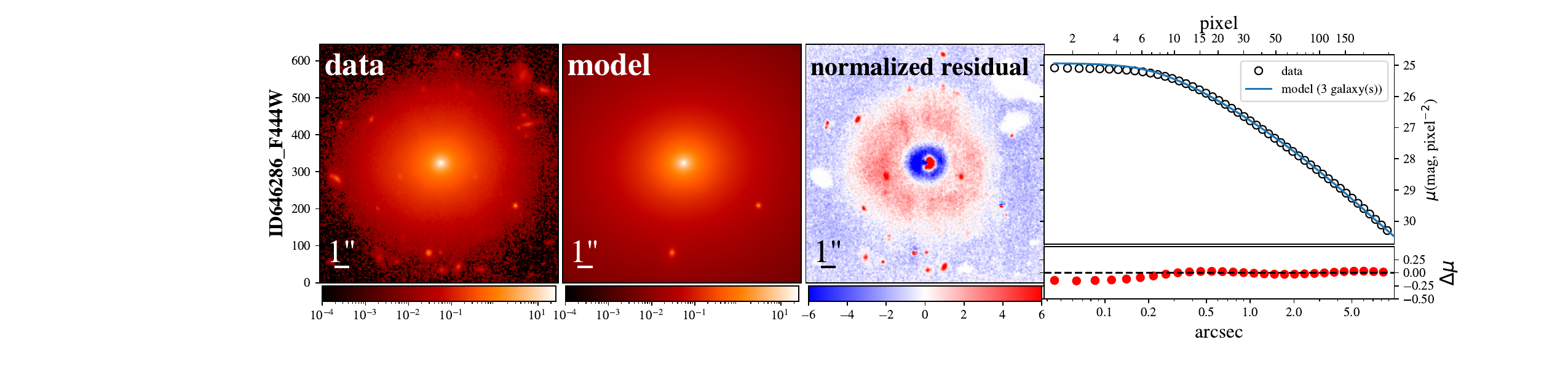}

    \caption{F444W Band: Sersic fit for six galaxies across the band.}
    \label{fig:F444W_band}
\end{figure}

Figures~\ref{fig:F150W_band}–\ref{fig:F444W_band} follow the same format as Figure~\ref{fig:F115W_band} and include, for each galaxy: the original data cutout, the best-fit Sérsic model, the normalized residual map, and the azimuthally averaged radial profile comparison between data and model. Residuals remain small and symmetric in most cases, indicating robust fits. The light profiles remain stable across filters, although subtle differences in morphology and concentration are apparent due to wavelength-dependent features such as dust attenuation or star-forming clumps.

These multi-band fits allow us to trace rest-frame optical sizes over a wide redshift range and ensure that our structural parameters are not biased by single-band anomalies. Combined, these figures demonstrate the reliability and consistency of the structural measurements used in our analysis.

\section{Size--Mass relations from alternative classification methods}
\label{app:alternative_classifications}

\begin{figure*}[h]
    \centering
    \includegraphics[width=0.5\linewidth]{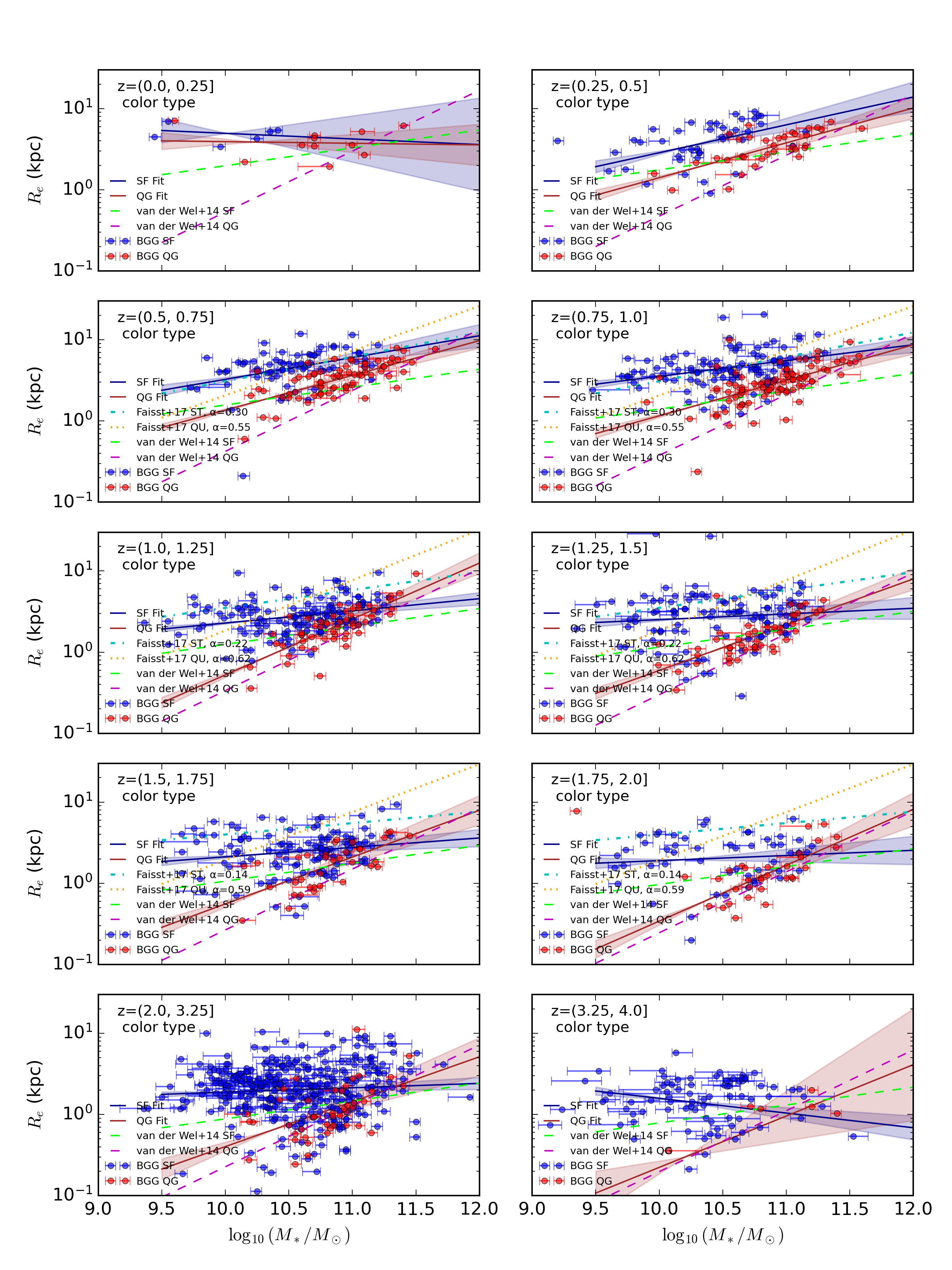}
    \caption{
        \textbf{Size--mass relation of BGGs using color-only classification.}
        Panels show BGGs across eight redshift bins from $z = 0$ to $z= 3.7$ ($\Delta z = 0.5$). Blue circles represent star-forming BGGs and red circles show quiescent BGGs, as identified using rest-frame NUV--$r$--$J$ colors. The dashed and solid lines represent best-fit size--mass relations for SFGs and QGs, respectively, with shaded regions indicating 1$\sigma$ uncertainties. SFGs show generally shallower slopes and larger sizes, while QGs are more compact, especially below $z \sim 2$. This classification is sensitive to dust reddening and intermediate colors. Results from \citet{Faisst2017} are shown as cyan dashed-dotted (SFGs) and orange dotted (QGs) lines in matching redshift bins for comparison.While dashed lime and dashed magenta lines show \citet{vdW2014} results.
    }
    \label{fig:size_mass_color_only}
\end{figure*}

\begin{figure*}[h]
    \centering
    \includegraphics[width=0.5\linewidth]{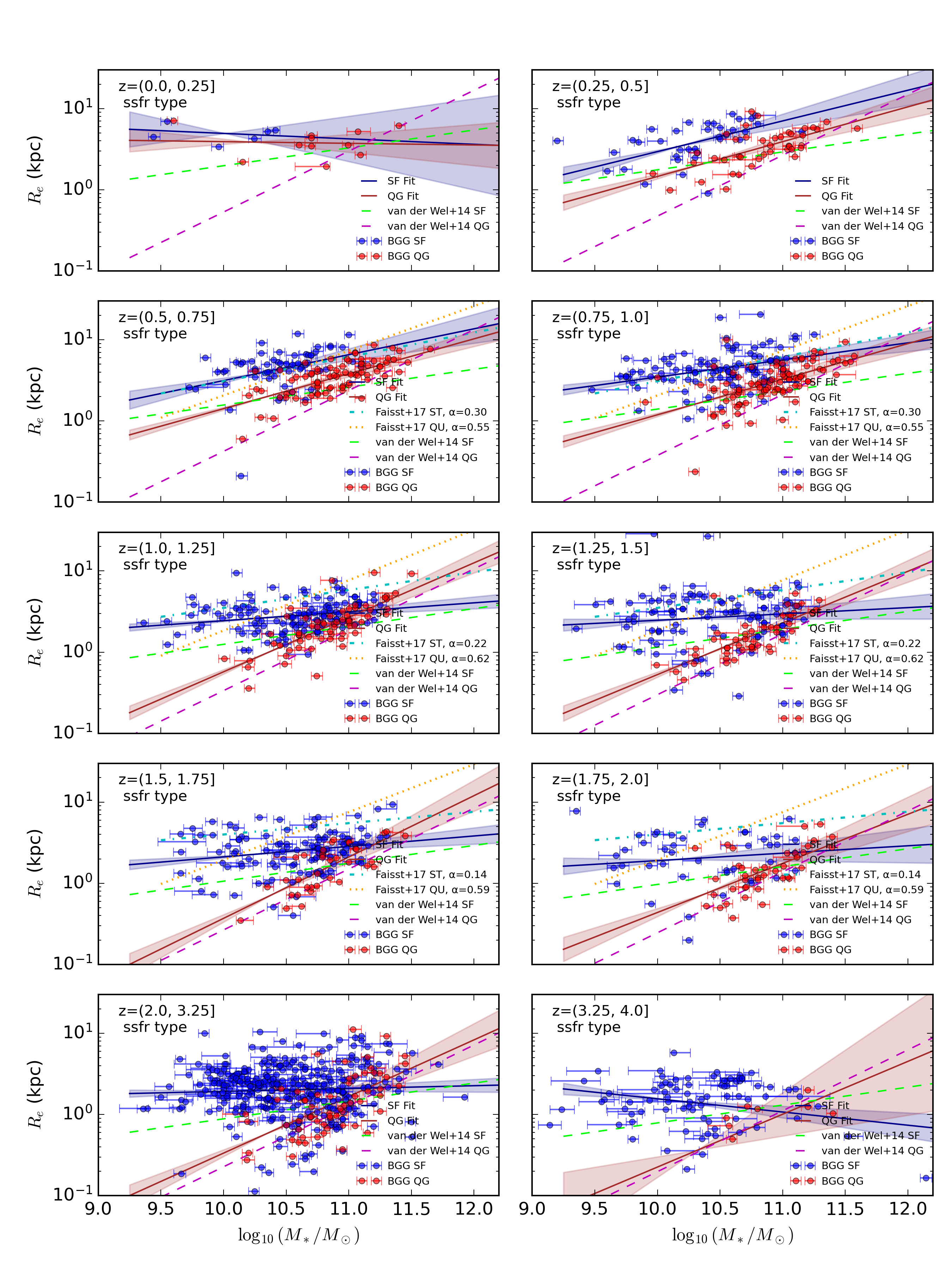}
    \caption{
        \textbf{Size--mass relation of BGGs using sSFR-only classification.}
        Each panel shows results for a redshift bin of width $\Delta z = 0.5$ spanning $z = 0$ to $z= 3.7$. Star-forming and quiescent BGGs are identified based on whether their $\log_{10}(\mathrm{sSFR})$ falls above or below the redshift-dependent threshold $\log_{10}(0.2 / t_{\mathrm{obs}})$, where $t_{\mathrm{obs}}$ is the age of the universe in Gyr. As in previous figures, dashed (SFGs) and solid (QGs) lines show best-fit power-law relations, and shaded regions represent 1$\sigma$ confidence intervals. The sSFR-based method yields similar trends but may misclassify dusty star-forming galaxies at intermediate redshift. Results from \citet{Faisst2017} are shown as cyan dashed-dotted (SFGs) and orange dotted (QGs) lines in matching redshift bins for comparison. While dashed lime and dashed magenta lines show \citet{vdW2014} results.
    }
    \label{fig:size_mass_ssfr_only}
\end{figure*}

To assess the impact of galaxy classification on the measured size--mass relations of BGGs, we present here the results obtained using two individual classification approaches: (i) color--color (NUV--$r$--$J$) selection and (ii) redshift-dependent sSFR thresholds. These are displayed in Figures~\ref{fig:size_mass_color_only} and \ref{fig:size_mass_ssfr_only}, respectively.

In both figures, we show the effective radius $R_e$ as a function of stellar mass for BGGs in eight redshift bins from $z = 0$ to $z= 3.7$. Blue circles denote star-forming galaxies, and red markers indicate quiescent systems. Dashed and solid lines represent the best-fit power-law size--mass relations for star-forming and quiescent BGGs, respectively. The shaded regions indicate 1$\sigma$ uncertainties derived from Bayesian posterior distributions.

While the overall trends are consistent with the consensus-based classification (Figure~\ref{fig:size_mass_consensus}), there are modest differences in slope and normalization, particularly at intermediate redshifts ($1 \lesssim z \lesssim 2$), where classification uncertainty and contamination are higher. The sSFR-based classification tends to yield slightly steeper slopes for quiescent systems, while color-based classification shows a higher scatter among star-forming galaxies.

These comparisons validate the robustness of our results and highlight the improved purity and interpretability of the consensus classification used in the main analysis.

\section{Size Distributions in two redshift bins} \label{app:size_twozbins}

To complement the detailed redshift-binned analysis presented in Section~\ref{sec:size_distributions}, we investigate the overall size distributions of BGGs using a simplified two-bin redshift division: \(0 < z \leq 1.5\) (low redshift) and \(1.5 < z \leq 3.7\) (high redshift). This approach allows us to boost the statistical power, particularly for QGs at high redshift, and to highlight the broad contrast between early and late cosmic epochs.

\begin{figure*}[h]
    \centering
    \includegraphics[width=0.85\linewidth]{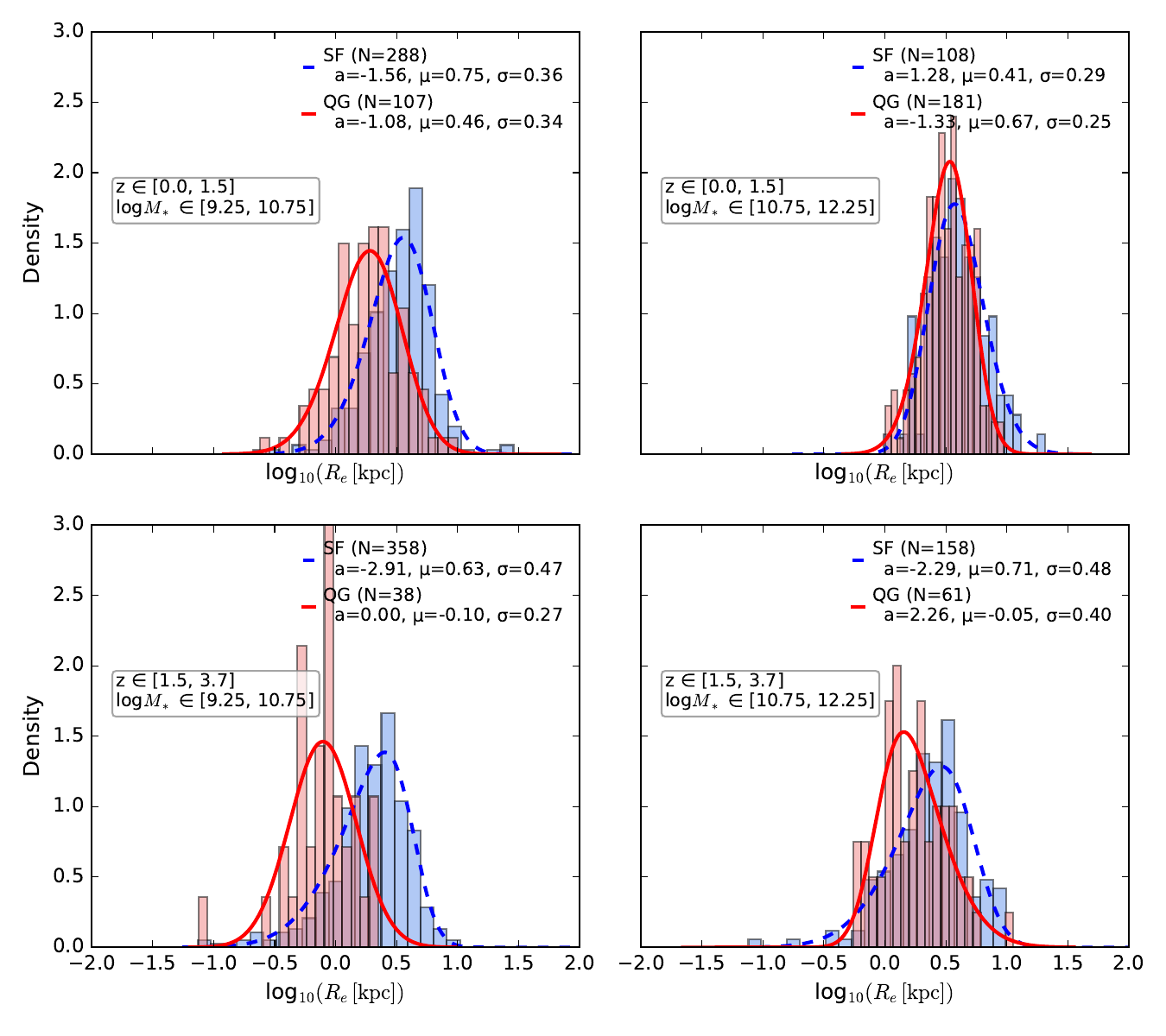}
    \caption{
        Distributions of \(\log_{10}(R_e/\mathrm{kpc})\) for BGGs in two stellar mass bins (\(9.25 \leq \log_{10}(M_\ast/M_\odot) < 10.75\) and \(10.75 \leq \log_{10}(M_\ast/M_\odot) < 12.25\)), split into two redshift bins (\(0 < z \leq 1.5\) and \(1.5 < z \leq 3.7\)). Blue and red histograms show SFG and QGs, respectively, with skew-normal fits overlaid (dashed for SFGs, solid for QGs). The legend reports the number of galaxies, best-fit skewness (\(a\)), mean (\(\mu\)), and standard deviation (\(\sigma\)) for each population. This overview highlights the increased structural diversity of SFGs and the persistently compact nature of QGs across cosmic time.
    }
    \label{fig:logRe_twozbins_massbins}
\end{figure*}

Figure~\ref{fig:logRe_twozbins_massbins} shows the distributions of \(\log_{10}(R_e/\mathrm{kpc})\) for BGGs split into two stellar mass bins (\(9.25 \leq \log_{10}(M_\ast/M_\odot) < 10.75\) and \(10.75 \leq \log_{10}(M_\ast/M_\odot) < 12.25\)). SFGs and QGs are shown in blue and red, respectively, with skew-normal fits overlaid. The legend in each panel reports the number of galaxies (\(N\)) and best-fit parameters for the skewness \(a\), mean \(\mu\), and standard deviation \(\sigma\).

At low redshift, both SFGs and QGs display relatively symmetric size distributions, with QGs typically more compact (lower \(\mu\)) and narrower (lower \(\sigma\)) than their star-forming counterparts. In contrast, the high-redshift bin reveals notable asymmetry and scatter, particularly among SFGs in the low-mass bin, where the size distribution is strongly negatively skewed (\(a \sim -2.9\)). QGs at high redshift remain small and relatively homogeneous, though their numbers are sparse, especially in the low-mass regime.

Overall, this two-bin analysis reinforces the results from the six-bin main analysis: BGG structural diversity increases at early epochs, with SFGs driving the broad and asymmetric size distributions, while QGs maintain compact, narrow profiles shaped by early quenching and passive evolution.

\end{document}